\PassOptionsToPackage{svgnames}{xcolor}
\documentclass[twocolumn]{aastex63}
\usepackage{fixfoot}
\usepackage[clockwise, figuresright]{rotating}
\usepackage{amsmath}
\usepackage{gensymb}
\usepackage{color,soul}
\usepackage{multirow}
\usepackage{IEEEtrantools}
\usepackage{bm}
\usepackage{float}
\usepackage{textcomp}
\usepackage{booktabs}
\usepackage{longtable}
\usepackage{csquotes}
\usepackage{morefloats}
\usepackage{appendix}
\accepted{2019 October 22, by The Astrophysical Journal}

\shorttitle{The $M_{BH}$--$\sigma$ and $L$--$\sigma$ Relations}

\shortauthors{Sahu, Graham, and Davis}

\begin{document}

\title{Revealing Hidden Substructures in the $M_{BH}$--$\sigma$ Diagram, and Refining the Bend in the $L$--$\sigma$ Relation}

\correspondingauthor{Nandini Sahu}
\email{nsahu@swin.edu.au}

\author[0000-0003-0234-6585]{Nandini Sahu}
\affil{OzGrav-Swinburne, Centre for Astrophysics and Supercomputing, Swinburne University of Technology, Hawthorn, VIC 3122, Australia}
\affil{Centre for Astrophysics and Supercomputing, Swinburne University of Technology, Hawthorn, VIC 3122, Australia}

\author[0000-0002-6496-9414]{Alister W. Graham}
\affil{Centre for Astrophysics and Supercomputing, Swinburne University of Technology, Hawthorn, VIC 3122, Australia}

\author[0000-0002-4306-5950]{Benjamin L. Davis}
\affil{Centre for Astrophysics and Supercomputing, Swinburne University of Technology, Hawthorn, VIC 3122, Australia}

\nocollaboration{3}


\begin{abstract}
Using 145 early- and late-type galaxies (ETGs and LTGs) with directly-measured super-massive black hole masses, $M_{BH}$, we build upon our previous discoveries that: (i) LTGs, most of which have been alleged to contain a pseudobulge, follow the relation 
$M_{BH}\propto\,M_{*,sph}^{2.16\pm0.32}$; and (ii) the ETG relation $M_{BH}\propto\,M_{*,sph}^{1.27\pm0.07}$ is an artifact of ETGs with/without disks following parallel $M_{BH}\propto\,M_{*,sph}^{1.9\pm0.2}$ relations which are offset by an order of
magnitude in the $M_{BH}$-direction. Here, we searched for substructure in the $M_{BH}$--(central velocity dispersion, $\sigma$)
diagram using our recently published, multi-component, galaxy decompositions; investigating divisions based on the presence of a
depleted stellar core (major dry-merger), a disk (minor wet/dry-merger, gas accretion), or a bar (evolved unstable disk). 
The S\'ersic and core-S\'ersic galaxies define two distinct relations: $M_{BH}\propto\sigma^{5.75\pm0.34}$ and $M_{BH}\propto\sigma^{8.64\pm1.10}$, with $\Delta_{rms|BH}=0.55$ and $0.46$~dex, respectively.  We also report on the consistency with the
slopes and bends in the galaxy luminosity ($L$)--$\sigma$ relation due to S\'ersic and core-S\'ersic ETGs, and
LTGs which all have S\'ersic light-profiles.  Two distinct relations (superficially) reappear in the $M_{BH}$--$\sigma$ diagram upon separating
galaxies with/without a disk (primarily for the ETG sample), while we find no significant offset between barred and non-barred galaxies, nor between galaxies with/without active galactic nuclei. We also address selection biases purported to affect the scaling relations for dynamically-measured $M_{BH}$ samples.  Our new, (morphological type)-dependent, $M_{BH}$--$\sigma$ relations more precisely estimate $M_{BH}$ in other galaxies, and hold implications for galaxy/black hole co-evolution theories, simulations, feedback, the pursuit of a black hole fundamental plane, and calibration of virial $f$-factors for reverberation-mapping.

\end{abstract}

\section{Introduction}

The  first observational works on the correlation between central black hole mass ($M_{BH}$) and the stellar velocity dispersion ($\sigma$)  of a galaxy \citep{Ferrarese:Merritt:2000, Gebhardt:2000} revealed a relation with little or no intrinsic scatter, suggesting that the $M_{BH}$--$\sigma$ relation could be the most fundamental of the black hole scaling relations. However, surprisingly, the slopes reported by the two studies were not in agreement and supported two competing feedback models between the super-massive black holes (SMBHs) and their host galaxies. \citet{Ferrarese:Merritt:2000} found $M_{BH} \propto \sigma^{4.80 \pm 0.50}$, which supported the prediction $M_{BH} \propto \sigma^{5}$ based on the energy-balancing feedback model of \citet{Silk:Rees:1998}. \citet{Gebhardt:2000} reported  $M_{BH} \propto \sigma^{3.75 \pm 0.30}$, supporting the feedback model of \citet{Fabian:1999} based upon momentum conservation, which predicted $M_{BH} \propto \sigma^{4}$.

\citet{Merritt:Ferrarese:2001} later revealed that \citet{Gebhardt:2000} had found a shallower slope due to the asymmetric linear regression routine that \citet{Gebhardt:2000} employed\footnote{\citet{Tremaine:ngc4742:2002} also used an asymmetric linear regression, ignoring the intrinsic scatter in the velocity dispersion direction \citep[see][his section titled \enquote{slippery slopes}]{Novak:2006, Graham:2016:Review}.}, plus  Gebhardt et al.'s relation was biased by the low-velocity dispersion which they had used for the Milky Way. \citet{Gebhardt:2000} had effectively solved the \enquote{Observer's Question} while \citet{Ferrarese:Merritt:2000} had effectively answered the \enquote{Theorist's Question,} as was later posed by \citet{Novak:2006}.
The reason behind obtaining almost zero intrinsic scatter in the $M_{BH}$--$\sigma$ relation was possibly the small sample size, or perhaps \citet{Ferrarese:Merritt:2000} had a \enquote{gold standard} of reliable black hole masses with well-resolved spheres-of-influence \citep{Ferrarese:Ford:2005}. Subsequent works on larger galaxy samples have found a non-zero intrinsic scatter.

With an increase in the number of barred galaxies with directly measured SMBH masses, some studies \citep{Graham:2007,Graham:2008:b, Graham:2008:a, Hu:2008} found that barred galaxies have a tendency to be offset, from the $M_{BH}$--$\sigma$ relation, towards higher $\sigma$ or lower $M_{BH}$, suggesting that the inclusion of barred galaxies may produce a steeper relation with larger scatter as warned by \citep{Graham:2011} and \citep{Graham:Scott:2013}. \citet{Hu:2008} claimed that the offset galaxies in their sample had \enquote{pseudo-bulges}\footnote{Pseudo-bulges are  difficult to identify \citep{Graham:2014}, and \citet{Graham:Grid:2019} explains why diagrams using  S\'ersic indices and \enquote{effective} half-light parameters cannot be used to  identify pseudo-bulges. Moreover, the range of diagnostics used to classify pseudo-bulges need to be subjectively applied \citep{Kormendy:Kennicutt:2004}, making it extremely problematic to distinguish pseudo-bulges from classical bulges. Furthermore, many galaxies contain both \citep{Erwin:Saglia:2015}. \label{foot_pseu}} with low-mass black holes, while according to \citet{Graham:2008:a}, the offset could be either because of a low black hole mass in pseudo-bulges or the elevated velocity dispersions in barred galaxies. Supporting the latter possibility, the simulation by \citet{Hartmann:2014} suggested that bars may cause increased velocity dispersion in galactic bulges whether they are classical or pseudo-bulges \citep[see also][]{Brown:Simulation:2013}. Interestingly, the recent observational work by \citet{Sahu:2019:I} found that barred galaxies are not offset in the black hole mass versus galaxy stellar mass ($M_{*, gal}$) diagram, nor in the black hole mass versus spheroid/bulge stellar mass ($M_{*,sph}$) diagram, eliminating under-massive black holes as the reason behind the apparent offset in the $M_{BH}$--$\sigma$ diagram and strengthening the prospect of barred galaxies having an increased velocity dispersion. However, as the number of barred galaxies in  \citet{Sahu:2019:I} is still quite small, this interpretation may require further confirmation.

In addition to the reported substructure in the $M_{BH}$--$\sigma$ diagram due to barred galaxies, some studies \citep[e.g.,][see their figure 5]{McConnell:Ma:2013, Bogdan:2018} have noticed that massive galaxies are offset towards the high-$M_{BH}$ side of their $M_{BH}$--$\sigma$ relation. These galaxies are mostly brightest cluster galaxies (BCGs) or central cluster galaxies (CCGs) which are considered to be a product of multiple dry mergers. 
Galaxies which have undergone dry mergers can have a deficit of light at their centers because the binary SMBHs formed from the two merging galaxies can scour out the stars from the center of the merged galaxy through the transfer of their orbital angular momentum \citep{Begelman:1980, Merritt:Milosavljevic:2005}. Such galaxies with a (partially) depleted core were discovered by \citet{King:Minkowski:1966, King:Minkowski:1972} and are referred to as core-S\'ersic  \citep{Graham:2003:CS} galaxies due to their flattened core relative to the inward extrapolation of their bulge's outer S\'ersic \citep{Sersic:1963} light profile. Galaxies which grow over time via gas-rich processes are likely to have bulges with S\'ersic light-profiles. 

Contrary to \citet{McConnell:Ma:2013}, the recent work by \citet{Savorgnan:Graham:2015} found that S\'ersic and core-S\'ersic galaxies broadly follow the same $M_{BH}$--$\sigma$ relation, and so was the case with slow and fast rotating galaxies in their sample. Thus, still,  debates over the substructures in the $M_{BH}$--$\sigma$ diagram due to barred and non-barred galaxies, S\'ersic and core-S\'ersic galaxies, and fast and slow rotating galaxies (galaxies with and without a rotating disk) persist. 

Using the hitherto largest sample of 145 galaxies, comprised of all early-type galaxies  (ETGs) and late-type galaxies (LTGs) with directly measured SMBH masses, our work investigates the underlying relationship between black hole mass and  central velocity dispersion for various sub-classes of the host galaxy. We classify these galaxies into S\'ersic, core-S\'ersic, barred, non-barred, and galaxies with and without a disk, based on our detailed multi-component decompositions (coupled with kinematical information) presented in \citet{Davis:2018:a} and \citet{Sahu:2019:I}, and also into galaxies with and without an Active Galactic Nucleus (AGN) identified using the catalog of \citet{AGN:Catalogue:2010}. 

We endeavor here to build upon our recent revelation that  ETGs superficially follow the relation $M_{BH} \propto M_{*,sph}^{1.27\pm0.07}$ \citep[][their Equation 10]{Sahu:2019:I}. We showed in \citet[][see their Figure 8]{Sahu:2019:I} that this single relation for ETGs is misleading because ETGs with and without a disk define two separate (parallel) $M_{BH} \propto M_{*,sph}^{1.9\pm0.2}$ relations which are offset by more than an order of magnitude (1.12 dex) in the $M_{BH}$-direction. This paradigm shifting discovery provided further impetus for us to re-examine old and search for new substructure in the $M_{BH}$-$\sigma$ diagram.

In order to provide a consistency check between the various scaling relations, this paper also establishes the galaxy luminosity ($L$)--$\sigma$ relation for our  ETG sample observed at $3.6\, \mu \rm m$, and for an updated V-band data-set of ETGs \citep{Lauer:Faber:2007}. We find a bend in the ETG $L$--$\sigma$ relation from both data-sets, which has been observed in  other bands  \citep[e.g.,][]{Matkovic:Guzman:2005, deRijcke:2005, Graham:Soria:2018}. Additionally, we explore the behavior of LTGs (spirals) with directly measured black hole masses in the $L$--$\sigma$ diagram. We mate these $L$--$\sigma$  relations with the $M_{BH}$--$L$ and  $M_{BH}$--$\sigma$ relations to investigate the consistency between the scaling relations.

Section \ref{Data} describes our data-set. In Section \ref{Results}, we briefly discuss the method of linear regression that we have used to establish our scaling relations, and the galaxy exclusions applied, along with the reasons for this. We further present the new $M_{BH}$--$\sigma$ relations that we have found for the various categories based on the morphological classes  mentioned above. This is accompanied by discussions on the behavior of the $M_{BH}$--$\sigma$ relation for each category.

In Section \ref{Consistency}, we  check on the internal consistency between our $M_{BH}$--$\sigma$ relations and the latest $M_{BH}$--$M_{*,gal}$ (and $M_{BH}$--$M_{*,sph}$) relations, while Section \ref{Modified Faber-Jackson} presents the bent $L$--$\sigma$ relations, based on different wavelength bands.
 Section \ref{selection bias} addresses a much-discussed selection bias regarding the spatial-resolution of the gravitational sphere-of-influence of the black holes, and  investigates the previously observed offset between galaxies with a dynamically measured black hole mass and galaxies without a dynamically measured black hole mass in the $L$--$\sigma$, or rather $\sigma$--$M_{*,gal}$, diagram \citep{Shankar:2016}.  This is followed by the main conclusions of our work summarized in Section \ref{conclusions} and a brief discussion on the implications of the new scaling relations.

\section{Data}
\label{Data}
We have identified 145 galaxies with directly measured super-massive black hole masses obtained from stellar dynamics, gas dynamics, kinematics of megamasers, proper motion, or recent direct imaging technique. This sample is comprised of 96 early-type and 49 late-type galaxies. Data for 84 ETGs came from \citet{Sahu:2019:I} and \citet{Savorgnan:2016:Slopes}. These 84 ETGs have been used in \citet{Sahu:2019:I} to establish the $M_{BH}$--$M_{*, sph}$ and $M_{BH}$--$M_{*, gal}$ relations for ETGs, based on the bulge and total galaxy stellar masses measured using state-of-the-art two dimensional (2D) isophotal modelling  \footnote{\citet{Davis:2018:a} and \citet{Sahu:2019:I} use \textsc{ISOFIT} \citep{Ciambur:2015:Ellipse} to generate a 2D model of each galaxy, and further use \textsc{Profiler} \citep{Ciambur:2016:Profiler}  to effectively realign the semi-major axis of each isophote. This 1D  surface brightness profile effectively encapsulates all of the key information about the galaxy structure and flux, including ellipticity gradients, position angle twists, and deviations from elliptical-shaped isophotes up to the  12th order Fourier harmonic coefficients. This major axis surface brightness profile is used for multi-component decomposition of the galaxy light. It should not be confused with a simple surface brightness profile obtained from a 1D cut of a galaxy image.}$^{,}$\footnote{\citet{Ciambur:2016:Profiler} provide a critical comparision between 1D and  2D decomposition techniques, concluding that multi-component galaxies may be easily modelled in 2D but gradients in the ellipticity, position angle, and structural perturbations are better captured in 1D. Furthermore, \citet{Savorgnan:Graham:2016:I} tried  both 1D  and 2D decompositions, and had more success using the 1D multi-component decomposition techniques.} and multi-component decompositions of  predominantly near infra-red (NIR) images. 
 
For the remaining 12 ETGs, data for two galaxies came from \citet{Nowak:2007} and \citet{Gultekin:2014}, who measured $M_{BH}$  using stellar dynamics. Another galaxy is taken from \citet{Hure:2011} with $M_{BH}$ measured using water masers, while the data for the remaining nine ETGs is taken from recent papers. Out of these nine, two ETGs are from \citet{Nguyen:2018} and six ETGs come from \citet{Thater:Krajnovic:2019}, where $M_{BH}$ is measured  using stellar dynamics. Data for the last  ETG is taken from \citet{Boizelle:2019} who measured $M_{BH}$ using gas dynamics.

Data for 44 of the 49 LTGs (spiral galaxies) is taken from \citet{Davis:2018:b} and \citet{Davis:2018:a}, where they also present the $M_{BH}$--$M_{*, sph}$,  $M_{BH}$--$M_{*, disk}$, and $M_{BH}$--$M_{*, gal}$ relations for spiral galaxies based on predominantly NIR imaging and multi-component decompositions.  Out of the remaining five LTGs, four are taken from \citet{Combes:2019:SMBH4}, and one from \citet{Nguyen:2019}, where the central SMBH masses have been measured using gas dynamics. 

Our galaxy sample is listed in Table \ref{Total Sample}, which includes information on the galaxy type, distance, updated morphology, presence of a bar, disk, depleted stellar core, AGN, $M_{BH}$, and the central stellar velocity dispersion. The morphologies reflect the presence, or not, of an intermediate or large-scale disk, and also bar, with types designated by the morphological galaxy classification grid given by \citet{Graham:Grid:2019}. 

The velocity dispersion has been measured in many ways in literature, for example: luminosity-weighted line-of-sight stellar velocity dispersion within one effective radius ($R_{e, sph}$) of the spheroid $\sigma_{e, sph}$ \citep[e.g.,][]{Gebhardt:2000};   luminosity-weighted line-of-sight stellar rotation and velocity dispersion (added in quadrature) within one effective radius of either the spheroid ($R_{e, sph}$) or the whole galaxy ($R_{e, gal}$) \citep{Gultekin:Richstone:2009}; or velocity dispersions within an aperture of radius equal to one-eighth \footnote{The velocity dispersion measurements available in Sloan Digital Sky Survey (SDSS) database use this aperture size.} of $R_{e, sph}$, $\sigma_{e/8}$ \citep[e.g.,][]{Ferrarese:Merritt:2000}.

It should be noted that the effective radius of the spheroid and the  effective radius of the whole galaxy are, in general, different quantities. Velocity dispersions measured using an aperture size equal to the effective radius of a galaxy is highly prone to contamination from the kinematics of the stellar disk in those galaxies with a (large-scale or intermediate-scale) disk. Whereas, studies \citep[e.g.,][]{Gultekin:2009} which use the luminosity-weighted average of both the stellar rotation and the velocity dispersion certainly represent a biased velocity dispersion. The use of the effective radius of the spheroid (bulge) as a scale of aperture size is also precarious as the measured velocity dispersion may also have contributions from the disk. Moreover, $R_{e, sph}$ does not have any physical significance, see \citet{Graham:Re:2019}  for a detailed study on  $R_{e, sph}$. 
The introduction of radii containing $50\%$ of the light reflects an arbitrary and physically meaningless percentage. The use of a different percentage, $x$, results in $R_e/R_x$ ratios that systematically change with luminosity, and in turn $\sigma_e/\sigma_x$ changes. There is nothing physically meaningful with $\sigma_e$, and $M_{BH}$--$\sigma_x$ relations are a function of the arbitrary percentage $x$.

\citet[][their Figure 1]{Bennert:2015} compare velocity dispersions based on different aperture sizes ($R_{e, sph}, R_{e,sph}/8, R_{e, gal}$) and conclude that different methods may produce velocity dispersion values different by up to $40 \%$. However, for most of their sample, the agreement between $\sigma_{SDSS}$ (aperture size $R_{e,sph}/8$) and  their $\sigma_{e, sph}$ (aperture size $R_{e, sph}$) values is much better than $40\%$. The radial variation of aperture velocity dispersions are a weak function of radius for ETGs, e.g., $\sigma_{R}= \sigma_{e}\times (R/R_e)^{-0.04}$ \citep{Jorgensen:1995},  and $\sigma_{R}= \sigma_{e}\times (R/R_e)^{-0.066}$ \citep{Cappellari:2006}. These empirical relations explain the reasonable agreement between $\sigma$ based on different apertures, however this might be true only for simple ETGs. Whereas for multi-component (barred-ETG, spiral) galaxies, $\sigma$ measurements are more complicated and large aperture sizes can introduce significant errors.

Given the inconsistency in the use of aperture size and contamination due to both disk rotation and velocity dispersion when using a large aperture size, we use the central velocity dispersion. Moreover, such data exists. The central velocity dispersions for the majority of our galaxies are taken from the \textsc{HyperLeda} database\footnote{\url{http://leda.univ-lyon1.fr/leda/param/vdis.html}} \citep{Paturel:2003}, as of October 2019. Galaxies for which we obtained velocity dispersions from other sources are indicated in Table \ref{Total Sample}. Velocity dispersions obtained from the \textsc{HyperLeda} database are homogenized for a uniform aperture size of $0.595 \,\rm h^{-1}\, kpc$. 
 
A source of error in the measured central velocity dispersions is broad line region (BLR) emission from AGNs and the movement of stars within the central black hole's sphere-of-influence. However, as our central velocity dispersions are based on an aperture size a few hundred times the typical radial extent of the  sphere-of-influence,  which is a few parsecs, the contamination in the luminosity-weighted velocity dispersion will be minimal. 

In the past, velocity dispersion observations have been obtained using long-slit spectroscopy.  Nowadays, we can get better measurements using integral field spectrographs equipped with Integral Field Units (IFUs), where a spatially resolved 2D spectrum gives an accurate measurement of  the stellar velocity dispersion of a galaxy. However, this measurement is not available for most of our galaxy sample; hence, we proceed with the central velocity dispersion measurements available on \textsc{HyperLeda}.   

For the majority of galaxies in our sample, the uncertainty in the velocity dispersion reported by \textsc{HyperLeda} is $\lesssim 10 \%$. Given that seeing and slit orientation can influence the measured velocity dispersion, we use a constant uncertainty of $10\%$,  whereas, for $M_{BH}$, we use the errors provided by the references, listed in Table \ref{Total Sample}. In addition, we check the robustness of our $M_{BH}$--$\sigma$ relations by using a $5\%$ to $15\%$ uncertainty on $\sigma$.

\startlongtable
\begin{deluxetable*}{llrcccccrcc}
\tablecolumns{11}
\tablecaption{Galaxy Sample}
\tabletypesize{\scriptsize}
\tablehead{
\colhead{Galaxy} &  \colhead{Type} &  \colhead{Distance} &  \colhead{Morph} &  \colhead{Bar} &  \colhead{Disk} &  \colhead{Core} & \colhead{AGN} & \colhead{$\log(M_{BH}/M_{\odot})$} &  \colhead{Source} &  \colhead{$\log(\sigma/\rm km\, s^{-1})$}  \\
\colhead{} & \colhead{} & \colhead{(Mpc)} & \colhead{} & \colhead{} & \colhead{} & \colhead{} & \colhead{} & \colhead{(dex)} & \colhead{} & \colhead{(dex)}\\
\colhead{(1)} & \colhead{(2)} & \colhead{(3)} & \colhead{(4)} & \colhead{(5)} & \colhead{(6)} & \colhead{(7)} & \colhead{(8)} & \colhead{(9)} & \colhead{(10)} & \colhead{(11)}
} 
\startdata
IC 1459 & ETG & 28.4 & E & no & no & yes & yes & 9.38 $\pm$ 0.20[S] & SG(2016) & 2.47 \\
NGC 0821 & ETG & 23.4 & E & no & no & no & no & 7.59 $\pm$ 0.17[S] & SG(2016) & 2.30 \\
NGC 1023 & ETG & 11.1& S0-bar & yes & yes & no & no & 7.62 $\pm$ 0.05[S] & SG(2016) & 2.29 \\
NGC 1316 & ETG & 18.6 & SAB0 (merger) & yes & yes & no & no & 8.18 $\pm$ 0.26[S] & SG(2016) & 2.35 \\
NGC 1332 & ETG & 22.3 & ES & no & yes & no & no & 9.16 $\pm$ 0.07[S] & SG(2016) & 2.47 \\
NGC 1399 & ETG & 19.4 & E & no & no & yes & no & 8.67 $\pm$ 0.06[S] & SG(2016) & 2.52 \\
NGC 2549 & ETG & 12.3 & S0 & yes & yes & no & no & 7.15 $\pm$ 0.60[S] & SG(2016) & 2.15 \\
NGC 2778 & ETG & 22.3 & S0 & yes & yes & no & no & 7.18 $\pm$ 0.34[S] & SG(2016) & 2.19 \\
NGC 3091 & ETG & 51.2  & E & no & no & yes & no & 9.56 $\pm$ 0.04[S] & SG(2016) & 2.49 \\
NGC 3115 & ETG & 9.4 & S0 & no & yes & no & no & 8.94 $\pm$ 0.25[S] & SG(2016) & 2.42 \\
NGC 3245 & ETG & 20.3 & S0 & yes & yes & no & no & 8.30 $\pm$ 0.12[G] & SG(2016) & 2.32 \\
NGC 3377 & ETG & 10.9 & E & no & no & no & no & 7.89 $\pm$ 0.04[S] & SG(2016) & 2.13 \\
NGC 3379 (M 105) & ETG & 10.3& E & no & no & yes & no & 8.60 $\pm$ 0.12[S] & SG(2016) & 2.31 \\
NGC 3384\tablenotemark{a} & ETG & 11.3 & S0 & yes & yes & no & no & 7.23 $\pm$ 0.05[S] & SG(2016) & 2.16 \\
NGC 3414 & ETG & 24.5 &  E & no & no & no & no & 8.38 $\pm$ 0.06[S] & SG(2016) & 2.38 \\
NGC 3489\tablenotemark{a} & ETG & 11.7 & S0 & yes & yes & no & no & 6.76 $\pm$ 0.07[S] & SG(2016) & 2.02 \\
NGC 3585 & ETG & 19.5 & E & no & no & no & no & 8.49 $\pm$ 0.13[S] & SG(2016) & 2.33 \\
NGC 3607 & ETG & 22.2 & E & no & no & no & yes & 8.11 $\pm$ 0.18[S] & SG(2016) & 2.35 \\
NGC 3608 & ETG & 22.3 & E & no & no & yes & no & 8.30 $\pm$ 0.18[S] & SG(2016) & 2.29 \\
NGC 3842 & ETG & 98.4 & E & no & no & yes & no & 9.99 $\pm$ 0.13[S] & SG(2016) & 2.49 \\
NGC 3998 & ETG & 13.7 & S0 & yes & yes & no & yes & 8.91 $\pm$ 0.11[S] & SG(2016) & 2.42 \\
NGC 4261 & ETG & 30.8 & E & no & no & yes & yes & 8.70 $\pm$ 0.09[S] & SG(2016) & 2.47 \\
NGC 4291 & ETG & 25.5 & E & no & no & yes & no & 8.52 $\pm$ 0.36[S] & SG(2016) & 2.47 \\
NGC 4374 (M 84) & ETG & 17.9 & E & no & no & yes & yes & 8.95 $\pm$ 0.05[S] & SG(2016) & 2.44 \\
NGC 4459 & ETG & 15.7 & S0 & no & yes & no & no & 7.83 $\pm$ 0.09[G] & SG(2016) & 2.24 \\
NGC 4472 (M 49) & ETG & 17.1 &E & no & no & yes & yes & 9.40 $\pm$ 0.05[S] & SG(2016) & 2.45 \\
NGC 4473 & ETG & 15.3 & E & no & no & no & no & 8.08 $\pm$ 0.36[S] & SG(2016) & 2.25 \\
NGC 4486 (M 87) & ETG & 16.8 & E & no & no & yes & yes & 9.81$\pm$0.05[DI]\tablenotemark{b}& SG(2016) & 2.51 \\
NGC 4564 & ETG & 14.6 & S0 & no & yes & no & no & 7.78 $\pm$ 0.06[S] & SG(2016) & 2.19 \\
NGC 4596 & ETG & 17.0 & S0 & yes & yes & no & no & 7.90 $\pm$ 0.20[G] & SG(2016) & 2.15 \\
NGC 4621 (M 59) & ETG & 17.8 & E & no & no & no & no & 8.59 $\pm$ 0.05[S] & SG(2016) & 2.36 \\
NGC 4697 & ETG & 11.4 & E & no & no & no & no & 8.26 $\pm$ 0.05[S] & SG(2016) & 2.22 \\
NGC 4889 & ETG & 103.2 & E & no & no & yes & no & 10.32 $\pm$ 0.44[S] & SG(2016) & 2.59 \\
NGC 5077 & ETG & 41.2 & E & no & no & yes & yes & 8.87 $\pm$ 0.22[G] & SG(2016) & 2.40 \\
NGC 5128 & ETG & 3.8 & S0 (merger) & no & yes & no & no & 7.65$\pm$0.13[SG] & SG(2016) & 2.01 \\
NGC 5576 & ETG & 24.8 & E & no & no & no & no & 8.20 $\pm$ 0.10[S] & SG(2016) & 2.26 \\
NGC 5846 & ETG & 24.2 & E & no & no & yes & no & 9.04 $\pm$ 0.05[S] & SG(2016) & 2.38 \\
NGC 6251 & ETG & 104.6 & E & no & no & yes & yes & 8.77 $\pm$ 0.16[G] & SG(2016) & 2.49 \\
NGC 7619 & ETG & 51.5 & E & no & no & yes & no & 9.40 $\pm$ 0.09[S] & SG(2016) & 2.50 \\
NGC 7768 & ETG & 112.8 & E & no & no & yes & no & 9.11 $\pm$ 0.15[S] & SG(2016) & 2.46 \\
NGC 1271 & ETG & 80.0 & ES & no & yes & no & no & 9.48 $\pm$ 0.16[S] & GCS(2016) & 2.44 [11a] \\
NGC 1277 & ETG & 72.5 & ES & no & yes & no & no & 9.08 $\pm$ 0.12[S] & G+7(2016) & 2.48 [11b]\\
A1836 BCG & ETG & 158.0  & E & no & no & yes & no & 9.59 $\pm$ 0.06[G] & SGD(2019) & 2.49 [11c] \\
A3565 BCG (IC 4296) & ETG & 40.7 & E & no & no & no & yes & 9.04 $\pm$ 0.09[G] & SGD(2019) & 2.52 \\
Mrk 1216 & ETG & 94.0 & S0 & no & yes & no & yes & 9.69 $\pm$ 0.16[S] & SGD(2019) & 2.51 \\
NGC 0307 & ETG & 52.8 & SAB0 & yes & yes & no & no & 8.34 $\pm$ 0.13[S] & SGD(2019) & 2.43 \\
NGC 0404 & ETG & 3.1 & S0 & no & yes & no & no & 4.85 $\pm$ 0.13[S] & SGD(2019) & 1.54 \\
NGC 0524 & ETG & 23.3 & SA0(rs) & no & yes & yes & no & 8.92 $\pm$ 0.10[S] & SGD(2019) & 2.37 \\
NGC 1194 & ETG & 53.2 & S0 (merger?) & no & yes & no & yes & 7.81 $\pm$ 0.04[M] & SGD(2019) & 2.17 [11d]\\
NGC 1275 & ETG & 72.9 & E & no & no & no & yes & 8.90 $\pm$ 0.20[G] & SGD(2019) & 2.39 \\
NGC 1374 & ETG & 19.2 & S0 & no & yes & no & no & 8.76 $\pm$ 0.05[S] & SGD(2019) & 2.25 \\
NGC 1407 & ETG & 28.0 & E & no & no & yes & no & 9.65 $\pm$ 0.08[S] & SGD(2019) & 2.42 \\
NGC 1550 & ETG & 51.6 & E & no & no & yes & no & 9.57 $\pm$ 0.06[S] & SGD(2019) & 2.48 \\
NGC 1600 & ETG & 64.0 & E & no & no & yes & no & 10.23 $\pm$ 0.05[S] & SGD(2019) & 2.52 \\
NGC 2787 & ETG\tablenotemark{a} & 7.3 & SB0(r) & yes & yes & no & yes & 7.60 $\pm$ 0.06[G] & SGD(2019) & 2.28 \\
NGC 3665 & ETG & 34.7 & S0 & no & yes & no & no & 8.76 $\pm$ 0.10[G] & SGD(2019) & 2.33 \\
NGC 3923 & ETG & 20.9 & E & no & no & yes & no & 9.45 $\pm$ 0.13[S] & SGD(2019) & 2.39 \\
NGC 4026 & ETG & 13.2 & SB0 & yes & yes & no & no & 8.26 $\pm$ 0.11[S] & SGD(2019) & 2.24 \\
NGC 4339 & ETG & 16.0 & S0 & no & yes & no & no & 7.63 $\pm$ 0.33[S] & SGD(2019) & 2.05 \\
NGC 4342 & ETG & 23.0 & ES & no & yes & no & no & 8.65 $\pm$ 0.18[S] & SGD(2019) & 2.38 \\
NGC 4350 & ETG & 16.8 & EBS & yes & yes & no & no & 8.86 $\pm$ 0.41[SG] & SGD(2019) & 2.26 \\
NGC 4371\tablenotemark{a} & ETG & 16.9 & SB(r)0 & yes & yes & no & no & 6.85 $\pm$ 0.08[S] & SGD(2019) & 2.11 \\
NGC 4429 & ETG & 16.5 & SB(r)0 & yes & yes & no & no & 8.18 $\pm$ 0.09[G] & SGD(2019) & 2.24 \\
NGC 4434 & ETG & 22.4 & S0 & no & yes & no & no & 7.85 $\pm$ 0.17[S] & SGD(2019) & 2.07 \\
NGC 4486B & ETG & 15.3 & E & no & no & no & no & 8.76 $\pm$ 0.24[S] & SGD(2019) & 2.22 \\
NGC 4526 & ETG & 16.9 & S0 & no & yes & no & no & 8.67 $\pm$ 0.05[G] & SGD(2019) & 2.35 \\
NGC 4552 & ETG & 14.9 & E & no & no & no & yes & 8.67 $\pm$ 0.05[S] & SGD(2019) & 2.40 \\
NGC 4578 & ETG & 16.3 & S0( r) & no & yes & no & no & 7.28 $\pm$ 0.35[S] & SGD(2019) & 2.05 \\
NGC 4649 & ETG & 16.4 & E & no & no & yes & no & 9.67 $\pm$ 0.10[S] & SGD(2019) & 2.52 \\
NGC 4742 & ETG & 15.5 & S0 & no & yes & no & no & 7.15 $\pm$ 0.18[S] & SGD(2019) & 2.01 \\
NGC 4751 & ETG & 26.9 & S0 & no & yes & yes & no & 9.15 $\pm$ 0.05[S] & SGD(2019) & 2.54 \\
NGC 4762 & ETG & 22.6 & SB0 & yes & yes & no & no & 7.36 $\pm$ 0.15[S] & SGD(2019) & 2.15 \\
NGC 5018 & ETG & 40.6 & S0 (merger) & no & yes & no & no & 8.02 $\pm$ 0.09[S] & SGD(2019) & 2.33 \\
NGC 5252 & ETG & 96.8 & S0 & no & yes & no & yes & 9.00 $\pm$ 0.40[G] & SGD(2019) & 2.27 \\
NGC 5328 & ETG & 64.1 & E & no & no & yes & no & 9.67 $\pm$ 0.15[S] & SGD(2019) & 2.50 \\
NGC 5419 & ETG & 56.2 & E & no & no & yes & no & 9.86 $\pm$ 0.14[S] & SGD(2019) & 2.54 \\
NGC 5516 & ETG & 58.4 & E & no & no & yes & no & 9.52 $\pm$ 0.06[S] & SGD(2019) & 2.49 \\
NGC 5813 & ETG & 31.3 & S0 & no & yes & yes & no & 8.83 $\pm$ 0.06[S] & SGD(2019) & 2.37 \\
NGC 5845 & ETG & 25.2 & ES & no & yes & no & no & 8.41 $\pm$ 0.22[S] & SGD(2019) & 2.36 \\
NGC 6086 & ETG & 138.0 & E & no & no & yes & no & 9.57 $\pm$ 0.17[S] & SGD(2019) & 2.51 \\
NGC 6861 & ETG & 27.3 & ES & no & yes & no & no & 9.30 $\pm$ 0.08[S] & SGD(2019) & 2.59 \\
NGC 7052 & ETG & 66.4 & E & no & no & yes & no & 8.57 $\pm$ 0.23[G] & SGD(2019) & 2.45 \\
NGC 7332 & ETG & 24.9 & SB0 & yes & yes & no & no & 7.11 $\pm$ 0.20[S] & SGD(2019) & 2.11 \\
NGC 7457 & ETG & 14.0 & S0 & no & yes & no & no & 7.00 $\pm$ 0.30[S] & SGD(2019) & 1.83 \\
NGC 4486A & ETG & 13.9 & E & no & no & no & no & 7.10 $\pm$ 0.32[S] & No+7(2007) & 2.12 \\
NGC 5102 & ETG & 3.2 & S0 & no & yes & no & no & 5.94 $\pm$ 0.38[S] & Ngu+10(2018) & 1.79 \\
NGC 5206 & ETG & 3.5 & dE/dS0 & no & no? & no & no & 5.67 $\pm$ 0.36[S] & Ngu+10(2018) & 1.62 \\
NGC 0584 & ETG & 19.1 & S0 & no & yes & yes & no & 8.11 $\pm$ 0.18[S] & Th+6(2019) & 2.33 [11e]\\
NGC 2784 & ETG & 9.6 & S0 & no & yes & no & no & 8.00 $\pm$ 0.31[S] & Th+6(2019) & 2.39 [11e]\\
NGC 3640 & ETG & 26.3 & E & no & no & yes & no & 7.89 $\pm$ 0.34[S] & Th+6(2019) & 2.24 [11e]\\
NGC 4281 & ETG & 24.4 & S0 & no & yes & no & no & 8.73 $\pm$ 0.08[S] & Th+6(2019) & 2.50 [11e]\\
NGC 4570 & ETG & 17.1 & S0 & no & yes & no & no & 7.83 $\pm$ 0.14[S] & Th+6(2019) & 2.32 [11e]\\
NGC 7049 & ETG & 29.9 & S0 & no & yes & no & no & 8.51 $\pm$ 0.12[S] & Th+6(2019) & 2.42 [11e]\\
NGC 3258 & ETG & 31.3 & E  & no & no & yes & no &  9.35 $\pm$ 0.05[G] & Bo+7(2019) & 2.41 \\
IC 1481 & ETG & 89.9 & E? (merger) & ... & ... & ... & ... & 7.15 $\pm$ 0.13[S] & Hu+4(2011) & ... \\
NGC 3706 & ETG& 46 & S0 & no & yes & yes & no & 8.78 $\pm$ 0.06[S] &Gu+6(2014) & 2.41 \\
Circinus\tablenotemark{a} & LTG & 4.2 & SABb & no & yes & no & yes & 6.25 $\pm$ 0.11[M] & DGC(2019) & 2.17 \\
Cygnus A & LTG & 258.8 & S & no & yes & no & yes & 9.44 $\pm$ 0.13[G] & DGC(2019) & 2.43 [11f] \\
ESO558-G009\tablenotemark{a} & LTG & 115.4 & Sbc & no & yes & no & no & 7.26 $\pm$ 0.04[M] & DGC(2019) & 2.23 [11g] \\
IC 2560\tablenotemark{a} & LTG & 31.0 & SBb & yes & yes & no & yes & 6.49 $\pm$ 0.20[M] & DGC(2019) & 2.14 \\
J0437+2456\tablenotemark{a} & LTG & 72.8 & SB & yes & yes & no & no & 6.51 $\pm$ 0.05[M] & DGC(2019) & 2.04 [11g]\\
Milky Way\tablenotemark{a} & LTG & 7.9 & SBbc & yes & yes & no & no & 6.60 $\pm$ 0.02[P] & DGC(2019) & 2.02 [11f]\\
Mrk 1029\tablenotemark{a} & LTG & 136.9 & S & no & yes & no & no & 6.33 $\pm$ 0.12[M] & DGC(2019) & 2.12 [11g]\\
NGC 0224 & LTG & 0.8 & SBb & yes & yes & no & no & 8.15 $\pm$ 0.16[S] & DGC(2019) & 2.19 \\
NGC 0253\tablenotemark{a} & LTG & 3.5 & SABc & yes & yes & no & no & 7.00 $\pm$ 0.30[G] & DGC(2019) & 1.98 \\
NGC 1068\tablenotemark{a} & LTG & 10.1 & SBb & yes & yes & no & yes & 6.75 $\pm$ 0.08[M] & DGC(2019) & 2.21 \\
NGC 1097\tablenotemark{a} & LTG & 24.9 & SBb & yes & yes & no & yes & 8.38 $\pm$ 0.04[G] & DGC(2019) & 2.29 [11h]\\
NGC 1300\tablenotemark{a} & LTG & 14.5 & SBbc & yes & yes & no & no & 7.71 $\pm$ 0.16[G] & DGC(2019) & 2.34 \\
NGC 1320\tablenotemark{a} & LTG & 37.7 & Sa & no & yes & no & no & 6.78 $\pm$ 0.29[M] & DGC(2019) & 2.04 \\
NGC 1398 & LTG & 24.8 & SBab & yes & yes & no & no & 8.03 $\pm$ 0.11[S] & DGC(2019) & 2.29 \\
NGC 2273\tablenotemark{a} & LTG & 31.6 & SBa & yes & yes & no & no & 6.97 $\pm$ 0.09[M] & DGC(2019) & 2.15 \\
NGC 2748\tablenotemark{a} & LTG & 18.2 & Sbc & no & yes & no & no & 7.54 $\pm$ 0.21[G] & DGC(2019) & 1.98 \\
NGC 2960\tablenotemark{a} & LTG& 71.1 & Sa (merger) & no & yes & no & no & 7.06 $\pm$ 0.17[M] & DGC(2019) & 2.22 [11i]\\
NGC 2974 & LTG & 21.5 & SB & yes & yes & no & yes & 8.23 $\pm$ 0.07[S] & DGC(2019) & 2.37 \\
NGC 3031 & LTG & 3.5 & SABab & no & yes & no & no & 7.83 $\pm$ 0.09[G] & DGC(2019) & 2.18 \\
NGC 3079\tablenotemark{a} & LTG & 16.5 & SBcd & yes & yes & no & yes & 6.38 $\pm$ 0.12[M] & DGC(2019) & 2.24 \\
NGC 3227\tablenotemark{a} & LTG & 21.1 & SABa & yes & yes & no & yes & 7.88 $\pm$ 0.14[SG] & DGC(2019) & 2.10 \\
NGC 3368\tablenotemark{a} & LTG & 10.7 & SABa & yes & yes & no & no & 6.89 $\pm$ 0.09[SG] & DGC(2019) & 2.07 \\
NGC 3393\tablenotemark{a} & LTG & 55.8 & SBa & yes & yes & no & yes & 7.49 $\pm$ 0.05[M] & DGC(2019) & 2.30 \\
NGC 3627\tablenotemark{a} & LTG & 10.6 & SBb & yes & yes & no & yes & 6.95 $\pm$ 0.05[S] & DGC(2019) & 2.10 \\
NGC 4151 & LTG & 19.0 & SABa & yes & yes & no & yes & 7.68 $\pm$ 0.37[SG] & DGC(2019) & 1.96 \\
NGC 4258 & LTG & 7.6 & SABb & yes & yes & no & yes & 7.60 $\pm$ 0.01[M] & DGC(2019) & 2.12 \\
NGC 4303\tablenotemark{a} & LTG & 12.3 & SBbc & yes & yes & no & yes & 6.58 $\pm$ 0.17[G] & DGC(2019) & 1.98 \\
NGC 4388\tablenotemark{a} & LTG & 17.8 & SBcd & yes & yes & no & yes & 6.90 $\pm$ 0.11[M] & DGC(2019) & 2.00 \\
NGC 4395 & LTG & 4.8 & SBm & yes & yes & no & yes & 5.64 $\pm$ 0.17[G] & DGC(2019) & 1.42 \\
NGC 4501\tablenotemark{a} & LTG & 11.2 & Sb & no & yes & no & yes & 7.13 $\pm$ 0.08[S] & DGC(2019) & 2.22 \\
NGC 4594 & LTG & 9.6 & Sa & no & yes & no & yes & 8.81 $\pm$ 0.03[S] & DGC(2019) & 2.35 \\
NGC 4699\tablenotemark{a} & LTG & 23.7 & SABb & yes & yes & no & no & 8.34 $\pm$ 0.10[S] & DGC(2019) & 2.28 \\
NGC 4736\tablenotemark{a} & LTG & 4.4 & SABab & no & yes & no & yes & 6.78 $\pm$ 0.10[S] & DGC(2019) & 2.03 \\
NGC 4826\tablenotemark{a} & LTG & 5.6 & Sab & no & yes & no & yes & 6.07 $\pm$ 0.15[S] & DGC(2019) & 1.99 \\
NGC 4945\tablenotemark{a} & LTG & 3.7 & SABc & no & yes & no & yes & 6.15 $\pm$ 0.30[M] & DGC(2019) & 2.07 \\
NGC 5055\tablenotemark{a} & LTG & 8.9 & Sbc & no & yes & no & no & 8.94 $\pm$ 0.10[G] & DGC(2019) & 2.00 \\
NGC 5495\tablenotemark{a} & LTG & 101.1 & SBc & yes & yes & no & no & 7.04 $\pm$ 0.08[M] & DGC(2019) & 2.22 [11g]\\
NGC 5765b\tablenotemark{a} & LTG & 133.9 & SABb & yes & yes & no & no & 7.72 $\pm$ 0.05[M] & DGC(2019) & 2.21 [11g]\\
NGC 6264\tablenotemark{a} & LTG & 153.9 & SBb & yes & yes & no & yes & 7.51 $\pm$ 0.06[M] & DGC(2019) & 2.20 [11f]\\
NGC 6323\tablenotemark{a} & LTG & 116.9 & SBab & yes & yes & no & no & 7.02 $\pm$ 0.14[M] & DGC(2019) & 2.20 [11f]\\
NGC 6926\tablenotemark{a} & LTG & 86.6 & SBc & yes & yes & no & yes & 7.74 $\pm$ 0.50[M] & DGC(2019) & ...\\
NGC 7582\tablenotemark{a} & LTG & 19.9 & SBab & yes & yes & no & yes & 7.67 $\pm$ 0.09[G] & DGC(2019) & 2.07 \\
UGC 3789\tablenotemark{a} & LTG & 49.6 & SABa & yes & yes & no & no & 7.06 $\pm$ 0.05[M] & DGC(2019) & 2.03 [11f]\\
UGC 6093\tablenotemark{a} & LTG & 152.8 & SBbc & yes & yes & no & no & 7.41 $\pm$ 0.03[M] & DGC(2019) & 2.19 [11f]\\
NGC 0613\tablenotemark{a} & LTG & 17.2 & SB(rs)bc & yes & yes & no & no & 7.57 $\pm$ 0.15[G] & Co+14(2019) & 2.09 \\
NGC 1365\tablenotemark{a} & LTG & 17.8 & SB(s)b & yes & yes & no & yes & 6.60 $\pm$ 0.30[G] & Co+14(2019) & 2.15 \\
NGC 1566\tablenotemark{a} & LTG & 7.2 & SAB(s)bc & yes & yes & no & yes & 6.83 $\pm$ 0.30[G] & Co+14(2019) & 1.99 \\
NGC 1672\tablenotemark{a} & LTG & 11.4 & SB(s)b & yes & yes & no & yes & 7.70 $\pm$ 0.10[G]& Co+14(2019) & 2.04 \\
NGC 3504 & LTG & 13.6 & SABab & yes & yes & no & yes & 7.01 $\pm$ 0.07[G] & Ngu+10(2019) & 2.08
\label{Total Sample}
\enddata
\tablecomments{Column: (1) Galaxy name. 
(2) Galaxy type: early-type or late-type.
(3) Distance to the galaxy.
(4) Galaxy Morphology.
(5) Presence of bar.
(6) Presence of a rotating intermediate-scale (ES) or large-scale (S0/Sp) disk.
(7) Presence of a depleted  stellar core.
(8) Presence of active galactic nucleus.
(9) Directly measured black hole mass along with measurement method indicated by [P]= proper motion, [S]= stellar-dynamical modeling, [G]= gas dynamical modeling, [SG]= stellar and gas dynamical modeling, [M]= megamaser kinematics, and [DI]= direct imaging.  
(10) Catalog references, where the information for column (2) to column (9) comes from SG(2016)= \citet{Savorgnan:2016:Slopes}, GCS(2016)= \citet{Graham:Ciambur:Savorgnan:2016}, G+7(2016)= \citet{Graham:Durr:Savorgnan:2016}, SGD(2019)= \citet{Sahu:2019:I}, No+7(2007)= \citep{Nowak:2007}, Ngu+10(2018)= \citet{Nguyen:2018}, Th+6(2019)= \citet{Thater:Krajnovic:2019}, Bo+7(2019)= \citet{Boizelle:2019}, Hu+4(2011)= \citet{Hure:2011}, Gu+6(2014)= \citet{Gultekin:2014}, DGC(2019)= \citet{Davis:2018:a}, Co+14(2019)= \citet{Combes:2019:SMBH4}, and Ngu+10(2019)=\citet{Nguyen:2019}. 
(11) Central velocity dispersion of galaxies mostly archived in \textsc{Hyperleda} \citep{Paturel:2003} unless otherwise specified: [11a]= \citet{Walsh:2015}; [11b]= \citet{Graham:Durr:Savorgnan:2016}, [11c]= \citet{DallaBonta:2009}; [11d]=\citet{Greene:2010}; [11e]=\citet{Thater:Krajnovic:2019}; [11f]=\citet{Kormendy:Ho:2013}; [11g]= \citet{Greene:2016}, [11h]= \citet{van_den_Bosch:2016}; [11i]= \citet{Saglia:2016}. Bulge and galaxy stellar masses can also be found in \citet{Savorgnan:Graham:2016:I}; \citet{Davis:2018:a}; and \citet{Sahu:2019:I}.}
\tablenotetext{a}{Alleged to host a pseudo-bulge according to \citet{Kormendy:Ho:2013}, \citet{Saglia:2016}, and the references mentioned in Table 1 of \citet{Davis:Graham:2017}. NGC~0613, NGC~1365, NGC~1566, and NGC~1672 are claimed to have pseudo-bulges by \citet{Combes:2019:SMBH4}.}
\tablenotetext{b}{Latest black hole mass measurement from the Event Horizon Telescope Collaboration through direct imaging \citep{EHT:M87:2019}.}
\end{deluxetable*}

\section{$M_{BH}$--$\sigma$ Relations}
\label{Results}
 
In this work, we use both the \textsc{BCES}\footnote{The BCES routine was used via the PYTHON
module written by Rodrigo Nemmen \citep{Nemmen:2012}, which is available
at \url{https://github.com/rsnemmen/BCES}.} \citep{Akritas:Bershady:1996} routine and the bisector line from the modified \textsc{FITEXY} \citep{Press:1992}  routine  \citep[\textsc{MPFITEXY},][]{Tremaine:ngc4742:2002, Novak:2006, Bedregal:2006, Williams:2010, Markwardt:2012}  to establish the $M$--$\sigma$ relations. Both the   \textsc{BCES} and \textsc{MPFITEXY}  regression routines take into account the measurement errors in the  X and Y coordinates and allow for intrinsic scatter in the data.
 
The \textsc{BCES} routine directly provides the forward regression  \textsc{BCES($Y | X$)} line, the inverse regression \textsc{BCES($X | Y$)} line, and the regression line which symmetrically bisects the two, i.e., \textsc{BCES(Bisector)\footnote{\textsc{BCES($Y | X$)} minimizes the offsets in  the Y-direction, and \textsc{BCES($X | Y$)} minimizes, the offsets in the X-direction. }}. 
However, to obtain a symmetrical treatment (\textsc{MPFITEXY(bisector)}) of the data with the \textsc{MPFITEXY} routine requires averaging the inclination of the best-fit lines obtained from the forward (\textsc{MPFITEXY($Y | X$)}) and inverse (\textsc{MPFITEXY($X | Y$)}) regressions as explained in \citet{Novak:2006}.   

We prefer the symmetric (bisector) regressions obtained from both the routines because we do not know whether the central SMBH mass fundamentally governs the central velocity dispersion of a galaxy or vice-versa, or indirectly through a third parameter. A symmetrical regression is also preferable for theoretical grounds, see \citet{Novak:2006}.

In our plots, we show the \textsc{BCES(Bisector)} regression line. These are also presented in Table \ref{fit parameters}. In addition, asymmetric (\textsc{BCES($Y | X$)} and \textsc{BCES($X | Y$)}) regression parameters are also provided in the Appendix (Table \ref{Extra_fit_parameters}). We do not provide the \textsc{MPFITEXY}  parameters for our relations as these were found to always be consistent with the parameters obtained from the \textsc{BCES} routine within the $\pm 1 \sigma$ confidence limits.

\subsection{Galaxy Exclusions}
\label{exclusion}
We identify and exclude the following eight galaxies which may bias the $M_{BH}$--$\sigma$ relation: NGC~404; NGC~5102; NGC~5206; NGC~7457; IC~1481; NGC~4395; NGC~5055; and NGC~6926; where the last three galaxies are LTGs.  

NGC~404 is the only galaxy anchoring the intermediate black hole mass end ($\lesssim 10^5 M_{\odot}$) of the relation, as such it may bias the best-fit line. Additionally, as we will see,  NGC~404, NGC~5102, and NGC~5206, for whom we obtained black hole masses from the same group \citep{Nguyen:2017, Nguyen:2018}, all  seem to lie above the $M_{BH}$--$\sigma$ relation defined by the remaining galaxies. As we have only a four galaxies (NGC~404, NGC~5102, NGC~5206, and NGC~4395) with $M_{BH} \lesssim 10^6 M_{\odot}$ (as can be seen in Figure \ref{ETG86_LTG44} and further in the left-hand panel of Figure \ref{ETG_LTG_130}), we do not include them in our primary regressions. As noted above, this also helps us detect possible departures at the low-mass end.

NGC~7457 has an unusually low-velocity dispersion, possibly because of a counter-rotating core \citep{Molaeinezhad:2019}, which makes it fall beyond the $\pm 2 \sigma$ scatter bounds of our single regression relation. Similarly, NGC~4395 and NGC 5055 have  lower velocity-dispersion values than expected from the $M_{BH}$--$\sigma$ relation defined by the bulk of the sample, which makes them stand out from the, soon to be seen, best-fit lines. These three (NGC~7457, NGC~4395, and NGC 5055) outlying galaxies significantly affect our best-fit lines; hence we exclude them from our regressions in order to obtain more stable relations reflective of the majority of the population. 

For IC~1481 and NGC~6926, we do not have a reliable measurement of their central velocity dispersion.  We have also provided regression parameters including all excluded galaxies (except IC~1481 and NGC~6926) in Table \ref{Total_sample_para} of the Appendix to show how much these few galaxies bias our best-fit lines.
Overall, we exclude a total of 8 galaxies, which leaves us with a reduced sample of 137 galaxies.  

In our reduced sample, five galaxies (NGC~1316, NGC~2960, NGC~5128, NGC~5018, and NGC~1194) are mergers identified by \citet[][their section 6.4 ]{Kormendy:Ho:2013}, \citet{Saglia:2016}, and  \citet[][see the light profile of NGC~1194 and references]{Sahu:2019:I}. A merger designation  refers to the stage when a galaxy is yet to reach a relaxed (stable) post merger configuration. \citet{Kormendy:Ho:2013} suggest excluding mergers from the black hole scaling relations as they may bias the results. However, given the small number of mergers in our sample, and given that they are not (significant) outliers in the $M_{BH}$--$\sigma$ relations,  we include them. 

Additionally, NGC~4342 \citep{Blom:Forbes:2014} and NGC~4486B \citep{Batcheldor:2010:b} are tidally stripped of their  stellar mass by the gravitational pull of their massive companion galaxies NGC~4365  and NGC~4486 (M87), respectively. However, stripping of the outer stellar mass should not considerably affect the central stellar velocity dispersions, hence we also  include these galaxies in our  $M_{BH}$--$\sigma$ relations. These seven (mergers and stripped) galaxies are displayed with a different color (yellow star) in our Figure \ref{ETG86_LTG44}, to show that these galaxies are neither significant outliers nor do they bias the relation. Excluding these mergers and stripped galaxies changes the slope and intercept of the best-fit-lines on an average by $~1\%$ and $~0.1\%$, respectively, which is insignificant compared to the error bars on the slopes and intercepts.  

In what follows, we divided our reduced sample of 137 galaxies into various categories, for example, early-type and late-type galaxies, S\'ersic and core-S\'ersic galaxies, galaxies with and without a disk,  galaxies with and without a bar, and galaxies with and without an AGN. The following subsections describe the scaling relations obtained for these sub-morphological classes.    

\subsection{Early-type Galaxies and Late-type Galaxies}
\begin{figure*}
\begin{center}
\includegraphics[clip=true,trim= 21mm 11mm 34mm 27mm,width=   0.7\textwidth]{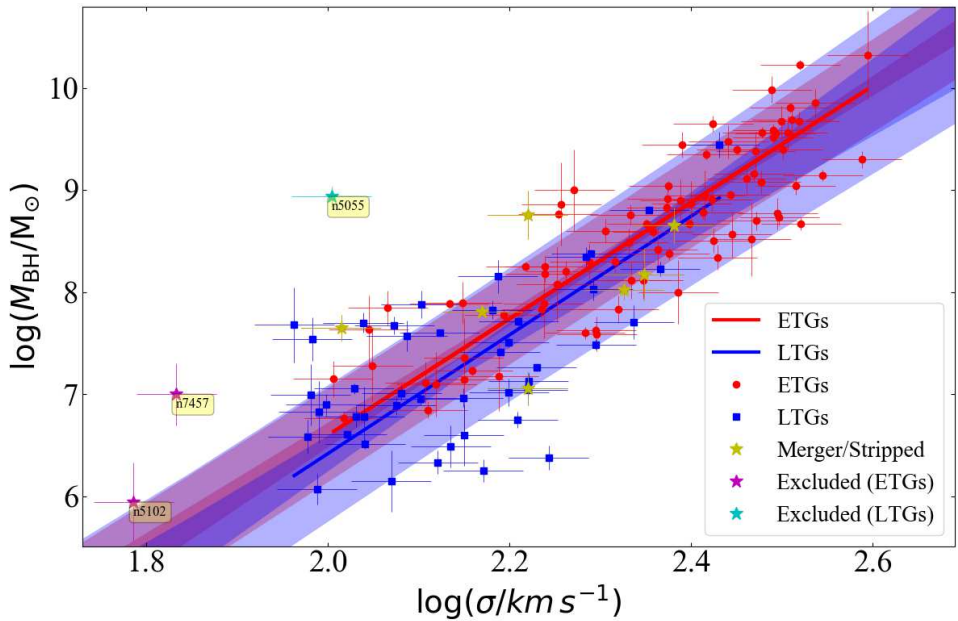}
\caption{Black hole mass versus central velocity dispersion relation followed by 91  ETGs (red circles) and 46 LTGs (blue squares).
Dark red and blue lines are the \textsc{BCES(bisector)} best-fit lines for ETGs and LTGs. The red and blue bands around these lines represent the $\pm 1 \sigma$ uncertainty limits in their slopes and the intercepts. Furthermore, the light red and light blue shaded regions depict the $\pm 1 \sigma$ scatter in the ETG and LTG samples, respectively. The yellow stars represent either merger or stripped galaxies. Labeled data-points represent galaxies excluded from the regressions, as noted in the inset legend. The best-fit lines for the two sub-populations are consistent (Equations \ref{ETGs} and \ref{LTGs}) with each other, suggesting a single $M_{BH}$--$\sigma$ relation as shown in Figure \ref{ETG_LTG_130}. }
\label{ETG86_LTG44}
\end{center}
\end{figure*}

\begin{figure*}
\begin{center}
\includegraphics[clip=true,trim= 8mm 3mm 12mm 12mm,width=   1\textwidth]{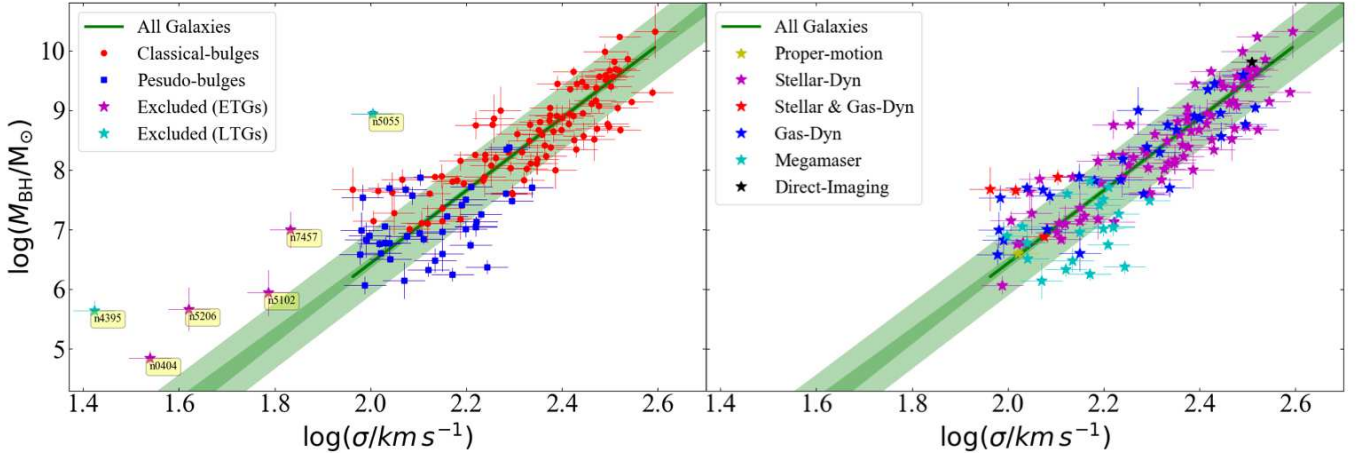}
\caption{Black hole mass versus central velocity dispersion relation obtained from a single regression on the sample of 137 ETGs and LTGs. The dark green line is the best-fit \textsc{BCES(bisector)} regression line (Equation \ref{single_reg}). The dark green band around the dark green line shows the $\pm 1 \sigma $ uncertainty in the slope and intercept of the best-fit line. The light green shaded region represents the $\pm 1 \sigma$ scatter in the data. This explanation of the dark and light-shaded regions around the best-fit line applies to all the subsequent figures in this paper. Labeled data-points in the left-hand panel represent all the excluded galaxies except for IC~1481 and NGC~6926, which cannot be included as they have no reliable $\sigma$ measurements (see Section \ref{exclusion}). The blue squares in the left-hand panel represent the galaxies which are alleged to contain pseudobulges  by \citet{Kormendy:Ho:2013}, \citet{Saglia:2016}, and the references mentioned in Table 1 of  \citet{Davis:Graham:2017}. This plot suggests that pseudobulges do follow the $M_{BH}$--$\sigma$ relation similar to classical bulges. Moreover, these pseudobulges are distributed above and below the best-fit line, albeit they are spread over a short range of $M_{BH}$ and $\sigma$. Right-hand panel shows the same plot but each galaxy is color coded according to the method used to measure its black hole mass.}
\label{ETG_LTG_130}
\end{center}
\end{figure*}

After excluding the  eight galaxies mentioned in Section \ref{exclusion}, our reduced sample is comprised of 91 ETGs and 46 LTGs\footnote{As noted in Section \ref{exclusion}, results including the six of these eight galaxies with velocity dispersions can be found in the Appendix.}. 
The \textsc{BCES(Bisector)} regression line for the ETGs can be expressed as,
\begin{IEEEeqnarray}{rCl}
\label{ETGs}
\log(M_{BH}/M_\odot) &=& (5.71\pm 0.33)\log\left(\frac{\sigma}{200\,\rm km\,s^{-1}}\right) \nonumber \\
&& +\> (8.32\pm 0.05),
\end{IEEEeqnarray}
with a total rms scatter of $\Delta_{rms | BH} = 0.44$ dex in the $\log M_{BH}$-direction. The relation followed by the LTGs can be formulated as,
 \begin{IEEEeqnarray}{rCl}
\label{LTGs}
\log(M_{BH}/M_\odot) &=& (5.82\pm 0.75)\log\left(\frac{\sigma}{200\,\rm km\,s^{-1}}\right) \nonumber \\
&& +\> (8.17\pm 0.14),
\end{IEEEeqnarray}
with   $\Delta_{rms | BH} = 0.63$ dex. The slopes and intercepts of both lines (see Figure \ref{ETG86_LTG44}) are consistent within the $\pm 1\sigma$ confidence limits, suggesting a single $M_{BH}$ versus $\sigma$ relation for both ETGs and LTGs is adequate. Therefore, we perform a single regression on the total sample of 137 galaxies, which is represented in Figure \ref{ETG_LTG_130}. 
The \textsc{BCES(Bisector)} best-fit line obtained from the single regression can be written as 
\begin{IEEEeqnarray}{rCl}
\label{single_reg}
\log(M_{BH}/M_\odot) &=& (6.10\pm 0.28)\log\left(\frac{\sigma}{200\,\rm km\,s^{-1}}\right) \nonumber \\
&& +\> (8.27\pm 0.04),
\end{IEEEeqnarray}
with   $\Delta_{rms | BH}= 0.53$ dex. However, as we will see in the following subsection, it is deceptive to think that one line is sufficient to understand the connection between super-massive black holes and the stellar velocity dispersion of the host galaxies.

Although we assigned a $10\%$ uncertainty to the measured velocity dispersions, as discussed in Section \ref{Data}, we find consistent results for our regressions when using either $5\%$ or  $15\%$ uncertainties on $\sigma$, or using the uncertainties provided in \textsc{HyperLeda} and the other corresponding sources (Column 11 of Table \ref{Total Sample}). 
In addition to the \textsc{BCES(Bisector)} regression line parameters, the slopes and intercepts of the best-fit lines from the BCES($M_{BH} | \sigma$) and BCES($\sigma | M_{BH}$) regressions, along with the scatter, Pearson correlation coefficient, and Spearman rank-order correlation coefficients are presented in Table \ref{Extra_fit_parameters} in the Appendix.  

In the left hand panel of Figure \ref{ETG_LTG_130}, we show the galaxies NGC~404, NGC~5102, NGC~5206, and NGC~4395 which are excluded from our regressions because they are the only data points in the low-mass ($M_{BH} \lesssim 10^6 M_{\odot}$) range. The first three galaxies are taken from \citet{Nguyen:2017, Nguyen:2018}. These galaxies depart from the line defined by galaxies with $M_{BH} \gtrsim 10^6 M_{\odot}$, perhaps revealing here a bend in the $M_{BH}$--$\sigma$ relation not detected by \citet{Nguyen:2017, Nguyen:2018}. Including these galaxies in the regression produces a shallower slope of $5.39 \pm 0.34$ (cf. $6.10\pm 0.28$ from Equation \ref{single_reg}), suggesting these four galaxies may have a significant effect on our best-fit line for the full sample, which is why  we decided to exclude them from our regressions. 

In the left-hand panel of Figure \ref{ETG_LTG_130}, we have additionally highlighted galaxies  alleged to have pseudo-bulges by \citet{Kormendy:Ho:2013}, \citet{Saglia:2016}, and  a few additional studies mentioned in \citet[][their Table 1]{Davis:Graham:2017}. These pseudo-bulges appear to follow the $M_{BH}$--$\sigma$ relation (see Figure \ref{ETG_LTG_130}); they are distributed about the best-fit (green) line, though with slightly more scatter than that of galaxies hosting classical bulges. However, 
given the difficulties  in assigning a bulge type (see Footnote~\ref{foot_pseu}), it is premature to draw conclusions about the co-evolution or not of black holes in pseudo-bulges.

In a recent work, \citet{van_den_Bosch:2016} fit a single $M_{BH}$--$\sigma$ line to all the morphological types of galaxies, and  reported  $M_{BH} \propto \sigma^{5.35 \pm 0.23}$, which is shallower than our relation (Equation \ref{single_reg}). We suspect that their best-fit line may be influenced by the inclusion of a few low-mass dwarf galaxies, the use of upper limits on $M_{BH}$ for many galaxies, and 24 reverberation-mapped black hole mass estimates \citep[pre-calibrated to a prior $M_{BH}$--$\sigma$ relation with a slope of $5.31\pm 0.33$ from][]{Woo:2013}.

\subsection{S{\'e}rsic and Core-S{\'e}rsic Galaxies}
\label{S_CS_subsection}

Out of the 91 ETGs in our reduced sample, 35 are core-S\'ersic, i.e., galaxies which have a deficit of stars at their center relative to the outer S\'ersic profile \citep{Graham:2003:CS}, while the remaining 56 ETGs, and all 46 LTGs, are S\'ersic galaxies. Core-S\'ersic or S\'ersic classifications for each of our galaxies are borrowed from their parent works, i.e., \citet{Savorgnan:2016:Slopes}, \citet{Davis:2018:a}, and \citet{Sahu:2019:I}, as mentioned in Table \ref{Total Sample} (Column 10).  

\begin{figure*}
\begin{center}
\includegraphics[clip=true,trim= 21mm 11mm 34mm 27mm,width=   0.7\textwidth]{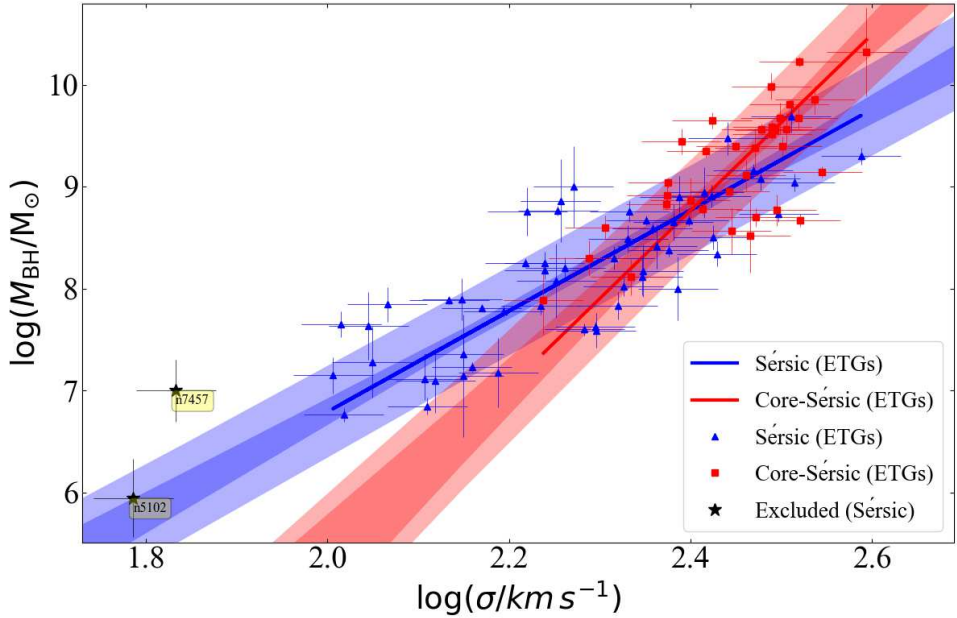}
\caption{Black hole mass versus central velocity dispersion relation for S\'ersic (blue triangles) and core-S\'ersic (red squares) ETGs. These two sub-populations follow two distinct relations (Equations \ref{Ser_ETG} and \ref{CS}), suggesting a broken $M_{BH}$--$\sigma$ relation.}
\label{CS_S_ETG}
\end{center}
\end{figure*}

\begin{figure*}
\begin{center}
\includegraphics[clip=true,trim= 21mm 11mm 34mm 27mm,width=   0.7\textwidth]{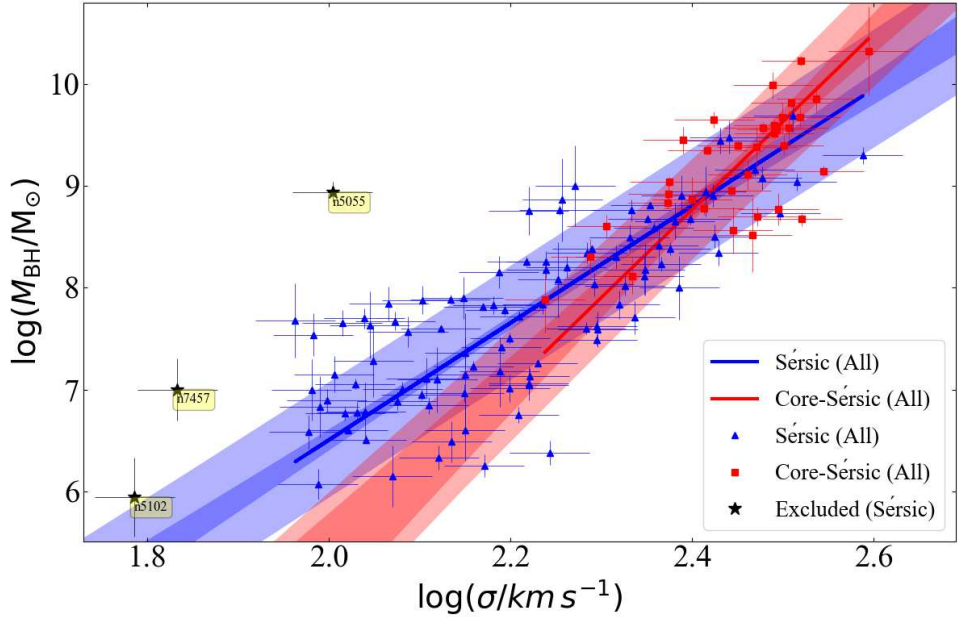}
\caption{Similar to Figure \ref{CS_S_ETG}, but including all early- and late-type S\'ersic galaxies in the same category (blue triangles) while all core-S\'ersic galaxies (red squares) are early-type galaxies. Upon including the LTGs (spirals) which are all S\'ersic galaxies, we still find two different $M_{BH}$--$\sigma$ relations followed by the S\'ersic and core-S\'ersic galaxies (Equations \ref{Ser_total} and \ref{CS}).}
\label{CS_S_total}
\end{center}
\end{figure*}

We first performed separate regressions for the S{\'e}rsic and core-S{\'e}rsic  ETGs, then on the combined sample of 137 galaxies. The $M_{BH}$--$\sigma$ plots for these two divisions are shown in Figure \ref{CS_S_ETG} and Figure \ref{CS_S_total}, respectively. 

S{\'e}rsic and core-S{\'e}rsic categorization reveals  two different relations followed by the two sub-populations. The symmetric best-fit line followed by the early-type S\'ersic galaxies can be expressed as 
\begin{IEEEeqnarray}{rCl}
\label{Ser_ETG}
\log(M_{BH}/M_\odot) &=& (4.95\pm 0.38)\log\left(\frac{\sigma}{200\,\rm km\,s^{-1}}\right) \nonumber \\
&& +\> (8.28\pm 0.06),
\end{IEEEeqnarray}
with   $\Delta_{rms | BH}= 0.42$ dex, represented by the dark blue line in Figure \ref{CS_S_ETG}. The total S\'ersic population, consisting of 102 early- and late-type S\'ersic galaxies, produces the relation 
\begin{IEEEeqnarray}{rCl}
\label{Ser_total}
\log(M_{BH}/M_\odot) &=& (5.75\pm 0.34)\log\left(\frac{\sigma}{200\,\rm km\,s^{-1}}\right) \nonumber \\
&& +\> (8.24\pm 0.05),
\end{IEEEeqnarray}
represented by the dark blue line in Figure \ref{CS_S_total}, with   $\Delta_{rms | BH}= 0.55$ dex. The best-fit lines obtained for only early-type S\'ersic galaxies and for all the S\'ersic galaxies are marginally consistent with each other within the $\pm 1\sigma$ bound of their slopes and intercepts. 

However, the core-S\'ersic galaxies follow a much steeper  $M_{BH}$--$\sigma$ relation, with   $\Delta_{rms | BH}= 0.46$ dex, as is shown by the dark red lines in both Figures \ref{CS_S_ETG} and \ref{CS_S_total}, which can be expressed as
\begin{IEEEeqnarray}{rCl}
\label{CS}
\log(M_{BH}/M_\odot) &=& (8.64\pm 1.10)\log\left(\frac{\sigma}{200\,\rm km\,s^{-1}}\right) \nonumber \\
&& +\> (7.91\pm 0.20).
\end{IEEEeqnarray}
The slope of this line is inconsistent with that of the S\'ersic galaxies. The difference in their slopes reveals that S\'ersic and core-S\'ersic galaxies follow two distinct relations, potentially linked to the evolutionary paths followed by these two type of galaxies, i.e., evolution via major dry-mergers versus gas-rich mergers and accretion events. Additionally, core-S\'ersic galaxies follow a steeper relation, that is, their $\sigma$ values do not appear to saturate or asymptote at the high black hole mass end. 

Core-S\'ersic galaxies are old, gas-poor, massive galaxies, many of which are BCGs which have undergone multiple major (equal mass) dissipation-less dry-mergers. During a dry-merger, their central SMBHs inspiral, expelling out stars from the center, thereby creating a deficit of light at the core of the resulting galaxy. The stellar mass deficit, relative to the central black hole mass, may be
a measure of the number of dry mergers a galaxy has undergone \citep{Merritt:Milosavljevic:2005, Savorgnan:Graham:2015}, with the radial size of the depleted core known to be correlated with the black hole mass \citep{Dullo:2014, Thomas:2016, Mehrgan:2019}.

The steeper $M_{BH}$--$\sigma$ relation for core-S\'ersic galaxies reveals that dry mergers do not increase the velocity dispersion, relative to the increased black hole mass, at the pace followed by S\'ersic galaxies (built through either gas-rich mergers or accretion of gas from their surroundings). This  has also been suggested by some theoretical studies \citep[e.g.,][]{Ciotti:vanAlbada:2001, Oser:2012, Shankar:2013, Hilz:2013}. Furthermore, \citet{Volonteri:2013} used their analytical and semi-analytical models to show that simulated BCGs are offset from the $M_{BH}$--$\sigma$ relation defined by non-BCGs because they undergo multiple gas-poor (dry) mergers resulting in over-massive black holes with only mildly increased velocity dispersion.

\subsection{Galaxies With a Disk (ES/S0/Sp) and Without a Disk (E)}
\label{EESS0}
ETGs include elliptical (E), ellicular (ES), and lenticular (S0) galaxies. Elliptical galaxies are pressure-supported, spheroid-dominated galaxies with minimal rotation. Ellicular galaxies host an intermediate-scale (rotating) stellar disk within their spheroids \citep{Liller:1966, Graham:Ciambur:Savorgnan:2016}, while lenticular galaxies have a large-scale disk extending beyond their bulges \citep[see][for a detailed morphological classification grid]{Graham:Grid:2019}. LTGs are spiral (Sp) galaxies  with a bulge, a large-scale disk, and spiral arms. The LTGs in our sample are predominantly early-type spirals (Sa--Sb).

Our reduced sample of 137 galaxies is comprised of 44 elliptical galaxies which do not have a rotating disk, plus 93 galaxies with a disk, which includes 47 ES or S0-types (ETGs) and 46 spirals (LTGs).
\begin{figure*}
\begin{center}
\includegraphics[clip=true,trim= 21mm 11mm 34mm 27mm,width=   0.7\textwidth]{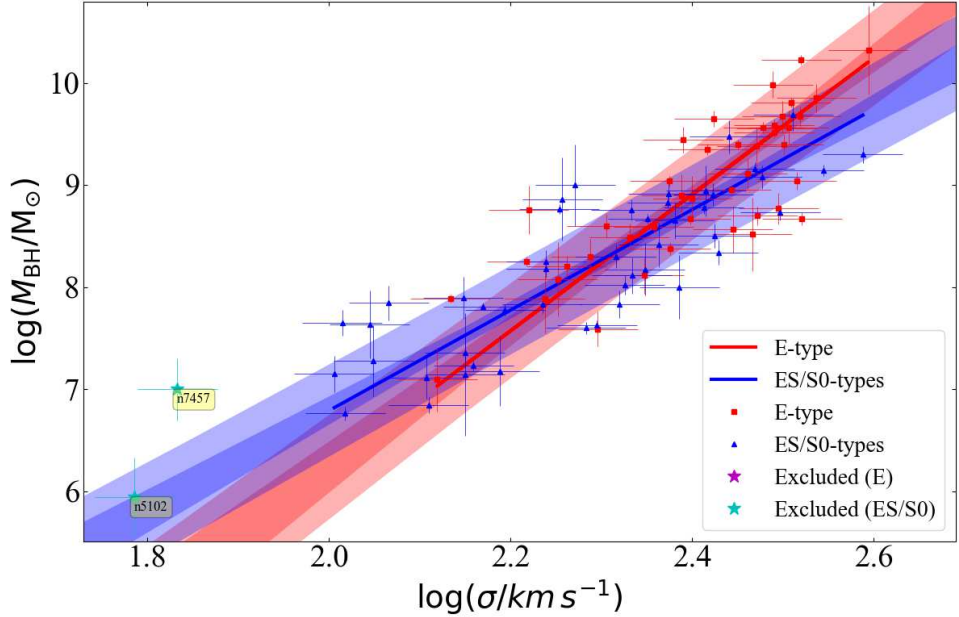}
\caption{Black hole mass versus central velocity dispersion relations for ETGs with a disk (ES/S0-types) and ETGs without a disk (E-type). We find two slightly different relations for galaxies with and without a disk, which is similar (but less pronounced) to the separation in the $M_{BH}$--$\sigma$ diagram due to S\'ersic and core-S\'ersic galaxies (see Figure \ref{CS_S_ETG}). This is not surprising as most of the elliptical galaxies in our sample are core-S\'ersic galaxies and most of the ETGs with a disk (ES/S0-types) are S\'ersic galaxies, hence the difference is caused by core-S\'ersic and S\'ersic galaxies.}
\label{rot_ETG}
\end{center}
\end{figure*}

\begin{figure*}
\begin{center}
\includegraphics[clip=true,trim= 21mm 11mm 34mm 27mm,width=   0.7\textwidth]{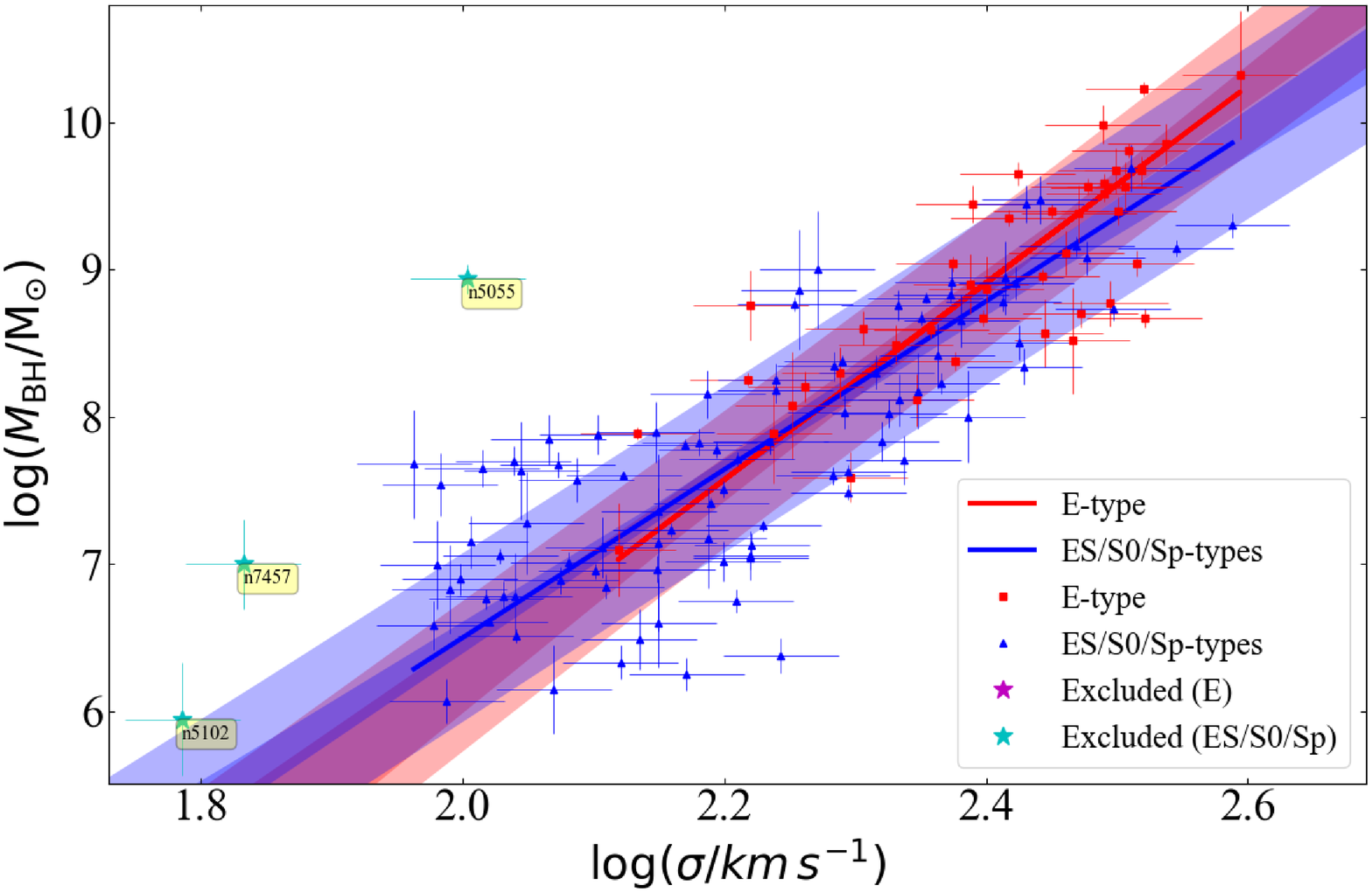}
\caption{Similar to Figure \ref{rot_ETG}, but now including spiral (Sp) galaxies---all of which have an extended rotating disk---along with the  ellicular (ES) and lenticular (S0) galaxies in the category of galaxies with a disk, while elliptical (E) galaxies without a disk are all ETGs. Here, we again find two slightly different relations in the $M_{BH}$--$\sigma$ diagram, but not as pronounced as between S\'ersic and core-S\'ersic galaxies. See Table \ref{fit parameters} for full equations of the two lines.}
\label{rot_total}
\end{center}
\end{figure*}

We first performed separate regressions on the ETGs with (ES/S0) and without (E) a disk, as shown in Figure \ref{rot_ETG} where the blue and red lines correspond to $M_{BH} \propto \sigma^{4.93\pm 0.39}$ and  $M_{BH} \propto \sigma^{6.69\pm 0.59}$, respectively. Then we performed regressions on all types of galaxies with a disk (ES/S0/Sp), and without a disk (E-types), as represented in Figure \ref{rot_total} where the blue line defines $M_{BH} \propto \sigma^{5.72\pm 0.34}$ and  the red line is the same as that in  Figure \ref{rot_ETG}, i.e., $M_{BH} \propto \sigma^{6.69\pm 0.59}$. Full equations of the best-fit lines can be found in our Table \ref{fit parameters}.

Not surprisingly, we find that galaxies with and without a disk seem to follow two slightly different relations in both cases (ETG-only, ETG+LTG). This is more apparent for the ETG sample (Figure \ref{rot_ETG}) than for the total sample (Figure \ref{rot_total}) because upon including spiral galaxies with ETGs with a disk (ES/S0), the apparent difference in slopes of the blue and red lines reduces. 
   
This difference in the $M_{BH}$--$\sigma$ relations due to galaxies with and without a disk is likely because most of the elliptical galaxies in our sample are (massive) core-S\'ersic galaxies and almost all the galaxies with a rotating disk are S\'ersic galaxies.
The extent of the difference between  the $M_{BH}$--$\sigma$ relation for core-S\'ersic and S\'ersic galaxies is greater than that of the relations followed by the galaxies with and without a disk. This suggests that the two distinct relations in the $M_{BH}$--$\sigma$ diagram are predominantly caused by core-S\'ersic versus S\'ersic galaxies. It should be noted that core-S\'ersic galaxies can also have disks \citep[e.g.][]{Dullo:Graham:2013, Dullo:2014, Dullo:S0:2014}, for example the lenticular galaxies NGC~524, NGC~584, NGC~3706, NGC~4751, and NGC~5813 in our sample have depleted stellar cores. 

We speculate that \citet{Savorgnan:Graham:2015} failed to detect different $M_{BH}$--$\sigma$ relations for core-S\'ersic and S\'ersic galaxies, or slow and fast rotators\footnote{Note: ES galaxies are both fast rotators and slow rotators \citep[e.g.,][]{Bellstedt:Graham:2017}.}, because of their smaller sample size. However, some of their core-S\'ersic galaxies can be spotted to be offset from their single $M_{BH}$--$\sigma$ relation at the high-mass end.

\subsection{Barred and Non-barred Galaxies}

In the past, some observational studies \citep{Graham:2007, Hu:2008, Graham:2008:b, Graham:2008:a} and simulations \citep{Brown:Simulation:2013, Hartmann:2014} have revealed that barred galaxies are offset towards the higher $\sigma$ side in the $M_{BH}$--$\sigma$ diagram. Based on that offset, these studies suggest that barred galaxies should be separated from non-barred galaxies in order to obtain $M_{BH}$--$\sigma$ relations for barred and non-barred galaxies. 

\begin{figure*}
\begin{center}
\includegraphics[clip=true,trim= 21mm 11mm 34mm 27mm,width=   0.7\textwidth]{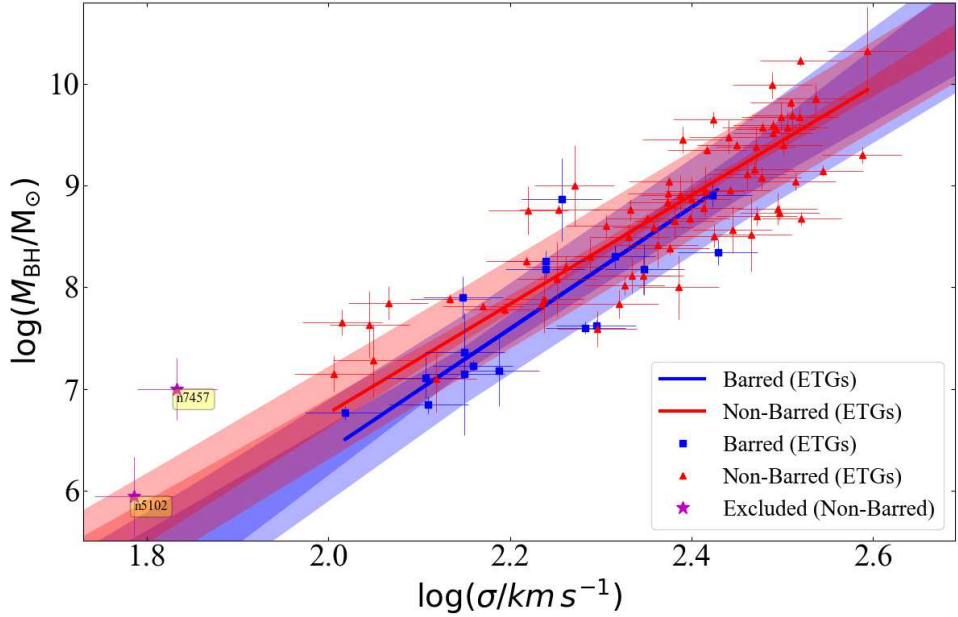}
\caption{Black hole mass versus central velocity dispersion relation for barred and non-barred ETGs. Although we only have a small sample of 17 barred ETGs, the consistency of the two regression lines (blue and red lines) suggests no offset between barred (Equation \ref{B_ETG}) and non-barred (Equation \ref{NB_ETG})  ETGs in the $M_{BH}$--$\sigma$ diagram.}
\label{bar_ETG}
\end{center}
\end{figure*}

\begin{figure*}
\begin{center}
\includegraphics[clip=true,trim= 21mm 11mm 34mm 27mm,width=   0.7\textwidth]{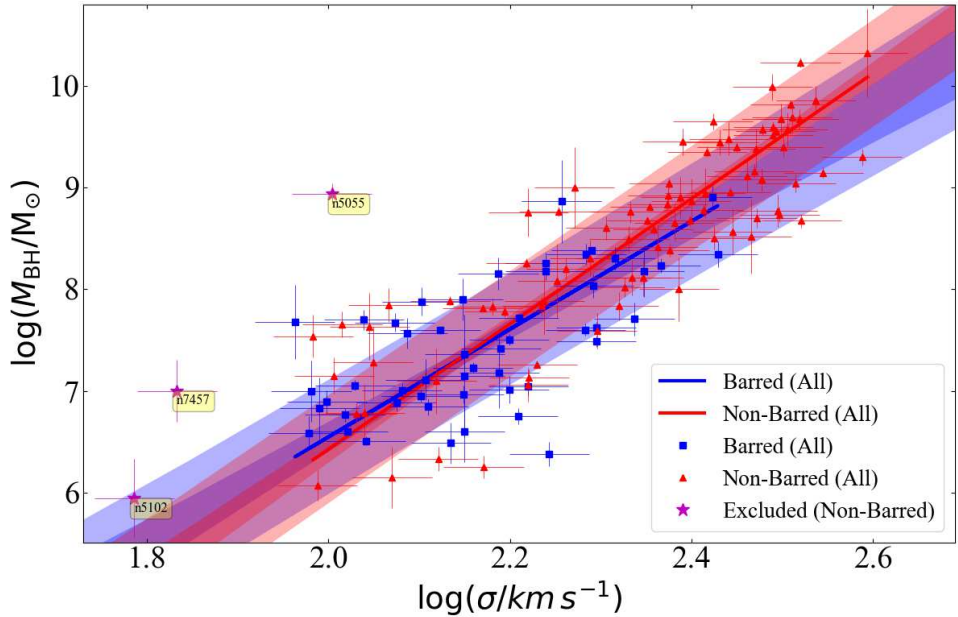}
\caption{Similar to Figure \ref{bar_ETG}, but including barred and non-barred  late-type galaxies as well. The regression lines obtained for the 50 barred (blue line, Equation \ref{all_bar}) and 87 non-barred (red line, Equation \ref{all_nonbar}) galaxies are overlapping and consistent with each other, implying no-offset between barred and non-barred galaxies.}
\label{bar_total}
\end{center}
\end{figure*}

To investigate the above offset using our larger data-set, accompanied with our revised classifications based upon multi-component decompositions, we also divided our sample into barred and non-barred galaxies, and performed separate regressions on both populations. This was first done for  barred and non-barred ETGs, then using the total (reduced) sample of 137 galaxies, as shown in Figures \ref{bar_ETG} and \ref{bar_total}, respectively. Our ETG sample consists of 17 barred and 74 non-barred galaxies, while the full  sample comprises  50 barred and 87 non-barred galaxies.

Surprisingly, we do not find any offset between barred and non-barred galaxies, in either case, i.e., only ETGs and the ETG + LTG sample. The best-fit line for the 17 barred ETGs is
\begin{IEEEeqnarray}{rCl}
\label{B_ETG}
\log(M_{BH}/M_\odot) &=& (5.98\pm 0.80)\log\left(\frac{\sigma}{200\,\rm km\,s^{-1}}\right) \nonumber \\
&& +\> (8.19\pm 0.14),
\end{IEEEeqnarray}
with   $\Delta_{rms | BH} = 0.41$ dex. However, we require a larger sample of barred ETGs for a robust relation.
The 74  non-barred ETGs define the following relation, with   $\Delta_{rms | BH} = 0.43$,
\begin{IEEEeqnarray}{rCl}
\label{NB_ETG}
\log(M_{BH}/M_\odot) &=& (5.35\pm 0.39)\log\left(\frac{\sigma}{200\,\rm km\,s^{-1}}\right) \nonumber \\
&& +\> (8.37\pm 0.06).
\end{IEEEeqnarray}

The 50 barred ETG + LTG population defines the line,
\begin{IEEEeqnarray}{rCl}
\label{all_bar}
\log(M_{BH}/M_\odot) &=& (5.30\pm 0.54)\log\left(\frac{\sigma}{200\,\rm km\,s^{-1}}\right) \nonumber \\
&& +\> (8.14\pm 0.10),
\end{IEEEeqnarray}
with   $\Delta_{rms | BH}= 0.53$ dex. The 87 non-barred galaxies define the relation
\begin{IEEEeqnarray}{rCl}
\label{all_nonbar}
\log(M_{BH}/M_\odot) &=& (6.16\pm 0.42)\log\left(\frac{\sigma}{200\,\rm km\,s^{-1}}\right) \nonumber \\
&& +\> (8.28\pm 0.06),
\end{IEEEeqnarray}
with $\Delta_{rms | BH}= 0.51$ dex. The best-fit lines for the barred and non-barred galaxies are consistent within the $\pm 1 \sigma$ bounds of their slopes and intercepts, suggesting no significant offset between barred and non-barred galaxies.

\subsubsection{Investigating Previous Offsets}
To find  the reason behind the offset observed by \citet{Graham:Scott:2013}, we have compared their regression lines with ours obtained using the latest $M_{BH}$, $\sigma$, and updated bar-morphologies. Their sample of 72 galaxies was comprised of 21 barred and 51 non-barred galaxies, according to the morphological classifications they adopted, which were obtained from the NASA/IPAC Extragalactic Database (NED). All of their galaxies are present in our current sample, and in order to make a comparison, we use only the galaxies present in the data-set of \citet{Graham:Scott:2013}.

Interestingly, out of those common 72 galaxies, we have classified 27 as barred, and 45 as non-barred. The barred and non-barred classifications for our current sample are based on the morphologies obtained from the multi-component decompositions of these galaxies presented in our recent works \citep{Savorgnan:Graham:2016:I, Davis:2018:a, Sahu:2019:I}. We notice that in the data-set of \citet{Graham:Scott:2013}, seven barred galaxies (NGC~224, NGC~2974, NGC~3245, NGC~3998, NGC~4026, NGC~4388, and NGC~6264) were misclassified as non-barred due to the presence of weak bars not detected in optical images \citep{Eskridge:2000}\footnote{\citet{Eskridge:2000} claim that bars are more detectable in NIR band than optical. However, see  \citet[][and references therein]{Buta:Sheth:2010} which suggest that bar-fraction is similar in the two wavelengths.}. Also, one non-barred galaxy (NGC~4945) in their sample appears to have been  misclassified as barred, with \citet{Davis:2018:a}
reporting only a nuclear bar too weak to include in their modelling.

\begin{figure*}
\begin{center}
\includegraphics[clip=true,trim= 21mm 11mm 34mm 27mm,width=   0.7\textwidth]{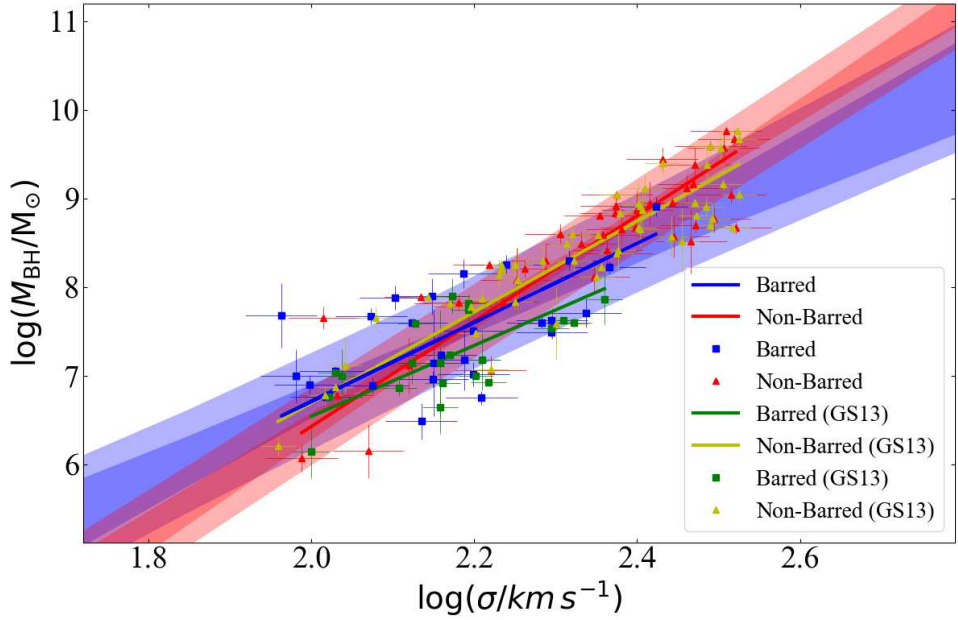}
\caption{Comparison of our $M_{BH}$--$\sigma$ relations for barred and non-barred galaxies with the relations reported in \citet[][GS13]{Graham:Scott:2013}. Their galaxy sample is a sub-set of our current sample, thus, for a comparison, we use our latest data  for the galaxies in their sample, applied with our new bar morphologies (blue and red points). The barred and non-barred data points (i.e., the green squares and yellow triangles, respectively) of \citet{Graham:Scott:2013} represent the $M_{BH}$, $\sigma$, and  bar classifications they used. Using the same galaxy sample as that of  \citet{Graham:Scott:2013}, we do not find any significant offset between barred and non-barred galaxies.}
\label{bar_GS13_Comp}
\end{center}
\end{figure*}

The green and yellow lines in Figure \ref{bar_GS13_Comp} are the BCES symmetric best-fit lines from \citet{Graham:Scott:2013} for the barred and non-barred galaxies, respectively. These two lines are offset by  $\sim 0.5$ dex at the median velocity dispersion of $200 \, \rm km\, s^{-1}$. The blue and red BCES bisector lines for the 72 reclassified barred and non-barred galaxies from our current data-set, are offset by only 0.16 dex. Moreover, on using the total (reduced) sample of 137 galaxies comprising 50 barred and 87 non-barred galaxies, as is represented in Figure \ref{bar_total}, the offset reduces to 0.14 dex (see Equations \ref{all_bar} and \ref{all_nonbar}). 

We find that there are two main reasons why \citet{Graham:Scott:2013} found an offset. First, they largely classified their galaxies as barred or non-barred based on the morphologies provided by NED, which are mainly from the RC3 catalog \citep{RC3:1991} and in many cases it failed to identify bars and some other galaxy structures as well.
The second reason is that their sample of 72 galaxies lacked (a sufficiently large sample of) barred galaxies residing above their regression line (the green line in Figure \ref{bar_GS13_Comp}). Another reason for the difference might have been the updated black hole masses and velocity dispersions. For example, the updated \citep{Greene:Ho:2006} velocity dispersion for the barred spiral galaxy NGC~4151 is  $91.8\pm 9.9  \, \rm km \,s^{-1}$ , which is notably different from the old value of $156\pm 7.8 \, \rm km \,s^{-1}$ reported in \textsc{HyperLeda}. However, we have found that, collectively, the updated velocity dispersions do not seem to have a significant effect on the offset between the regression lines for the barred and non-barred galaxies, because the latest $\sigma$ values are not particularly different for most of the galaxies.

\subsubsection{Strong versus Weak or Faint Bars}
We also investigated if weak/faint barred galaxies are biasing our barred $M_{BH}$--$\sigma$ relation (Equation \ref{all_bar}). There was a possibility that perhaps most of the weak/faint barred galaxies fall above the best-fit relation (blue line in Figure \ref{bar_total})  for the barred galaxies in our current sample, and thereby reduce the offset between the best-fit relation for barred and non-barred galaxies.

\begin{figure*}
\begin{center}
\includegraphics[clip=true,trim= 21mm 11mm 34mm 27mm,width=   0.7\textwidth]{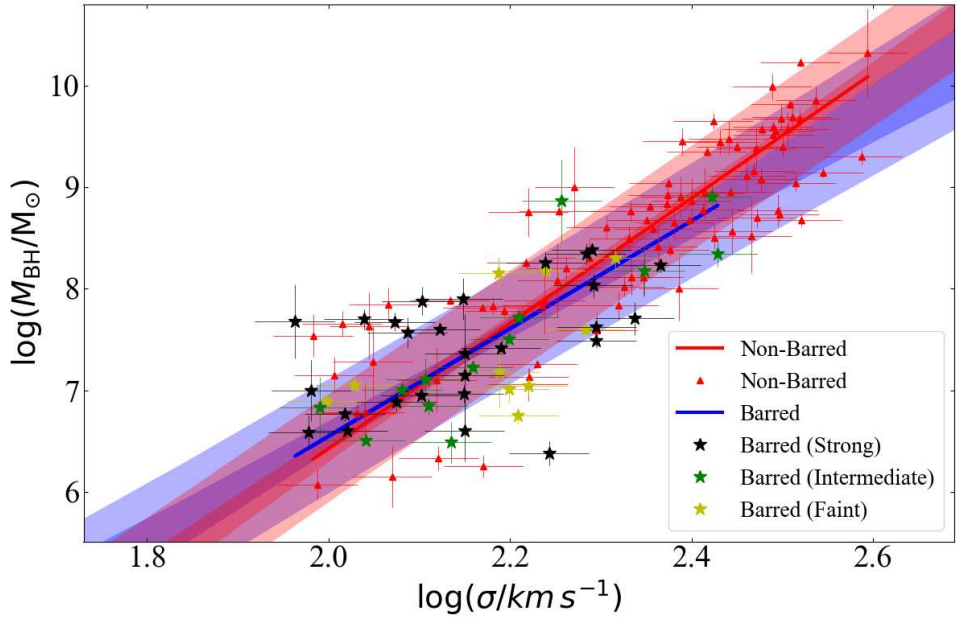}
\caption{Similar to Figure \ref{bar_total}, but now categorizing our barred galaxies into strong, intermediate, and faint barred galaxies.}
\label{3bar}
\end{center}
\end{figure*}

For this investigation, we used the bar-to-total (galaxy) luminosity ($L_{bar}/L_{tot}$) ratio to categorize our barred galaxies into strong and weak/faint categories. However, as we were not sure of where to make the cut, we performed this test twice, first making the division at $L_{bar}/L_{tot} = 0.05$, then at $L_{bar}/L_{tot} = 0.1$. Figure \ref{3bar} shows the barred galaxies color coded as black strong-barred  ($L_{bar}/L_{tot} \geq 0.1$), yellow faint-barred ($L_{bar}/L_{tot} \leq 0.05$), and green with intermediate bar strength ($0.05 <  L_{bar}/L_{tot} < 0.1$).  For 14 barred-galaxies, 9 of which are from \citep{Savorgnan:Graham:2016:I}, 4 are from \citet{Combes:2019:SMBH4}, and one is from \citet{Nguyen:2019}, we do not have the luminosity of the bar. Hence, we categorized them on the basis of their multi-component decomposition profile, the morphological bar classification provided by the literature, and a visual inspection of their images which was also performed for all the other barred galaxies. Overall, our total sample of 50 barred galaxies consists of 27 strong, 10 weak/faint, and 13 intermediate-strength barred galaxies.

For the first test, i.e., for the division at $L_{bar}/L_{tot} = 0.05$, all the strong (and intermediate) barred galaxies  are distributed almost uniformly about the best-fit (blue) line for the barred galaxies, and many of the faint barred galaxies are below the best-fit line (see Figure \ref{3bar}). This suggests that galaxies with faint-bars do not minimize the offset between  barred and non-barred galaxies. 
As for the second cut at $L_{bar}/L_{tot} = 0.1$, we can see in Figure \ref{3bar}, that most of the intermediate and faint barred galaxies  are below the best-fit line for barred-galaxies, again indicating that weak/faint- barred, or even intermediate-barred galaxies, do not take part in reducing the offset between barred and non-barred galaxies. Strongly-barred galaxies are distributed above and below the best-fit line for barred galaxies.

\subsection{Galaxies with and without an AGN}

Our reduced sample of 137 galaxies includes 41 galaxies hosting an AGN. We identified the AGN hosts using the 13th edition of the catalog of quasars and active nuclei presented by \citet{AGN:Catalogue:2010}. Interestingly, these AGN hosts are spread almost uniformly about the best-fit bisector regression line (for the sample of 137 galaxies) for the range of $M_{BH}$ and $\sigma$ that we have, indicating that galaxies with and without an AGN follow a single relation. 

\begin{figure*}
\begin{center}
\includegraphics[clip=true,trim= 21mm 11mm 34mm 27mm,width=   0.7\textwidth]{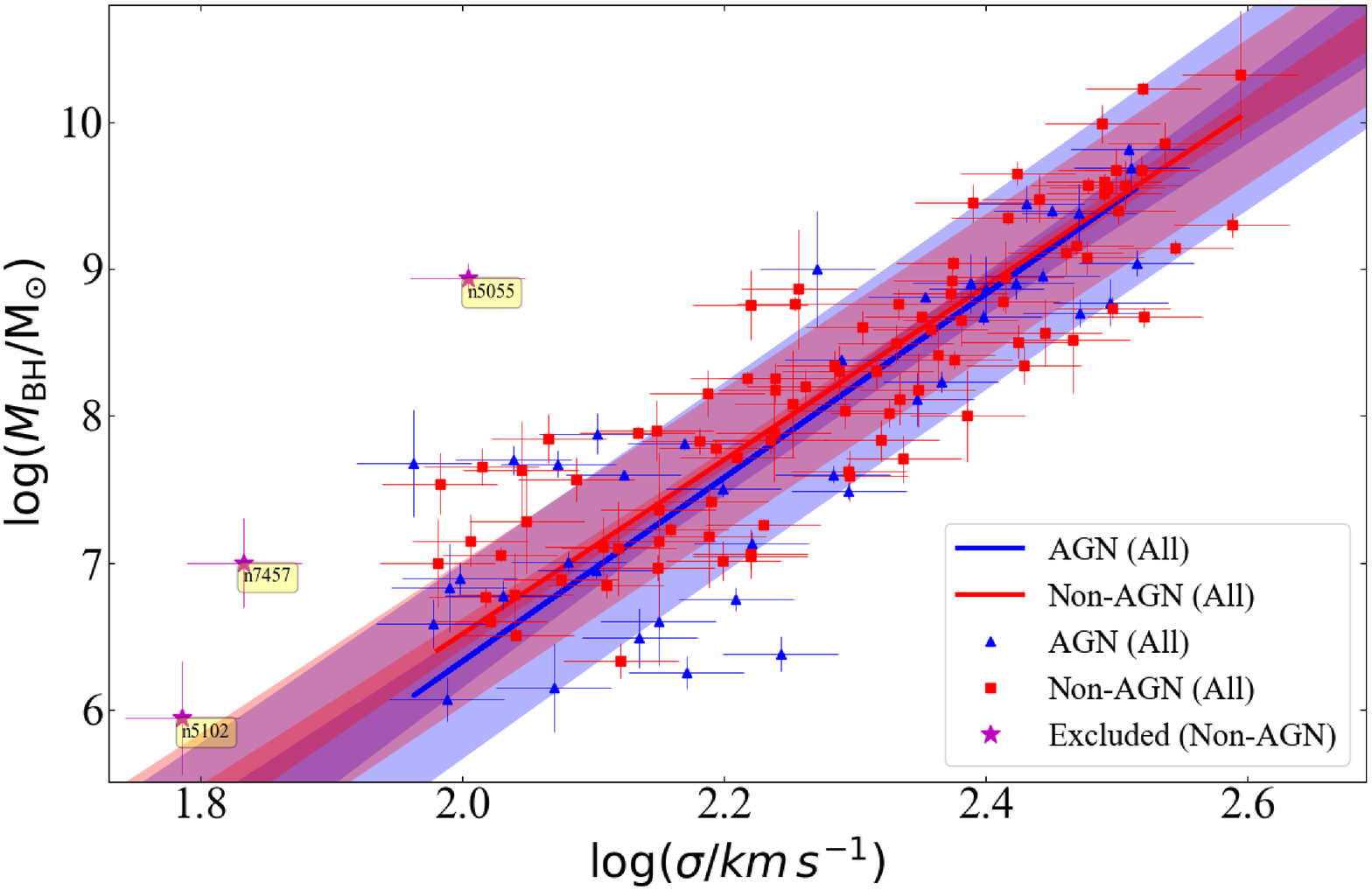}
\caption{Black hole mass versus velocity dispersion followed by galaxies hosting an AGN and galaxies without an AGN.}
\label{AGN}
\end{center}
\end{figure*}

Also, upon performing separate regressions on AGN hosts and galaxies without AGN, we obtain almost overlapping regression lines for the two categories, such that their slopes and intercept are consistent with each other within the $\pm 1\sigma$ confidence bounds (Figure \ref{AGN}).
The regression parameters for the best-fit lines for galaxies with and without AGNs are given in Table \ref{fit parameters}. 
 A galaxy hosting an AGN can be S\'ersic or core-S\'ersic, as can a galaxy without an AGN; hence, regardless of whether a galaxy hosts an AGN or not, the $M_{BH}$--$\sigma$ relations defined by S\'ersic and core-S\'ersic galaxies remain applicable, and should be used depending on the presence or absence of a  core  (deficit of star light, not due to dust obscuration).

\clearpage
\startlongtable
\begin{deluxetable*}{lcrrccrr}
\tabletypesize{\footnotesize}
\tablecolumns{8}
\tablecaption{Linear Regressions\label{fit parameters} [ $\log(M_{\rm BH}/{\rm M_{\sun}})=\alpha\log(\sigma/200)+\beta$ ] }
\tablehead{
\colhead{ \textbf{Category} } & \colhead{ \textbf{Number} } & \colhead{ \bm{$\alpha$} } & \colhead{ \bm{$\beta$ } } & \colhead{ \bm{$\epsilon$} } & \colhead{ \bm{ $\Delta_{rms | BH}$} }  & \colhead{ \bm{$r$ } } & \colhead{ \bm{$r_s$}}  \\
\colhead{} & \colhead{} & \colhead{} & \colhead{ \textbf{(dex)}} & \colhead{\textbf{(dex)}} & \colhead{\textbf{(dex)}} & \colhead{} & \colhead{} \\
\colhead{ \textbf{(1)}} & \colhead{\textbf{(2)}} & \colhead{\textbf{(3)}} & \colhead{\textbf{(4)}} & \colhead{\textbf{(5)}} & \colhead{\textbf{(6)}} & \colhead{\textbf{(7)}} & \colhead{\textbf{(8)}} 
}
\startdata
Early-Type Galaxies & 91 & $5.71\pm 0.33$ & $8.32\pm 0.05$ & $0.32$ & $0.44$ & $0.86$  & $0.85$ \\
Late-Type Galaxies & 46 & $5.82\pm 0.75$ & $8.17\pm 0.14$ & $0.57$ & $0.63$ & $0.59$  & $0.49$  \\
\hline
All Galaxies & 137 & $6.10\pm0.28$ & $8.27\pm0.04$ & $0.43$ & $0.53$ & $0.86$ & $0.87$  \\
\hline
S{\'e}rsic Galaxies & 102 & $5.75\pm0.34$ & $8.24\pm 0.05$ & $0.46$ & $0.55$ & $0.78$ & $0.78$ \\
Core-S{\'e}rsic Galaxies & 35 & $8.64\pm 1.10$ & $7.91\pm0.20$ & $0.25$ & $0.46$ & $0.73$ & $0.65$ \\
\hline
Galaxies with a disk (ES, S0, Sp-types) & 93 & $5.72\pm0.34$ & $8.22\pm0.06$ & $0.47$ & $0.56$ & $0.79$ & $0.78$ \\
Galaxies without a disk (E-type) & 44 & $6.69\pm0.59$ & $8.25\pm0.10$ & $0.30$ & $0.43$ & $0.82$ & $0.80$  \\
\hline
Barred Galaxies & 50 & $5.30\pm0.54$ & $8.14\pm0.10$ & $0.45$ & $0.53$ & $0.65$  & $0.61$  \\
Non-Barred Galaxies & 87 & $6.16\pm0.42$ & $8.28\pm 0.06$ & $0.40$ & $0.51$ & $0.86$ & $0.86$  \\
\hline
AGN host Galaxies & 41 & $6.26\pm0.49$ & $8.21\pm0.09$ & $0.55$ & $0.63$ & $0.83$  & $0.79$  \\
Galaxies without AGN & 96 & $5.92\pm0.31$ & $8.30\pm 0.05$ & $0.37$ & $0.48$ & $0.87$ & $0.88$  \\
\enddata
\tablecomments{
Columns:
(1) Subclass of galaxies.
(2) Number of galaxies in a subclass.
(3) Slope of the line obtained from the \textsc{BCES(Bisector)} regression.
(4) Intercept of the line line obtained from the \textsc{BCES(Bisector)} regression.
(5) Intrinsic scatter in the $\log M_{\rm BH}$-direction \citep[using Equation 1 from][]{Graham:Driver:2007}.
(6) Total root mean square (rms) scatter in the $\log M_{\rm BH}$ direction.
(7) Pearson correlation coefficient.
(8) Spearman rank-order correlation coefficient.
}
\end{deluxetable*}

\section{Internal consistency between the $M_{BH}$--$M_{*,gal}$, $M_{BH}$--$M_{*,sph}$, and $M_{BH}$--$\sigma$ relations}
\label{Consistency}

Recent studies by \citet{Sahu:2019:I}  and \citet{Davis:2018:a} established robust $M_{BH}$--$M_{*,gal}$ and $M_{BH}$--$M_{*,sph}$ correlations for ETGs and LTGs, using a (reduced) sample of 76 ETGs and 40 LTGs, respectively. 
As elaborated above in Section \ref{Results},  we also observe a strong correlation between black hole mass and the central stellar velocity dispersion, along with the discovery of two distinct relations in the $M_{BH}$--$\sigma$ diagram due to S\'ersic and core-S\'ersic galaxies.

The $M_{BH}$--$M_{*,gal}$ (and $M_{BH}$--$M_{*,sph}$) relations combined with our $M_{BH}$--$\sigma$ relations can predict the $M_{*,gal}$--$\sigma$ and $M_{*,sph}$--$\sigma$ relations. They should be compared with the observed $M_{*,gal}$--$\sigma$ and $M_{*,sph}$--$\sigma$ relations to check for internal consistency of our relations. The ETGs and LTGs of  \citet{Sahu:2019:I}  and \citet{Davis:2018:a}, respectively, constitute  $ 85 \%$ of the sample used in this work to obtain the $M_{BH}$--$\sigma$ relations, hence their $M_{BH}$--$M_{*,gal}$ and $M_{BH}$--$M_{*,sph}$ relations are appropriate for internal consistency checks.
To derive the $M_{*,gal}$--$\sigma$ and $M_{*,sph}$--$\sigma$ relations, we used the galaxy and spheroid stellar masses measured in \citet{Davis:2018:b}, \citet{Davis:2018:a} and \citet{Sahu:2019:I}.

S\'ersic and core-S\'ersic ETGs have been found to follow the same $M_{BH}$--$M_{*,gal}$ and $M_{BH}$--$M_{*,sph}$ relations in \citet{Sahu:2019:I}, such that  $M_{BH} \propto M_{*,gal}^{1.65 \pm 0.11}$ and $M_{BH} \propto M_{*,sph}^{1.27 \pm 0.07}$ for all ETGs, i.e., when combining those with a disk and those without a disk. Whereas, the LTGs in \citet{Davis:2018:a}, all of which are S\'ersic galaxies, define the  relations $M_{BH} \propto M_{*,gal}^{ 3.05 \pm 0.70}$ and $M_{BH} \propto M_{*,sph}^{2.16 \pm 0.32}$, with slopes almost twice that of the (single regression) slopes for ETGs in \citet[][see their Figure 11]{Sahu:2019:I}.
However, separating the ETGs into those with and without a disk reveals that they follow two different $M_{BH}$--$M_{*,sph}$ relations with slopes of approximately $1.9 \pm 0.2$ but with intercepts offset by more than a factor of $10$ in the $M_{BH}$-direction \citep[][their Figure 8]{Sahu:2019:I}. While in the $M_{BH}$--$M_{*,gal}$ diagram, the two relations for ETGs with and without a disk agree with each other much more closely, suggesting that the $M_{BH}$--$M_{*,gal}$ relation obtained from the single regression is a reasonable approximation  for ETGs with and without a disk. In the $M_{BH}$--$\sigma$ diagram,  S\'ersic and core-S\'ersic galaxies in our total (ETG+LTG) sample define two distinct relations, see Equations \ref{Ser_total} and \ref{CS}, respectively.

Theoretically, to  check on the consistency between all of these  $M_{BH}$--$M_{*,sph}$, $M_{BH}$--$\sigma$, and $M_{*,sph}$--$\sigma$ relations for ETGs, we should use the two distinct $M_{BH}$--$M_{*,sph}$ relations for ETGs with and without a disk with the two $M_{BH}$--$\sigma$ relations for core-S\'ersic and S\'ersic  ETGs (Section \ref{S_CS_subsection}), to predict different $M_{*,sph}$--$\sigma$ relations for core-S\'ersic ETGs with and without a disk and S\'ersic ETGs with and without a disk. However, if we separate the core-S\'ersic (or S\'ersic) ETGs into galaxies with and without a disk, each sub-population will be too small to derive a robust $M_{*,sph}$--$\sigma$ relation for comparison with the predicted relation. Hence, for the current consistency checks, we have used the following single regression relation for  ETGs: $M_{BH} \propto M_{*,sph}^{1.27\pm 0.07}$.

Using $M_{BH} \propto \sigma^{8.64 \pm 1.10}$ (Equation \ref{CS}) for our core-S\'ersic galaxies, all of which are ETGs, and the  $M_{BH}$--$M_{*,gal}$ (and $M_{BH}$--$M_{*,sph}$) relations for the ETGs from \citet{Sahu:2019:I}, we expect the relations $M_{*,gal} \propto \sigma^{5.24\pm 0.75}$ and $M_{*,sph} \propto \sigma^{6.80\pm 0.94}$ for core-S\'ersic galaxies. These two relations are found to be consistent with the directly derived relations $M_{*,gal} \propto \sigma^{6.07\pm 1.04}$ and $M_{*,sph} \propto \sigma^{6.41\pm 1.31}$, obtained for our core-S\'ersic galaxies using the \textsc{BCES(bisector)} regression.

Using the single relation for all (ETG+LTG) S\'ersic galaxies, $M_{BH} \propto \sigma^{5.75 \pm 0.34}$ (Equation \ref{Ser_total}), and the $M_{BH}$--$M_{*,gal}$ (and $M_{BH}$--$M_{*,sph}$) relations for the ETGs from \citet{Sahu:2019:I},  S\'ersic ETGs are expected to follow $M_{*,gal} \propto \sigma^{3.48\pm 0.31}$ and $M_{*,sph} \propto \sigma^{4.52\pm 0.36}$. These are consistent with the directly-derived relations $M_{*,gal} \propto \sigma^{2.90\pm 0.36}$ and $M_{*,sph} \propto \sigma^{3.85\pm 0.46}$ using the \textsc{BCES(Bisector)} regression.

Similarly, for S\'ersic LTGs, using our Equation \ref{Ser_total} and the $M_{BH}$--$M_{*,gal}$ (and $M_{BH}$--$M_{*,sph}$) relations for LTGs from \citet{Davis:2018:a}, we predict the relations $M_{*,gal} \propto \sigma^{1.88\pm 0.45}$ and $M_{*,sph} \propto \sigma^{2.66\pm 0.42}$, which are consistent with the directly-derived relations $M_{*,gal} \propto \sigma^{2.00\pm 0.38}$ and $M_{*,sph} \propto \sigma^{2.96\pm 0.55}$. In the same way, the relations for all the other subcategories, as described in the above subsections, have been found to be internally consistent. In the following sections, we turn our attention to matters of external consistency.

\section{The $L$--$\sigma$ diagram}
\label{Modified Faber-Jackson}

For half a century, astronomers have been studying the correlation between the total luminosity of a galaxy and the velocity dispersion of the stars in it  \citep{Minknowski:1962}. However, with the increase in the number of reliable measurements at high and low luminosities, various studies found different relations when using different samples \citep{Faber:Jackson:1976, Schechter:1980, Malumuth:Kirshner:1981, Tonry:1981, Binney:1982, Farouki:1983, Davies:1983, Held:1992, deRijcke:2005, Matkovic:Guzman:2005, Lauer:Faber:2007}, which collectively suggested a broken or curved $L$--$\sigma$ relation \citep[see][for a brief overview of previous studies]{Graham:2016:Review, Graham:Soria:2018}. Here, we re-investigate the bend or curve in the $L$--$\sigma$ diagram.

\subsection{V-band Data-set}
Using elliptical galaxies from the  V-band data-set of \citet{Lauer:Faber:2007}, with several modifications, \citet{Kormendy:Bender:2013} reported a steep $L _V \propto \sigma^{8}$ relation for the core (core-S\'ersic) elliptical galaxies, and $L_V \propto \sigma^{4}$  for the core-less (S\'ersic)  elliptical galaxies. Although they specifically mention the use of a symmetric least squares regression routine from \citet[][modified FITEXY]{Tremaine:ngc4742:2002}, the slopes they report seem to be obtained from an asymmetric regression, i.e., a least squares minimization of the offsets in the $\sigma$-direction over V-band absolute magnitude ($\mathfrak M_{V}$) which produces a steep $L_V$--$\sigma$ slope\footnote{$\mathfrak{M}= -2.5 \log(L)$}. The modified \textsc{FITEXY} routine from \citep{Tremaine:ngc4742:2002} does not directly provide a symmetric regression line: one first needs to obtain the forward ($Y|X$) and inverse ($X|Y$)  regression lines using this routine, and then find the  bisector line.  For the data used by \citet{Kormendy:Bender:2013}, we report here that the symmetric application of the modified \textsc{FITEXY} regression routine gives $L_V \propto \sigma^{4.39 \pm 0.61}$  for the core-S\'ersic elliptical galaxies, and $L_V \propto \sigma^{2.98 \pm 0.31}$ for the S\'ersic elliptical galaxies.

We have used all of the 178 ETGs (for which $\sigma$ is available) from \citet{Lauer:Faber:2007} to revisit the V-band  $\mathfrak M_V$--$\sigma$ relations\footnote{\citet{Kormendy:Bender:2013} pruned the data sample from \citet{Lauer:Faber:2007}  by excluding many dwarf ETGs which define the low-mass slope, and by excluding some lenticular galaxies while including other lenticular galaxies which had been misclassified as elliptical galaxies (see Graham 2019b).}, except for the stripped M32-type\footnote{These M32-type compact elliptical galaxies are M32, VCC~1192 (NGC~4467), VCC~1199, VCC~1297 (NGC~4486B), VCC~1440 (IC~798), VCC~1545 (IC~3509), and VCC~1627.} compact elliptical galaxies which can bias the relation \citep[][see their Figure 11]{Graham:Soria:2018}.  We updated the core designation for the galaxies NGC~4458, NGC~4473, NGC~4478, and NGC~4482  according to \citet[][their Table 1]{Kormendy:Fisher:2009}, and the core designation of NGC~524, NGC~821, NGC~1374, NGC~3607, and NGC~5576 according to our Table \ref{Total Sample}. We also changed the designation of NGC~4552 from core-S\'ersic to S\'ersic following \citet{Bonfini:2018}, who claimed that the apparent core detected in this galaxy is because of the dust rings obstructing the light from the galactic center. 

\begin{figure*}
\begin{center}
\includegraphics[clip=true,trim= 14mm 11mm 33mm 25mm,width=   0.7\textwidth]{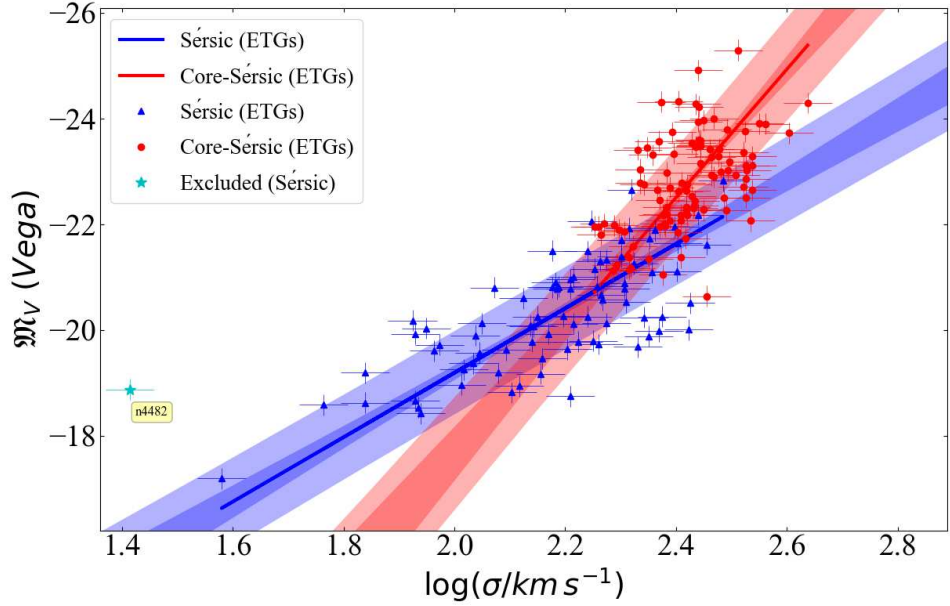}
\caption{ V-band absolute magnitude versus velocity dispersion diagram for S\'ersic and core-S\'ersic ETGs taken from the sample of \citet{Lauer:Faber:2007}. The \textsc{BCES(bisector)} regression provides the relations $L_V \propto \sigma^{2.44 \pm 0.18}$ (Equation \ref{V_mag_S}) and $L_V \propto \sigma^{4.86 \pm 0.54}$ (Equation \ref{V_mag_CS}) for S\'ersic and core-S\'ersic ETGs, respectively. This diagram suggests a broken $L$--$\sigma$ relation with the bend point at $\mathfrak M_V \approx -20.7\, \rm mag$ (Vega). }
\label{Mag_V_sigma}
\end{center}
\end{figure*}

We used a constant $10\%$ error on the velocity dispersion, and a 0.2 mag uncertainty on the absolute magnitude, i.e., a $20\%$ error in the luminosity. Before performing the regression on the updated data-set, we checked to see if any single galaxies might bias the underlying relation defined by the bulk of the sample. This led us to  exclude the S\'ersic galaxy NGC~4482  from our regressions as it appears to have an underestimated velocity dispersion (Figure \ref{Mag_V_sigma}). 

Figure \ref{Mag_V_sigma} shows the V-band magnitude versus the velocity dispersion relation for  S\'ersic and core-S\'ersic ETGs from the updated sample of \citet{Lauer:Faber:2007}. We obtain the bend-point at $\mathfrak M_V = -20.7 \, \rm mag$ (Vega), with 97 core-S\'ersic ETGs defining the  relation
\begin{IEEEeqnarray}{rCl}
\label{V_mag_CS}
\log(L_V) &=& (4.86\pm 0.54)\log\left(\frac{\sigma}{200\, \rm km\,s^{-1}}\right) \nonumber \\
&& +\> (8.52\pm 0.07),
\end{IEEEeqnarray}
with $\Delta_{rms | L_{V}}= 0.37$ dex in the $\log L_{V}$-direction, and 80 S\'ersic ETGs defining a shallower relation given by,
\begin{IEEEeqnarray}{rCl}
\label{V_mag_S}
\log(L_V) &=& (2.44\pm 0.18)\log\left(\frac{\sigma}{200\,\rm km\,s^{-1}}\right) \nonumber \\
&& +\> (8.41\pm 0.04),
\end{IEEEeqnarray}
with $\Delta_{rms | L_{V}}= 0.31$ dex, obtained using the \textsc{BCES(Bisector)} regression\footnote{Including NGC~4482 changes the S\'ersic slope to $2.18 \pm 0.25$, revealing that this single galaxy has a significant leverage on the slope of S\'ersic population, hence it is better to exclude NGC~4482.}. 

\subsection{$3.6 \,\mu \rm m$ Data-set}

\begin{figure*}
\begin{center}
\includegraphics[clip=true,trim=14mm 11mm 33mm 25mm,width=   0.7\textwidth]{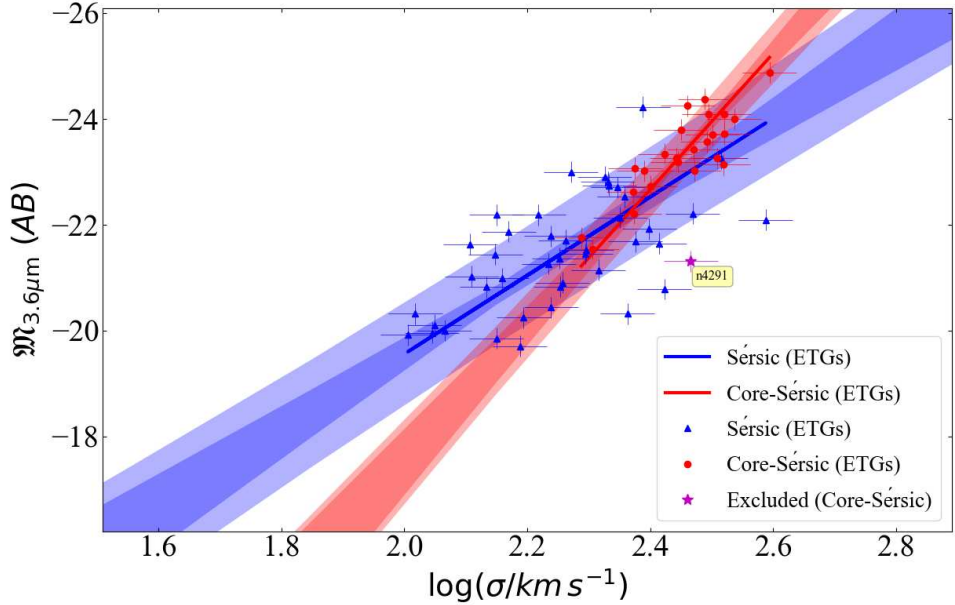}
\caption{$3.6\, \mu \rm m$ absolute magnitude versus velocity dispersion  for the S\'ersic and core-S\'ersic ETGs in our sample. We find the bend in the relation at $\mathfrak M_{3.6 \mu \rm m} \approx -22.3$ mag (AB) with S\'ersic and core-S\'ersic galaxies following the best-fit lines $L_{3.6 \mu \rm m} \propto \sigma^{2.97 \pm 0.43}$ (Equation \ref{L_Sigma_S}) and $L_{3.6 \mu \rm m} \propto \sigma^{5.16 \pm 0.53}$ (Equation \ref{L_Sigma_CS}), respectively. The color-magnitude relation for S\'ersic ETGs explains the different slope of $\sim 2.44 \pm 0.18$ in Figure \ref{Mag_V_sigma} for the $L_V$--$\sigma$ relation.}
\label{Mag_3.6_Sigma}
\end{center}
\end{figure*}

To probe the behavior of S\'ersic and core-S\'ersic ETGs in the $L$--$\sigma$ diagram using near-infrared 3.6 $\mu \rm m$-derived luminosities, we obtained the 3.6 $\mu \rm m$ absolute magnitudes ($\mathfrak M_{3.6 \mu m}$) for 73 ETGs from \citet{Sahu:2019:I}. This sample of 73 ETGs with 3.6 $\mu \rm m$ absolute magnitudes, has two galaxies (NGC~404, NGC~7457) common to our excluded sample (Section \ref{exclusion})  and five galaxies (NGC~404, NGC~1316, NGC~2787, NGC~4342 and NGC~5128) common to the exclusions applied in \citet[][their Section 4]{Sahu:2019:I}. Hence, to maintain a consistency we exclude those galaxies in the $L_{3.6 \mu m}$--$\sigma$ as well, which leaves us with a reduced 3.6 $\mu \rm m$ data-set of 67 ETGs. Checking  for considerable outliers, we found that the core-S\'ersic ETG NGC~4291 (shown in  Figure \ref{Mag_3.6_Sigma} by a magenta-colored star), is a more than $2\sigma$ outlier, and significantly biases (changes the slope for) the  best-fit line for core-S\'ersic galaxies, hence we exclude NGC~4291 from the regression. The reduced 3.6 $\mu \rm m$ ETG sample is comprised of 42 S\'ersic and 24 core-S\'ersic ETGs. 

Using our 3.6 $\mu \rm m$ data for ETGs, we recover the bend in the $L$--$\sigma$ relation  (Figure \ref{Mag_3.6_Sigma}). 
Our core-S\'ersic galaxies follow the relation 
\begin{IEEEeqnarray}{rCl}
\label{L_Sigma_CS}
\log(L_{3.6 \mu \rm m}) &=& (5.16\pm 0.53)\log\left(\frac{\sigma}{200\,\rm km\,s^{-1}}\right) \nonumber \\
&& +\> (8.56\pm 0.08),
\end{IEEEeqnarray}
with   $\Delta_{rms | L_{3.6 \mu \rm m}}= 0.19$ dex (in the $\log L_{3.6 \mu \rm m}$-direction) and S\'ersic galaxies follow the shallower relation,
\begin{IEEEeqnarray}{rCl}
\label{L_Sigma_S}
\log(L_{3.6 \mu \rm m}) &=& (2.97\pm 0.43)\log\left(\frac{\sigma}{200\,\rm km\,s^{-1}}\right) \nonumber \\
&& +\> (8.72\pm 0.07),
\end{IEEEeqnarray}
with  $\Delta_{rms | L_{3.6 \mu \rm m}}= 0.36$ dex\footnote{Including NGC~4291 in the regression changes the slope for the  core-S\'ersic galaxies to $5.94 \pm 1.00$, proving that this one single outlier does affect the relation and hence it should remain excluded.}.

The different exponent of the  relations $L_{B} \propto \sigma^{2}$ \citep{Graham:Soria:2018},  $L_{V} \propto \sigma^{2.5}$ (Figure \ref{Mag_V_sigma}, Equation \ref{V_mag_S}), and $L_{3.6\, \mu \rm m} \propto \sigma^{3}$ (Figure \ref{Mag_3.6_Sigma}, Equation \ref{L_Sigma_S}) followed by S\'ersic ETGs in different wavelength bands is consistent with the fact that they also follow a color-magnitude relation. Core-S\'ersic ETGs, on the other hand, have roughly a constant color, suggesting similar slopes of the $L$--$\sigma$ relation for all wavelength bands. The observed $L$--$\sigma$ relations for core-S\'ersic ETGs in different bands, i.e., $L_{B} \propto \sigma^{4 - 6}$ \citep{Graham:Soria:2018},  $L_{V} \propto \sigma^{4.9}$ (Figure \ref{Mag_V_sigma}, Equation \ref{V_mag_CS}), and $L_{3.6\, \mu \rm m} \propto \sigma^{5.2}$ (Figure \ref{Mag_3.6_Sigma}, Equation \ref{L_Sigma_CS}),  are consistent as expected. 

In the $3.6\, \mu \rm m$ magnitude ($\mathfrak M_{3.6\, \mu \rm m}$) versus velocity dispersion diagram, we observe the bend-point at $\mathfrak M_{3.6\, \mu \rm m}  \approx -22.3\, \rm mag $ in the AB magnitude system, which is $\mathfrak M_{3.6\, \mu \rm m}  \approx -25.1\, \rm mag $ in the Vega magnitude system. 
Assuming a $B-3.6\,\mu \rm m$ color of $\sim 5$ (based on $B-K \approx 4$ and $K-3.6\, \mu \rm m \approx 1$), it seems to be consistent with the bend-point reported by previous studies at  $\mathfrak M_B \approx -20.5\, \rm mag$ \citep{Graham:Soria:2018},  $\mathfrak M_V \approx -21\, \rm mag$ \citep{Lauer:Faber:2007}, and  $\mathfrak M_R \approx -22.17 \, \rm mag$ \citep{Matkovic:Guzman:2005}.

In \citet{Sahu:2019:I}, we found that S\'ersic and core-S\'ersic ETGs follow the same $M_{BH} \propto M_{*,gal}^{1.65\pm 0.11}$ relation.  The relations $M_{BH} \propto \sigma^{4.95\pm 0.38}$ for S\'ersic ETGs (Equation \ref{Ser_ETG}) and $M_{BH} \propto \sigma^{8.64\pm 1.10}$ for core-S\'ersic galaxies (Equation \ref{CS}), all of which are ETGs, combined with the above $M_{BH}$--$M_{*,gal}$  relation from \citet{Sahu:2019:I} predict $M_{*,gal} \propto \sigma^{3.00 \pm 0.30}$ and $ M_{*,gal} \propto \sigma^{5.24\pm 0.75}$ for S\'ersic and core-S\'ersic ETGs, respectively. These two expected relations are consistent with what we have obtained (Equations \ref{L_Sigma_S} and \ref{L_Sigma_CS}, respectively) given that a constant stellar mass-to-light ratio of $0.6\pm 0.1$ \citep{Meidt:2014} was used for $3.6\, \mu \rm m$ data in \citet{Sahu:2019:I}.

\begin{figure*}
\begin{center}
\includegraphics[clip=true,trim= 15mm 5mm 35mm 20mm,width=   0.7\textwidth]{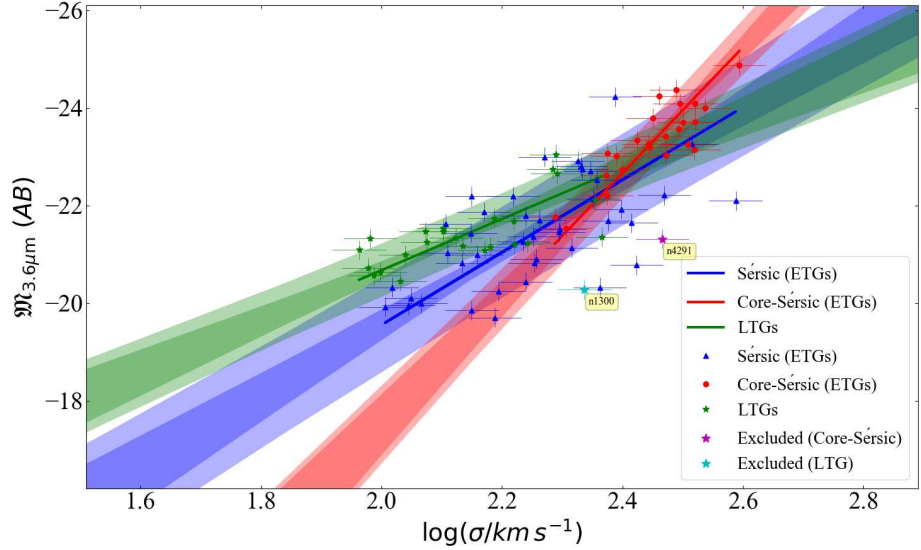}
\caption{Similar to Figure \ref{Mag_3.6_Sigma} but  including LTGs (spirals). All the spirals in our sample are S\'ersic galaxies, and they also seem to define a tight correlation in the $L$--$\sigma$ diagram (Equation \ref{L_Sigma_Sp}). }
\label{Mag_3.6_Sigma2}
\end{center}
\end{figure*}

We have also plotted and performed regressions on our 26 LTGs (with $3.6\, \mu \rm m$ data from \citet{Davis:2018:b}) in the $L_{3.6\mu \rm m}$--$\sigma$ diagram, as shown in Figure \ref{Mag_3.6_Sigma2}.  This sample of 26 LTGs, includes only one galaxy (NGC~5055) common to exclusions applied for our $M_{BH}$--$\sigma$ relations (described in Section \ref{exclusion}). In addition to NGC~5055, we also exclude  NGC~1300 as it is a considerable (more than $2\sigma$) outlier which can bias the relation for LTGs, as can be seen in Figure \ref{Mag_3.6_Sigma2} with a cyan-colored star. 

The reduced $3.6\, \mu \rm m$ sample of 24 LTGs define the relation 
\begin{IEEEeqnarray}{rCl}
\label{L_Sigma_Sp}
\log(L_{3.6 \mu \rm m}) &=& (2.10\pm 0.41)\log\left(\frac{\sigma}{200\,\rm km\,s^{-1}}\right) \nonumber \\
&& +\> (8.90\pm 0.09),
\end{IEEEeqnarray}
with  $\Delta_{rms | L_{3.6 \mu \rm m}}= 0.20$ dex\footnote{Including NGC~1300 in the regression changes the slope to $1.88 \pm 0.48$.}, consistent with the expected $M_{*, gal} \propto \sigma^{1.88 \pm 0.45}$ relation, derived from the relations $M_{BH} \propto M_{*,gal}^{3.05\pm 0.70}$ \citep{Davis:2018:a} and $M_{BH} \propto \sigma^{5.75\pm 0.34}$ (Equation \ref{LTGs}). The slope of the $L$--$\sigma$  relation that we derived for the LTGs, is also consistent with the B-band slope of 2.13 reported by \citet[][see their Figure 7]{Graham:Soria:Davis:2018}. 

The parameters obtained from the asymmetric regression routines (\textsc{BCES($Y|X$)} and \textsc{BCES($X|Y$)}), for all the $L$--$\sigma$ relations discussed above, are  presented in Table \ref{L_sigma_parameters} in the Appendix.

\section{Some Musings on Selection biases}
\label{selection bias}

The lack of directly measured low-mass SMBHs, due to the technological limitations to resolve their spheres-of-influence, poses a possible selection bias on the black hole mass scaling relations. In the past, several studies have discussed the consequences of, and possible solutions  to,  this sample selection bias \citep[e.g.,][]{Batcheldor:2010:a, Graham:2011, Shankar:2016}. 

\citet{Batcheldor:2010:a} obtained an artificial $M_{BH}$--$\sigma$ relation using simulated random $M_{BH}$ and $\sigma$ data, selected through the constraint of a best available resolution limit of $0\farcs1$ attainable from the \textit{Hubble Space Telescope (HST)}, for a maximum distance of 100 Mpc. The fake data produced the relation $ \log(M_{BH}/M_\odot) = (4.0\pm 0.3)\log\left(\sigma/200\,\rm km\,s^{-1} \right) + (8.3\pm 0.2)$, which was nearly consistent with the then observed $M_{BH}$--$\sigma$ relation of \citet{Gultekin:2009}.    
\citet{Batcheldor:2010:a}  highlighted a crucial point for assessing the credibility of the observed black hole scaling relations. However, his relation with a slope of around 4 is lower than the steeper $M_{BH}$--$\sigma$ relations  based on  larger samples of dynamically measured $M_{BH}$ data \citep{Graham:2011, McConnell:Ma:2013, Graham:Scott:2013, Savorgnan:Graham:2015, Sabra:2015}. 

\begin{figure*}
\begin{center}
\includegraphics[clip=true,trim= 10mm 6mm 15mm 10mm,width=1\textwidth]{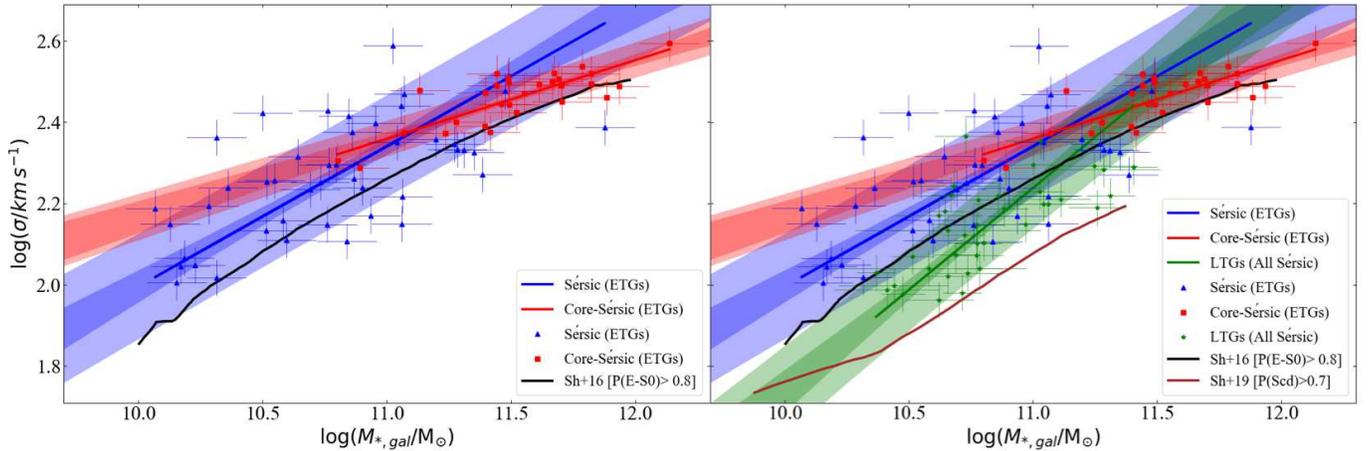}
\caption{Velocity dispersion versus total galaxy stellar mass  for S\'ersic and core-S\'ersic ETGs (left panel), and including LTGs, which are all S\'ersic galaxies, in a separate  panel for clarity.  The mean $\sigma$--$M_{*,gal}$ distribution for (i) SDSS early-type galaxies from \citet[][their Figure 1]{Shankar:2016} (black curve) and (ii) late spiral galaxies (P(Scd)$>$0.7) from \citet[][their Figure 1]{Shankar:2019} (brown curve) are shown. 
The brown curve resides below the relation defined by our LTG sample which are predominantly early (Sa-Sb) spirals. 
The black curve may reside below the relation defined by our ETG sample because of contamination by early spirals.   
}
\label{Shankar16}
\end{center}
\end{figure*}

\citet{Shankar:2016} claim that galaxies which host a directly measured central SMBH have a higher velocity dispersion in comparison to other galaxies of similar stellar mass but without a direct SMBH measurement. Their claim is based on the offset they observed in the velocity dispersion versus galaxy stellar mass diagram ($\sigma$--$M_{STAR}$, their Figure-1), between several samples of  local ETGs with dynamically measured SMBH masses and a larger data-set of galaxies from Data Release-7 of the Sloan Digital Sky Survey \citep[SDSS,][]{York:2000, Abazajian:2009}. This is restated in  \citet{Shankar:2019} with a slight change in their galaxy stellar masses based on the SDSS data they used.

\citet{Shankar:2016} suggest that the offset they obtain is a consequence of a sample selection effect in which galaxies with low-mass BHs are excluded because it is not possible to resolve their spheres-of-influence due to technological limitations. They performed the comparison with the data from four different observational studies and provided a unified conclusion that galaxies hosting a directly-measured SMBH are offset in the $\sigma$--$M_{*, gal}$ relation, such that they have  a higher $\sigma$ relative to other similar mass galaxies. However, this is not completely true for all the data-sets they used and all of the galaxy stellar mass range in their plots.  In their Figure-1, a significant number of data points from \citet{Savorgnan:2016:Slopes} overlap with the grey $\pm 1\sigma$ dispersion bands around the mean  curve of the SDSS data, especially in the high-mass range $ 11 \lesssim \log(M_{*,gal}/M_\odot) \lesssim 12 \,\rm$. This can similarly be observed in Figure 1 of \citet{Shankar:2019}.    

Interestingly, as described in Section \ref{Consistency}, we have shown that  S\'ersic and  core-S\'ersic ETGs follow two distinct $M_{*,gal}$--$\sigma$ relations,  consistent with S\'ersic and core-S\'ersic ETGs following two different $M_{BH}$--$\sigma$ relations (Section \ref{S_CS_subsection}), but a single $M_{BH}$--$M_{*,gal}$ relation \citep{Sahu:2019:I}. Thus, we have two different relations in the $\sigma$--$M_{*,gal}$ diagram for S\'ersic and core-S\'ersic ETGs as shown in the left panel of Figure \ref{Shankar16}. The mean (black) curve from \citet{Shankar:2016} lays within the $\pm 1 \sigma$ scatter of the two relations followed by our S\'ersic and core-S\'ersic ETGs with directly-measured black hole masses, but outside of the more relevant darker (red and blue) bands denoting the $\pm 1 \sigma$ uncertainty on the $\sigma$--$M_{*,gal}$ relations for ETGs with directly-measured black hole masses. 
  
Upon inclusion of our LTGs (in the right panel of Figure \ref{Shankar16}), all of which are S\'ersic galaxies, along with the  (core-S\'ersic and S\'ersic) ETGs,  we find that at the low-mass range,  $ 10 \lesssim \log(M_{*,gal}/M_\odot) \lesssim  11 \,\rm$, their (black) curve resides  between the two relations followed by our S\'ersic ETGs (blue line) and LTGs (green line) which are primarily early-type (Sa-Sc) spiral galaxies. This suggests that their galaxy sample of ETGs may contain LTGs which could (partly) cause the offset. 

In \citet{Shankar:2016}, the criteria for selecting only ETGs out of the exhaustive SDSS data-set was based upon having a probability of greater than 0.8 for a galaxy being an E- or S0-type ($P(E-S0) \geq 0.8$). From the probabilities of galaxy types made available by \citet{Meert:Vikram:2015}, we have calculated a $\sim$10$\%$ contamination by spiral galaxies (LTGs) in the Shankar et al.'s ETG sample.
Their best-fit $\sigma$--$M_{*,gal}$ relation's  position in-between the relation followed by our S\'ersic  ETGs and LTGs (right panel of Figure \ref{Shankar16}), coupled with their ETG selection criteria based on probability, supports the  suspicion that some of  the offset may be due to spiral galaxy contamination in their SDSS ETG sample. 

In the right-hand panel of Figure \ref{Shankar16}, we also include the brown curve for late spiral galaxies ($P(Scd) \geq 0.7$) from \citet[][see the left panel in their Figure 1]{Shankar:2019}, which lies below the relation defined by our  predominantly early spiral galaxies (Sa-Sb), simply referred to as LTGs in this paper.
The various curves in Figure \ref{Shankar16}  represent the major morphological types.  Their layering suggests that the apparent offset between galaxies with and without a directly measured black hole mass, as observed by \citet{Shankar:2016, Shankar:2019}, could simply be a reflection of the difference in the dominant morphological type  in each sample. However, this is not conclusive and further investigation is required as their may yet be a selection bias or a discrepancy in the way that velocity dispersions are measured.

\section{Conclusions and Implications}
\label{conclusions}

Using the reduced sample of 137 galaxies with updated black hole masses and  central stellar velocity dispersions, our work reveals sub-structure in the $M_{BH}$--$\sigma$ diagram due to galaxies with and without a core. 
Our previous galaxy decompositions \citep{Savorgnan:Graham:2016:I, Davis:2018:a, Sahu:2019:I} have enabled us to accurately identify various structural components, such as intermediate or extended disks, bars, and partially-depleted stellar cores. This allowed us to search for substructures in the $M_{BH}$--$\sigma$ diagram, based on galaxy morphology, and also enabled us to clarify the situation regarding offset barred galaxies found in previous observational studies.

We performed and reported both symmetric \textsc{BCES(Bisector)} and asymmetric \textsc{BCES($Y|X$)} and \textsc{BCES($X|Y$)} regressions. The best-fit  line obtained from the symmetric \textsc{BCES(Bisector)} regression is preferred because we are looking for a fundamental relation between two quantities \citep{Feigelson:Babu:1992, Novak:2006}. For all our relations, we also obtained a symmetric (bisector) regression line using the \textsc{MPFITEXY} (modified \textsc{FITEXY}) routine, which are consistent with the corresponding \textsc{BCES(Bisector)} best-fit lines within the $\pm 1 \sigma$ limits of the slopes and intercepts.

Our main results can be summarized as follows:

\begin{itemize}

\item{The consistency between the best-fit lines for ETGs and LTGs in the $M_{BH}$ versus $\sigma$ diagram (Figure \ref{ETG86_LTG44}),  suggests that ETGs and LTGs follow the same $M_{BH} \propto \sigma^{6.10 \pm 0.28} $ relation with a total scatter of $\Delta_{rms | BH} = 0.53$ dex, obtained using a single regression (Equation \ref{single_reg}). However, this result depends on the galaxy sample and is somewhat misleading or limited. It is a fusion of substructures caused by (massive) core-S\'ersic and (low-mass) S\'ersic galaxies following two different  $M_{BH}$--$\sigma$ relations. }

\item{Core-S\'ersic galaxies define the relation $M_{BH} \propto \sigma^{8.64 \pm 1.10}$ (Equation \ref{CS}) and S\'ersic galaxies define the relation $M_{BH} \propto \sigma^{5.75 \pm 0.34}$ (Equation \ref{Ser_total}),  with  $\Delta_{rms | BH} = 0.46$ dex and  $\Delta_{rms | BH} = 0.55$ dex, respectively. The inconsistency between the slopes of these two relations suggests two distinct relations in the $M_{BH}$--$\sigma$ diagram. The two lines intersect at $\sigma \approx 255 \, \rm km  \,s^{-1}$ in Figure \ref{CS_S_total} .}

\item{We also detect a substructure in the $M_{BH}$--$\sigma$ diagram upon dividing our sample into galaxies with and without a stellar disk (Figures \ref{rot_ETG} and \ref{rot_total}). However, this is likely  because most of the elliptical ETGs are massive core-S\'ersic galaxies, while most of the galaxies with a disk (ES, S0, and Sp-types) are S\'ersic galaxies.}

\item{We do not find any offset between the slope or intercept of the best-fit lines for barred and non-barred galaxies (Figures \ref{bar_ETG} and \ref{bar_total}). We reveal that some previous studies noticed an offset in the intercepts between the $M_{BH}$--$\sigma$ relations for barred and non-barred galaxies partly because they relied on incomplete bar morphologies for several galaxies which failed to identify weak bars. Our previous image analysis improved upon this situation, and  in our current larger sample we also have new galaxies with bars. Given that bars are known to elevate the velocity dispersion \citep{Hartmann:2014}, this result begs further investigation, possibly folding in disc inclination, bar orientation to our line-of-sight, and rotational velocity.}

\item{Galaxies with and without an AGN follow consistent relations in the $M_{BH}$--$\sigma$ diagram (Figure \ref{AGN}). Hence, the  $M_{BH}$--$\sigma$ relations defined by S\'ersic and core-S\'ersic galaxies should be valid for a galaxy irrespective of whether or not its nucleus is active.}

\item{Analyzing the $L$--$\sigma$ relation, based on  V-band data from \citet{Lauer:Faber:2007}, our $3.6\, \mu$m data from Spitzer, and previously reported  $L$--$\sigma$ relations using B- and R-bands, we investigated the  $L$--$\sigma$ relation (Figures \ref{Mag_V_sigma} and \ref{Mag_3.6_Sigma}). We found that the relation between the luminosity of a galaxy and its central stellar velocity dispersion is bent due to core-S\'ersic and S\'ersic galaxies, analogous and consistent with the bend found in the $M_{BH}$--$\sigma$ relation and the $L$--$\mu_{0}$ relation \citep{Graham:Guzman:2003}.  Core-S\'ersic galaxies follow the relation  $L_{V} \propto \sigma^{4.86\pm 0.54}$ and $L_{3.6 \,\mu \rm m} \propto \sigma^{5.16\pm 0.53}$ (Equations \ref{V_mag_CS} and \ref{L_Sigma_CS}), whereas S\'ersic galaxies follow the relation $L_V \propto \sigma^{2.44 \pm 0.18}$ and $L_{3.6 \,\mu \rm m} \propto \sigma^{2.97 \pm 0.43}$ (Equations \ref{V_mag_S} and \ref{L_Sigma_S}).  The bend-point  is consistent in the B-, V-, and $3.6 \,\mu \rm m$ bands, corresponding to a stellar mass of $\approx 11 \, M_{\odot}$.}

\item{The LTGs in our sample follow the relation $L_{3.6 \,\mu \rm m} \propto \sigma^{2.10 \pm 0.41}$ (Equation \ref{L_Sigma_Sp}), and  the $L_{3.6 \,\mu \rm m}$--$\sigma$ relations for S\'ersic ETGs, core-S\'ersic ETGs, and LTGs  are internally consistent with our $M_{BH}$--$\sigma$ relations, and  the $M_{BH}$--$M_{*,gal}$ relations from  \citep{Sahu:2019:I}.}

\end{itemize}
  
Our $M_{BH}$--$\sigma$ (and $M_{BH}$--$M_{*,gal}$, and $M_{BH}$--$M_{*,sph}$) relations hold insights for theoretical studies into the co-evolution of black holes with their host galaxy properties \citep[e.g.,][]{Volonteri:2013, Heckman:2014}, AGN feedback \citep{Marconi:2008}, and the connection between black hole growth and star formation rates which have been found to depend on  galaxy morphology \citep{Calvi:SFR:2018}. Black hole mass scaling relations are also used to determine virial $f$-factors, for calculating AGN (black hole) masses  \citep[e.g.,][]{Onken:2004, Graham:2011, Bennert:2011, Bentz:2015, Yu:2019}. Our $M_{BH}$--$\sigma$ relation due to S\'ersic and core-S\'ersic galaxies can be used to improve the virial $f$-factor based upon the galaxy core-type. 

The new black hole mass scaling relations can be used to estimate the black hole masses of other galaxies using their easily measured properties, i.e., their galaxy stellar mass, spheroid/bulge stellar mass, or stellar velocity dispersion. 
These scaling relations, based on high resolution images of local ($z \sim 0$) galaxies, provide a benchmark for  studies attempting to determine the evolution of the $M_{BH}$--$\sigma$ (or $M_{BH}$--$M_{*,gal}$ and $M_{BH}$--$M_{*,sph}$) relations \citep{Woo:2006, Salviander:2007, Bennert:2011, Sexton:2019}. Moreover, given the different scaling relations based on the galaxy sub-morphologies, care should be taken in regard to the galaxy types present in one's sample. For distant galaxies where it is difficult to perform multi-component decompositions to obtain bulge masses and extract detailed morphologies, $M_{BH}$--$M_{*,gal}$ relations can be used provided ETG or LTG classifications are known because ETGs and LTGs follow two different $M_{BH}$--$M_{*,gal}$ relations \citep{Sahu:2019:I}. Similarly, as it might be difficult to detect the (depleted) core in  distant galaxies, the single regression $M_{BH}$--$\sigma$ relation  presented in this paper  (Equation \ref{single_reg}) can be used. However, if one is primarily sampling massive distant galaxies, with $\sigma \gtrsim 255 \rm km s^{-1}$, it would be preferable to compare that data with the core-S\'ersic $M_{BH}$--$\sigma$ relation, or  risk inferring a false evolution if using the shallower relation.

Our scaling relations can be used to estimate black hole masses for a large data-set of galaxies to obtain the black hole mass function in the local Universe \citep{McLure:Dunlop:2004, Shankar:Salucci:2004, Graham:MGC:2007}.  This can be used to improve the predictions of the amplitude and frequency of ground-based detections of long-wavelength gravitational waves, produced by merging SMBHs, using pulsar timing arrays \citep{Shannon:2015, Hobbs:2017} and also MeerKAT \citep{Jonas:2007}. Furthermore, these scaling relations can also be used to constrain the space-based detection of  long-wavelength gravitational waves by the  Laser Interferometer Space Antenna \citep[LISA,][]{Danzmann:2017}, and  beyond LISA  \citep[bLISA,][]{Baker:2019}.

\acknowledgements
We thank the anonymous referee whose comments helped to increase the clarity of this paper.
This research was conducted with the Australian Research Council Centre of Excellence for Gravitational Wave Discovery (OzGrav), through project number CE170100004. AWG was supported under the Australian Research Council's funding scheme DP17012923. This work has made use of the NASA/IPAC Infrared Science Archive and the NASA/IPAC Extragalactic Database (NED). This research has also made use of the Two Micron All Sky Survey and Sloan Digital Sky Survey database. We also acknowledge the use of the   \textsc{HyperLeda}  database \url{http://leda.univ-lyon1.fr}.

\bibliography{bibliography}

\begin{thebibliography}{}
\expandafter\ifx\csname natexlab\endcsname\relax\def\natexlab#1{#1}\fi
\providecommand{\url}[1]{\href{#1}{#1}}
\providecommand{\dodoi}[1]{doi:~\href{http://doi.org/#1}{\nolinkurl{#1}}}
\providecommand{\doeprint}[1]{\href{http://ascl.net/#1}{\nolinkurl{http://ascl.net/#1}}}
\providecommand{\doarXiv}[1]{\href{https://arxiv.org/abs/#1}{\nolinkurl{https://arxiv.org/abs/#1}}}

\bibitem[{{Abazajian} {et~al.}(2009){Abazajian}, {Adelman-McCarthy},
  {Ag{\"u}eros}, {Allam}, {Allende Prieto}, {An}, {Anderson}, {Anderson},
  {Annis}, {Bahcall}, \& et~al.}]{Abazajian:2009}
{Abazajian}, K.~N., {Adelman-McCarthy}, J.~K., {Ag{\"u}eros}, M.~A., {et~al.}
  2009, \apjs, 182, 543, \dodoi{10.1088/0067-0049/182/2/543}

\bibitem[{{Akritas} \& {Bershady}(1996)}]{Akritas:Bershady:1996}
{Akritas}, M.~G., \& {Bershady}, M.~A. 1996, \apj, 470, 706,
  \dodoi{10.1086/177901}

\bibitem[{{Baker} {et~al.}(2019){Baker}, {Barke}, {Bender}, {Berti},
  {Caldwell}, {Conklin}, {Cornish}, {Ferrara}, {Holley-Bockelmann}, {Kamai},
  {Larson}, {Livas}, {McWilliams}, {Mueller}, {Natarajan}, {Rioux}, {Sankar},
  {Schnittman}, {Shoemaker}, {Slutsky}, {Stebbins}, {Thorpe}, \&
  {Ziemer}}]{Baker:2019}
{Baker}, J., {Barke}, S.~F., {Bender}, P.~L., {et~al.} 2019, arXiv e-prints,
  arXiv:1907.11305.
\newblock \doarXiv{1907.11305}

\bibitem[{{Batcheldor}(2010)}]{Batcheldor:2010:a}
{Batcheldor}, D. 2010, \apjl, 711, L108, \dodoi{10.1088/2041-8205/711/2/L108}

\bibitem[{{Batcheldor} {et~al.}(2010){Batcheldor}, {Robinson}, {Axon},
  {Perlman}, \& {Merritt}}]{Batcheldor:2010:b}
{Batcheldor}, D., {Robinson}, A., {Axon}, D.~J., {Perlman}, E.~S., \&
  {Merritt}, D. 2010, \apj, 717, L6, \dodoi{10.1088/2041-8205/717/1/L6}

\bibitem[{{Bedregal} {et~al.}(2006){Bedregal}, {Arag{\'o}n-Salamanca}, \&
  {Merrifield}}]{Bedregal:2006}
{Bedregal}, A.~G., {Arag{\'o}n-Salamanca}, A., \& {Merrifield}, M.~R. 2006,
  \mnras, 373, 1125, \dodoi{10.1111/j.1365-2966.2006.11031.x}

\bibitem[{{Begelman} {et~al.}(1980){Begelman}, {Blandford}, \&
  {Rees}}]{Begelman:1980}
{Begelman}, M.~C., {Blandford}, R.~D., \& {Rees}, M.~J. 1980, \nat, 287, 307,
  \dodoi{10.1038/287307a0}

\bibitem[{{Bellstedt} {et~al.}(2017){Bellstedt}, {Graham}, {Forbes},
  {Romanowsky}, {Brodie}, \& {Strader}}]{Bellstedt:Graham:2017}
{Bellstedt}, S., {Graham}, A.~W., {Forbes}, D.~A., {et~al.} 2017, \mnras, 470,
  1321, \dodoi{10.1093/mnras/stx1348}

\bibitem[{{Bennert} {et~al.}(2011){Bennert}, {Auger}, {Treu}, {Woo}, \&
  {Malkan}}]{Bennert:2011}
{Bennert}, V.~N., {Auger}, M.~W., {Treu}, T., {Woo}, J.-H., \& {Malkan}, M.~A.
  2011, \apj, 726, 59, \dodoi{10.1088/0004-637X/726/2/59}

\bibitem[{{Bennert} {et~al.}(2015){Bennert}, {Treu}, {Auger}, {Cosens}, {Park},
  {Rosen}, {Harris}, {Malkan}, \& {Woo}}]{Bennert:2015}
{Bennert}, V.~N., {Treu}, T., {Auger}, M.~W., {et~al.} 2015, \apj, 809, 20,
  \dodoi{10.1088/0004-637X/809/1/20}

\bibitem[{{Bentz} \& {Katz}(2015)}]{Bentz:2015}
{Bentz}, M.~C., \& {Katz}, S. 2015, \pasp, 127, 67, \dodoi{10.1086/679601}

\bibitem[{{Binney}(1982)}]{Binney:1982}
{Binney}, J. 1982, \araa, 20, 399, \dodoi{10.1146/annurev.aa.20.090182.002151}

\bibitem[{{Blom} {et~al.}(2014){Blom}, {Forbes}, {Foster}, {Romanowsky}, \&
  {Brodie}}]{Blom:Forbes:2014}
{Blom}, C., {Forbes}, D.~A., {Foster}, C., {Romanowsky}, A.~J., \& {Brodie},
  J.~P. 2014, \mnras, 439, 2420, \dodoi{10.1093/mnras/stu095}

\bibitem[{{Bogd{\'a}n} {et~al.}(2018){Bogd{\'a}n}, {Lovisari}, {Volonteri}, \&
  {Dubois}}]{Bogdan:2018}
{Bogd{\'a}n}, {\'A}., {Lovisari}, L., {Volonteri}, M., \& {Dubois}, Y. 2018,
  \apj, 852, 131, \dodoi{10.3847/1538-4357/aa9ab5}

\bibitem[{{Boizelle} {et~al.}(2019){Boizelle}, {Barth}, {Walsh}, {Buote},
  {Baker}, {Darling}, \& {Ho}}]{Boizelle:2019}
{Boizelle}, B.~D., {Barth}, A.~J., {Walsh}, J.~L., {et~al.} 2019, arXiv
  e-prints, arXiv:1906.06267.
\newblock \doarXiv{1906.06267}

\bibitem[{{Bonfini} {et~al.}(2018){Bonfini}, {Gonz{\'a}lez-Mart{\'\i}n},
  {Fritz}, {Bitsakis}, {Bruzual}, \& {Cervantes Sodi}}]{Bonfini:2018}
{Bonfini}, P., {Gonz{\'a}lez-Mart{\'\i}n}, O., {Fritz}, J., {et~al.} 2018,
  \mnras, 478, 1161, \dodoi{10.1093/mnras/sty1087}

\bibitem[{{Brown} {et~al.}(2013){Brown}, {Valluri}, {Shen}, \&
  {Debattista}}]{Brown:Simulation:2013}
{Brown}, J.~S., {Valluri}, M., {Shen}, J., \& {Debattista}, V.~P. 2013, \apj,
  778, 151, \dodoi{10.1088/0004-637X/778/2/151}

\bibitem[{{Buta} {et~al.}(2010){Buta}, {Sheth}, {Regan}, {Hinz}, {Gil de Paz},
  {Men{\'e}ndez-Delmestre}, {Munoz-Mateos}, {Seibert}, {Laurikainen}, {Salo},
  {Gadotti}, {Athanassoula}, {Bosma}, {Knapen}, {Ho}, {Madore}, {Elmegreen},
  {Masters}, {Comer{\'o}n}, {Aravena}, \& {Kim}}]{Buta:Sheth:2010}
{Buta}, R.~J., {Sheth}, K., {Regan}, M., {et~al.} 2010, \apjs, 190, 147,
  \dodoi{10.1088/0067-0049/190/1/147}

\bibitem[{{Calvi} {et~al.}(2018){Calvi}, {Vulcani}, {Poggianti}, {Moretti},
  {Fritz}, \& {Fasano}}]{Calvi:SFR:2018}
{Calvi}, R., {Vulcani}, B., {Poggianti}, B.~M., {et~al.} 2018, \mnras, 481,
  3456, \dodoi{10.1093/mnras/sty2476}

\bibitem[{{Cappellari} {et~al.}(2006){Cappellari}, {Bacon}, {Bureau}, {Damen},
  {Davies}, {de Zeeuw}, {Emsellem}, {Falc{\'o}n-Barroso}, {Krajnovi{\'c}},
  {Kuntschner}, {McDermid}, {Peletier}, {Sarzi}, {van den Bosch}, \& {van de
  Ven}}]{Cappellari:2006}
{Cappellari}, M., {Bacon}, R., {Bureau}, M., {et~al.} 2006, \mnras, 366, 1126,
  \dodoi{10.1111/j.1365-2966.2005.09981.x}

\bibitem[{{Ciambur}(2015)}]{Ciambur:2015:Ellipse}
{Ciambur}, B.~C. 2015, \apj, 810, 120, \dodoi{10.1088/0004-637X/810/2/120}

\bibitem[{{Ciambur}(2016)}]{Ciambur:2016:Profiler}
---. 2016, \pasa, 33, e062, \dodoi{10.1017/pasa.2016.60}

\bibitem[{{Ciotti} \& {van Albada}(2001)}]{Ciotti:vanAlbada:2001}
{Ciotti}, L., \& {van Albada}, T.~S. 2001, \apjl, 552, L13,
  \dodoi{10.1086/320260}

\bibitem[{{Combes} {et~al.}(2019){Combes}, {Garc{\'{\i}}a-Burillo}, {Audibert},
  {Hunt}, {Eckart}, {Aalto}, {Casasola}, {Boone}, {Krips}, {Viti}, {Sakamoto},
  {Muller}, {Dasyra}, {van der Werf}, \& {Martin}}]{Combes:2019:SMBH4}
{Combes}, F., {Garc{\'{\i}}a-Burillo}, S., {Audibert}, A., {et~al.} 2019, \aap,
  623, A79, \dodoi{10.1051/0004-6361/201834560}

\bibitem[{{Dalla Bont{\`a}} {et~al.}(2009){Dalla Bont{\`a}}, {Ferrarese},
  {Corsini}, {Miralda-Escud{\'e}}, {Coccato}, {Sarzi}, {Pizzella}, \&
  {Beifiori}}]{DallaBonta:2009}
{Dalla Bont{\`a}}, E., {Ferrarese}, L., {Corsini}, E.~M., {et~al.} 2009, \apj,
  690, 537, \dodoi{10.1088/0004-637X/690/1/537}

\bibitem[{{Danzmann}(2017)}]{Danzmann:2017}
{Danzmann}, K. 2017, in Society of Photo-Optical Instrumentation Engineers
  (SPIE) Conference Series, Vol. 10566, \procspie, 1056610,
  \dodoi{10.1117/12.2308272}

\bibitem[{{Davies} {et~al.}(1983)}]{Davies:1983}
{Davies}, R.~L., {et~al.} 1983, \apj, 266, 41, \dodoi{10.1086/160757}

\bibitem[{{Davis} {et~al.}(2018){Davis}, {Graham}, \& {Cameron}}]{Davis:2018:b}
{Davis}, B.~L., {Graham}, A.~W., \& {Cameron}, E. 2018, \apj, 869, 113,
  \dodoi{10.3847/1538-4357/aae820}

\bibitem[{Davis {et~al.}(2019)Davis, Graham, \& Cameron}]{Davis:2018:a}
Davis, B.~L., Graham, A.~W., \& Cameron, E. 2019, \apj, 873, 85,
  \dodoi{10.3847/1538-4357/aaf3b8}

\bibitem[{{Davis} {et~al.}(2017){Davis}, {Graham}, \&
  {Seigar}}]{Davis:Graham:2017}
{Davis}, B.~L., {Graham}, A.~W., \& {Seigar}, M.~S. 2017, \mnras, 471, 2187,
  \dodoi{10.1093/mnras/stx1794}

\bibitem[{{de Rijcke} {et~al.}(2005){de Rijcke}, {Michielsen}, {Dejonghe},
  {Zeilinger}, \& {Hau}}]{deRijcke:2005}
{de Rijcke}, S., {Michielsen}, D., {Dejonghe}, H., {Zeilinger}, W.~W., \&
  {Hau}, G.~K.~T. 2005, \aap, 438, 491, \dodoi{10.1051/0004-6361:20042213}

\bibitem[{{de Vaucouleurs} {et~al.}(1991){de Vaucouleurs}, {de Vaucouleurs},
  {Corwin}, {Buta}, {Paturel}, \& {Fouque}}]{RC3:1991}
{de Vaucouleurs}, G., {de Vaucouleurs}, A., {Corwin}, Herold~G., J., {et~al.}
  1991, {Third Reference Catalogue of Bright Galaxies}

\bibitem[{{Dullo}(2014)}]{Dullo:S0:2014}
{Dullo}, B.~T. 2014, in Astronomical Society of the Pacific Conference Series,
  Vol. 480, Structure and Dynamics of Disk Galaxies, ed. M.~S. {Seigar} \&
  P.~{Treuthardt}, 75

\bibitem[{{Dullo} \& {Graham}(2013)}]{Dullo:Graham:2013}
{Dullo}, B.~T., \& {Graham}, A.~W. 2013, \apj, 768, 36,
  \dodoi{10.1088/0004-637X/768/1/36}

\bibitem[{{Dullo} \& {Graham}(2014)}]{Dullo:2014}
---. 2014, \mnras, 444, 2700, \dodoi{10.1093/mnras/stu1590}

\bibitem[{{Erwin} {et~al.}(2015){Erwin}, {Saglia}, {Fabricius}, {Thomas},
  {Nowak}, {Rusli}, {Bender}, {Vega Beltr{\'a}n}, \&
  {Beckman}}]{Erwin:Saglia:2015}
{Erwin}, P., {Saglia}, R.~P., {Fabricius}, M., {et~al.} 2015, \mnras, 446,
  4039, \dodoi{10.1093/mnras/stu2376}

\bibitem[{{Eskridge} {et~al.}(2000){Eskridge}, {Frogel}, {Pogge}, {Quillen},
  {Davies}, {DePoy}, {Houdashelt}, {Kuchinski}, {Ram{\'{\i}}rez}, {Sellgren},
  {Terndrup}, \& {Tiede}}]{Eskridge:2000}
{Eskridge}, P.~B., {Frogel}, J.~A., {Pogge}, R.~W., {et~al.} 2000, \aj, 119,
  536, \dodoi{10.1086/301203}

\bibitem[{{Event Horizon Telescope Collaboration} {et~al.}(2019){Event Horizon
  Telescope Collaboration}, {Akiyama}, {Alberdi}, {Alef}, {Asada}, {Azulay},
  {Baczko}, {Ball}, {Balokovi{\'c}}, {Barrett}, {Bintley}, {Blackburn},
  {Boland}, {Bouman}, {Bower}, {Bremer}, {Brinkerink}, {Brissenden}, {Britzen},
  {Broderick}, {Broguiere}, {Bronzwaer}, {Byun}, {Carlstrom}, {Chael}, {Chan},
  {Chatterjee}, {Chatterjee}, {Chen}, {Chen}, {Cho}, {Christian}, {Conway},
  {Cordes}, {Crew}, {Cui}, {Davelaar}, {De Laurentis}, {Deane}, {Dempsey},
  {Desvignes}, {Dexter}, {Doeleman}, {Eatough}, {Falcke}, {Fish}, {Fomalont},
  {Fraga-Encinas}, {Friberg}, {Fromm}, {G{\'o}mez}, {Galison}, {Gammie},
  {Garc{\'\i}a}, {Gentaz}, {Georgiev}, {Goddi}, {Gold}, {Gu}, {Gurwell},
  {Hada}, {Hecht}, {Hesper}, {Ho}, {Ho}, {Honma}, {Huang}, {Huang}, {Hughes},
  {Ikeda}, {Inoue}, {Issaoun}, {James}, {Jannuzi}, {Janssen}, {Jeter}, {Jiang},
  {Johnson}, {Jorstad}, {Jung}, {Karami}, {Karuppusamy}, {Kawashima},
  {Keating}, {Kettenis}, {Kim}, {Kim}, {Kim}, {Kino}, {Koay}, {Koch}, {Koyama},
  {Kramer}, {Kramer}, {Krichbaum}, {Kuo}, {Lauer}, {Lee}, {Li}, {Li},
  {Lindqvist}, {Liu}, {Liuzzo}, {Lo}, {Lobanov}, {Loinard}, {Lonsdale}, {Lu},
  {MacDonald}, {Mao}, {Markoff}, {Marrone}, {Marscher}, {Mart{\'\i}-Vidal},
  {Matsushita}, {Matthews}, {Medeiros}, {Menten}, {Mizuno}, {Mizuno}, {Moran},
  {Moriyama}, {Moscibrodzka}, {M{\"u}ller}, {Nagai}, {Nagar}, {Nakamura},
  {Narayan}, {Narayanan}, {Natarajan}, {Neri}, {Ni}, {Noutsos}, {Okino},
  {Olivares}, {Oyama}, {{\"O}zel}, {Palumbo}, {Patel}, {Pen}, {Pesce},
  {Pi{\'e}tu}, {Plambeck}, {PopStefanija}, {Porth}, {Prather},
  {Preciado-L{\'o}pez}, {Psaltis}, {Pu}, {Ramakrishnan}, {Rao}, {Rawlings},
  {Raymond}, {Rezzolla}, {Ripperda}, {Roelofs}, {Rogers}, {Ros}, {Rose},
  {Roshanineshat}, {Rottmann}, {Roy}, {Ruszczyk}, {Ryan}, {Rygl},
  {S{\'a}nchez}, {S{\'a}nchez-Arguelles}, {Sasada}, {Savolainen}, {Schloerb},
  {Schuster}, {Shao}, {Shen}, {Small}, {Sohn}, {SooHoo}, {Tazaki}, {Tiede},
  {Tilanus}, {Titus}, {Toma}, {Torne}, {Trent}, {Trippe}, {Tsuda}, {van
  Bemmel}, {van Langevelde}, {van Rossum}, {Wagner}, {Wardle}, {Weintroub},
  {Wex}, {Wharton}, {Wielgus}, {Wong}, {Wu}, {Young}, {Young}, {Younsi},
  {Yuan}, {Yuan}, {Zensus}, {Zhao}, {Zhao}, {Zhu}, {Farah}, {Meyer-Zhao},
  {Michalik}, {Nadolski}, {Nishioka}, {Pradel}, {Primiani}, {Souccar},
  {Vertatschitsch}, \& {Yamaguchi}}]{EHT:M87:2019}
{Event Horizon Telescope Collaboration}, {Akiyama}, K., {Alberdi}, A., {et~al.}
  2019, \apj, 875, L6, \dodoi{10.3847/2041-8213/ab1141}

\bibitem[{{Faber} \& {Jackson}(1976)}]{Faber:Jackson:1976}
{Faber}, S.~M., \& {Jackson}, R.~E. 1976, \apj, 204, 668,
  \dodoi{10.1086/154215}

\bibitem[{{Fabian}(1999)}]{Fabian:1999}
{Fabian}, A.~C. 1999, \mnras, 308, L39,
  \dodoi{10.1046/j.1365-8711.1999.03017.x}

\bibitem[{{Farouki} {et~al.}(1983){Farouki}, {Shapiro}, \&
  {Duncan}}]{Farouki:1983}
{Farouki}, R.~T., {Shapiro}, S.~L., \& {Duncan}, M.~J. 1983, \apj, 265, 597,
  \dodoi{10.1086/160704}

\bibitem[{{Feigelson} \& {Babu}(1992)}]{Feigelson:Babu:1992}
{Feigelson}, E.~D., \& {Babu}, G.~J. 1992, \apj, 397, 55,
  \dodoi{10.1086/171766}

\bibitem[{{Ferrarese} \& {Ford}(2005)}]{Ferrarese:Ford:2005}
{Ferrarese}, L., \& {Ford}, H. 2005, \ssr, 116, 523,
  \dodoi{10.1007/s11214-005-3947-6}

\bibitem[{{Ferrarese} \& {Merritt}(2000)}]{Ferrarese:Merritt:2000}
{Ferrarese}, L., \& {Merritt}, D. 2000, \apj, 539, L9, \dodoi{10.1086/312838}

\bibitem[{{Gebhardt} {et~al.}(2000){Gebhardt}, {Bender}, {Bower}, {Dressler},
  {Faber}, {Filippenko}, {Green}, {Grillmair}, {Ho}, {Kormendy}, {Lauer},
  {Magorrian}, {Pinkney}, {Richstone}, \& {Tremaine}}]{Gebhardt:2000}
{Gebhardt}, K., {Bender}, R., {Bower}, G., {et~al.} 2000, \apj, 539, L13,
  \dodoi{10.1086/312840}

\bibitem[{{Graham}(2007)}]{Graham:2007}
{Graham}, A. 2007, in Bulletin of the American Astronomical Society, Vol.~39,
  American Astronomical Society Meeting Abstracts, 759

\bibitem[{{Graham}(2008{\natexlab{a}})}]{Graham:2008:b}
{Graham}, A.~W. 2008{\natexlab{a}}, \pasa, 25, 167, \dodoi{10.1071/AS08013}

\bibitem[{{Graham}(2008{\natexlab{b}})}]{Graham:2008:a}
---. 2008{\natexlab{b}}, \apj, 680, 143, \dodoi{10.1086/587473}

\bibitem[{{Graham}(2014)}]{Graham:2014}
{Graham}, A.~W. 2014, in Astronomical Society of the Pacific Conference Series,
  Vol. 480, Structure and Dynamics of Disk Galaxies, ed. M.~S. {Seigar} \&
  P.~{Treuthardt}, 185.
\newblock \doarXiv{1311.7207}

\bibitem[{{Graham}(2016)}]{Graham:2016:Review}
{Graham}, A.~W. 2016, in Astrophysics and Space Science Library, Vol. 418,
  Galactic Bulges, ed. E.~{Laurikainen}, R.~{Peletier}, \& D.~{Gadotti}, 263,
  \dodoi{10.1007/978-3-319-19378-6_11}

\bibitem[{{Graham}(2019{\natexlab{a}})}]{Graham:Grid:2019}
---. 2019{\natexlab{a}}, \mnras, 1547, \dodoi{10.1093/mnras/stz1623}

\bibitem[{{Graham}(2019{\natexlab{b}})}]{Graham:Re:2019}
---. 2019{\natexlab{b}}, \pasa, 36, e035, \dodoi{10.1017/pasa.2019.23}

\bibitem[{{Graham} {et~al.}(2016{\natexlab{a}}){Graham}, {Ciambur}, \&
  {Savorgnan}}]{Graham:Ciambur:Savorgnan:2016}
{Graham}, A.~W., {Ciambur}, B.~C., \& {Savorgnan}, G. A.~D. 2016{\natexlab{a}},
  \apj, 831, 132, \dodoi{10.3847/0004-637X/831/2/132}

\bibitem[{{Graham} \& {Driver}(2007)}]{Graham:Driver:2007}
{Graham}, A.~W., \& {Driver}, S.~P. 2007, \apj, 655, 77, \dodoi{10.1086/509758}

\bibitem[{{Graham} {et~al.}(2016{\natexlab{b}}){Graham}, {Durr{\'e}},
  {Savorgnan}, {Medling}, {Batcheldor}, {Scott}, {Watson}, \&
  {Marconi}}]{Graham:Durr:Savorgnan:2016}
{Graham}, A.~W., {Durr{\'e}}, M., {Savorgnan}, G. A.~D., {et~al.}
  2016{\natexlab{b}}, \apj, 819, 43, \dodoi{10.3847/0004-637X/819/1/43}

\bibitem[{{Graham} \& {Guzm{\'a}n}(2003)}]{Graham:Guzman:2003}
{Graham}, A.~W., \& {Guzm{\'a}n}, R. 2003, \aj, 125, 2936,
  \dodoi{10.1086/374992}

\bibitem[{{Graham} \& {Scott}(2013)}]{Graham:Scott:2013}
{Graham}, A.~W., \& {Scott}, N. 2013, \apj, 764, 151,
  \dodoi{10.1088/0004-637X/764/2/151}

\bibitem[{{Graham} \& {Soria}(2019)}]{Graham:Soria:2018}
{Graham}, A.~W., \& {Soria}, R. 2019, \mnras, 484, 794,
  \dodoi{10.1093/mnras/sty3398}

\bibitem[{{Graham} {et~al.}(2019){Graham}, {Soria}, \&
  {Davis}}]{Graham:Soria:Davis:2018}
{Graham}, A.~W., {Soria}, R., \& {Davis}, B.~L. 2019, \mnras, 484, 814,
  \dodoi{10.1093/mnras/sty3068}

\bibitem[{{Graham} {et~al.}(2003)}]{Graham:2003:CS}
{Graham}, A.~W., {et~al.} 2003, \aj, 125, 2951, \dodoi{10.1086/375320}

\bibitem[{{Graham} {et~al.}(2007)}]{Graham:MGC:2007}
---. 2007, \mnras, 378, 198, \dodoi{10.1111/j.1365-2966.2007.11770.x}

\bibitem[{{Graham} {et~al.}(2011)}]{Graham:2011}
---. 2011, \mnras, 412, 2211, \dodoi{10.1111/j.1365-2966.2010.18045.x}

\bibitem[{{Greene} \& {Ho}(2006)}]{Greene:Ho:2006}
{Greene}, J.~E., \& {Ho}, L.~C. 2006, \apjl, 641, L21, \dodoi{10.1086/500507}

\bibitem[{{Greene} {et~al.}(2010){Greene}, {Peng}, {Kim}, {Kuo}, {Braatz},
  {Impellizzeri}, {Condon}, {Lo}, {Henkel}, \& {Reid}}]{Greene:2010}
{Greene}, J.~E., {Peng}, C.~Y., {Kim}, M., {et~al.} 2010, \apj, 721, 26,
  \dodoi{10.1088/0004-637X/721/1/26}

\bibitem[{{Greene} {et~al.}(2016){Greene}, {Seth}, {Kim}, {L{\"a}sker},
  {Goulding}, {Gao}, {Braatz}, {Henkel}, {Condon}, {Lo}, \&
  {Zhao}}]{Greene:2016}
{Greene}, J.~E., {Seth}, A., {Kim}, M., {et~al.} 2016, \apj, 826, L32,
  \dodoi{10.3847/2041-8205/826/2/L32}

\bibitem[{{G{\"u}ltekin} {et~al.}(2014){G{\"u}ltekin}, {Gebhardt}, {Kormendy},
  {Lauer}, {Bender}, {Tremaine}, \& {Richstone}}]{Gultekin:2014}
{G{\"u}ltekin}, K., {Gebhardt}, K., {Kormendy}, J., {et~al.} 2014, \apj, 781,
  112, \dodoi{10.1088/0004-637X/781/2/112}

\bibitem[{{G{\"u}ltekin} {et~al.}(2009{\natexlab{a}}){G{\"u}ltekin},
  {Richstone}, {Gebhardt}, {Lauer}, {Tremaine}, {Aller}, {Bender}, {Dressler},
  {Faber}, {Filippenko}, {Green}, {Ho}, {Kormendy}, {Magorrian}, {Pinkney}, \&
  {Siopis}}]{Gultekin:Richstone:2009}
{G{\"u}ltekin}, K., {Richstone}, D.~O., {Gebhardt}, K., {et~al.}
  2009{\natexlab{a}}, \apj, 698, 198, \dodoi{10.1088/0004-637X/698/1/198}

\bibitem[{{G{\"u}ltekin} {et~al.}(2009{\natexlab{b}}){G{\"u}ltekin},
  {Richstone}, {Gebhardt}, {Lauer}, {Pinkney}, {Aller}, {Bender}, {Dressler},
  {Faber}, {Filippenko}, {Green}, {Ho}, {Kormendy}, \&
  {Siopis}}]{Gultekin:2009}
---. 2009{\natexlab{b}}, \apj, 695, 1577, \dodoi{10.1088/0004-637X/695/2/1577}

\bibitem[{{Hartmann} {et~al.}(2014){Hartmann}, {Debattista}, {Cole}, {Valluri},
  {Widrow}, \& {Shen}}]{Hartmann:2014}
{Hartmann}, M., {Debattista}, V.~P., {Cole}, D.~R., {et~al.} 2014, \mnras, 441,
  1243, \dodoi{10.1093/mnras/stu627}

\bibitem[{{Heckman} \& {Best}(2014)}]{Heckman:2014}
{Heckman}, T.~M., \& {Best}, P.~N. 2014, \araa, 52, 589,
  \dodoi{10.1146/annurev-astro-081913-035722}

\bibitem[{{Held} {et~al.}(1992){Held}, {de Zeeuw}, {Mould}, \&
  {Picard}}]{Held:1992}
{Held}, E.~V., {de Zeeuw}, T., {Mould}, J., \& {Picard}, A. 1992, \aj, 103,
  851, \dodoi{10.1086/116106}

\bibitem[{{Hilz} {et~al.}(2013){Hilz}, {Naab}, \& {Ostriker}}]{Hilz:2013}
{Hilz}, M., {Naab}, T., \& {Ostriker}, J.~P. 2013, \mnras, 429, 2924,
  \dodoi{10.1093/mnras/sts501}

\bibitem[{{Hiner}(2012)}]{Hiner:2012}
{Hiner}, K.~D. 2012, PhD thesis, University of California, Riverside

\bibitem[{{Hobbs} \& {Dai}(2017)}]{Hobbs:2017}
{Hobbs}, G., \& {Dai}, S. 2017, arXiv e-prints, arXiv:1707.01615.
\newblock \doarXiv{1707.01615}

\bibitem[{{Hu}(2008)}]{Hu:2008}
{Hu}, J. 2008, \mnras, 386, 2242, \dodoi{10.1111/j.1365-2966.2008.13195.x}

\bibitem[{{Hur{\'e}} {et~al.}(2011){Hur{\'e}}, {Hersant}, {Surville}, {Nakai},
  \& {Jacq}}]{Hure:2011}
{Hur{\'e}}, J.~M., {Hersant}, F., {Surville}, C., {Nakai}, N., \& {Jacq}, T.
  2011, \aap, 530, A145, \dodoi{10.1051/0004-6361/201015062}

\bibitem[{{Jonas}(2007)}]{Jonas:2007}
{Jonas}, J. 2007, in From Planets to Dark Energy: the Modern Radio Universe, 7

\bibitem[{{Jorgensen} {et~al.}(1995){Jorgensen}, {Franx}, \&
  {Kjaergaard}}]{Jorgensen:1995}
{Jorgensen}, I., {Franx}, M., \& {Kjaergaard}, P. 1995, \mnras, 276, 1341,
  \dodoi{10.1093/mnras/276.4.1341}

\bibitem[{{King} \& {Minkowski}(1966)}]{King:Minkowski:1966}
{King}, I.~R., \& {Minkowski}, R. 1966, \apj, 143, 1002, \dodoi{10.1086/148580}

\bibitem[{{King} \& {Minkowski}(1972)}]{King:Minkowski:1972}
{King}, I.~R., \& {Minkowski}, R. 1972, in IAU Symposium, Vol.~44, External
  Galaxies and Quasi-Stellar Objects, ed. D.~S. {Evans}, D.~{Wills}, \& B.~J.
  {Wills}, 87

\bibitem[{{Kormendy} \& {Bender}(2013)}]{Kormendy:Bender:2013}
{Kormendy}, J., \& {Bender}, R. 2013, \apj, 769, L5,
  \dodoi{10.1088/2041-8205/769/1/L5}

\bibitem[{{Kormendy} {et~al.}(2009){Kormendy}, {Fisher}, {Cornell}, \&
  {Bender}}]{Kormendy:Fisher:2009}
{Kormendy}, J., {Fisher}, D.~B., {Cornell}, M.~E., \& {Bender}, R. 2009, \apjs,
  182, 216, \dodoi{10.1088/0067-0049/182/1/216}

\bibitem[{{Kormendy} \& {Ho}(2013)}]{Kormendy:Ho:2013}
{Kormendy}, J., \& {Ho}, L.~C. 2013, Annual Review of Astronomy and
  Astrophysics, 51, 511, \dodoi{10.1146/annurev-astro-082708-101811}

\bibitem[{{Kormendy} \& {Kennicutt}(2004)}]{Kormendy:Kennicutt:2004}
{Kormendy}, J., \& {Kennicutt}, Robert~C., J. 2004, \araa, 42, 603,
  \dodoi{10.1146/annurev.astro.42.053102.134024}

\bibitem[{{Lauer} {et~al.}(2007){Lauer}, {Faber}, {Richstone}, {Gebhardt},
  {Tremaine}, {Postman}, {Dressler}, {Aller}, {Filippenko}, {Green}, {Ho},
  {Kormendy}, {Magorrian}, \& {Pinkney}}]{Lauer:Faber:2007}
{Lauer}, T.~R., {Faber}, S.~M., {Richstone}, D., {et~al.} 2007, \apj, 662, 808,
  \dodoi{10.1086/518223}

\bibitem[{{Liller}(1966)}]{Liller:1966}
{Liller}, M.~H. 1966, \apj, 146, 28, \dodoi{10.1086/148857}

\bibitem[{{Malumuth} \& {Kirshner}(1981)}]{Malumuth:Kirshner:1981}
{Malumuth}, E.~M., \& {Kirshner}, R.~P. 1981, \apj, 251, 508,
  \dodoi{10.1086/159490}

\bibitem[{{Marconi} {et~al.}(2008){Marconi}, {Axon}, {Maiolino}, {Nagao},
  {Pastorini}, {Pietrini}, {Robinson}, \& {Torricelli}}]{Marconi:2008}
{Marconi}, A., {Axon}, D.~J., {Maiolino}, R., {et~al.} 2008, \apj, 678, 693,
  \dodoi{10.1086/529360}

\bibitem[{{Markwardt}(2012)}]{Markwardt:2012}
{Markwardt}, C. 2012, {MPFIT: Robust non-linear least squares curve fitting}.
\newblock \doeprint{1208.019}

\bibitem[{{Matkovi{\'c}} \& {Guzm{\'a}n}(2005)}]{Matkovic:Guzman:2005}
{Matkovi{\'c}}, A., \& {Guzm{\'a}n}, R. 2005, \mnras, 362, 289,
  \dodoi{10.1111/j.1365-2966.2005.09298.x}

\bibitem[{{McConnell} \& {Ma}(2013)}]{McConnell:Ma:2013}
{McConnell}, N.~J., \& {Ma}, C.-P. 2013, \apj, 764, 184,
  \dodoi{10.1088/0004-637X/764/2/184}

\bibitem[{{McLure} \& {Dunlop}(2004)}]{McLure:Dunlop:2004}
{McLure}, R.~J., \& {Dunlop}, J.~S. 2004, \mnras, 352, 1390,
  \dodoi{10.1111/j.1365-2966.2004.08034.x}

\bibitem[{{Meert} {et~al.}(2015){Meert}, {Vikram}, \&
  {Bernardi}}]{Meert:Vikram:2015}
{Meert}, A., {Vikram}, V., \& {Bernardi}, M. 2015, \mnras, 446, 3943,
  \dodoi{10.1093/mnras/stu2333}

\bibitem[{{Mehrgan} {et~al.}(2019){Mehrgan}, {Thomas}, {Saglia}, {Mazzalay},
  {Erwin}, {Bender}, {Kluge}, \& {Fabricius}}]{Mehrgan:2019}
{Mehrgan}, K., {Thomas}, J., {Saglia}, R., {et~al.} 2019, arXiv e-prints,
  arXiv:1907.10608.
\newblock \doarXiv{1907.10608}

\bibitem[{{Meidt} {et~al.}(2014){Meidt}, {Schinnerer}, {van de Ven},
  {Zaritsky}, {Peletier}, {Knapen}, {Sheth}, {Regan}, {Querejeta},
  {Mu{\~n}oz-Mateos}, {Kim}, {Hinz}, {Gil de Paz}, {Athanassoula}, {Bosma},
  {Buta}, {Cisternas}, {Ho}, {Holwerda}, {Skibba}, {Laurikainen}, {Salo},
  {Gadotti}, {Laine}, {Erroz-Ferrer}, {Comer{\'o}n}, {Men{\'e}ndez-Delmestre},
  {Seibert}, \& {Mizusawa}}]{Meidt:2014}
{Meidt}, S.~E., {Schinnerer}, E., {van de Ven}, G., {et~al.} 2014, \apj, 788,
  144, \dodoi{10.1088/0004-637X/788/2/144}

\bibitem[{{Merritt} \& {Ferrarese}(2001)}]{Merritt:Ferrarese:2001}
{Merritt}, D., \& {Ferrarese}, L. 2001, \apj, 547, 140, \dodoi{10.1086/318372}

\bibitem[{{Merritt} \& {Milosavljevi{\'c}}(2005)}]{Merritt:Milosavljevic:2005}
{Merritt}, D., \& {Milosavljevi{\'c}}, M. 2005, Living Reviews in Relativity,
  8, 8, \dodoi{10.12942/lrr-2005-8}

\bibitem[{{Minkowski}(1962)}]{Minknowski:1962}
{Minkowski}, R. 1962, in IAU Symposium, Vol.~15, Problems of Extra-Galactic
  Research, ed. G.~C. {McVittie}, 112

\bibitem[{{Molaeinezhad} {et~al.}(2019){Molaeinezhad}, {Zhu},
  {Falc{\'o}n-Barroso}, {van de Ven}, {M{\'e}ndez-Abreu}, {Balcells},
  {Aguerri}, {Vazdekis}, {Khosroshahi}, \& {Peletier}}]{Molaeinezhad:2019}
{Molaeinezhad}, A., {Zhu}, L., {Falc{\'o}n-Barroso}, J., {et~al.} 2019, \mnras,
  488, 1012, \dodoi{10.1093/mnras/stz1776}

\bibitem[{{Nemmen} {et~al.}(2012){Nemmen}, {Georganopoulos}, {Guiriec},
  {Meyer}, {Gehrels}, \& {Sambruna}}]{Nemmen:2012}
{Nemmen}, R.~S., {Georganopoulos}, M., {Guiriec}, S., {et~al.} 2012, Science,
  338, 1445, \dodoi{10.1126/science.1227416}

\bibitem[{{Nguyen} {et~al.}(2017){Nguyen}, {Seth}, {den Brok}, {Neumayer},
  {Cappellari}, {Barth}, {Caldwell}, {Williams}, \& {Binder}}]{Nguyen:2017}
{Nguyen}, D.~D., {Seth}, A.~C., {den Brok}, M., {et~al.} 2017, \apj, 836, 237,
  \dodoi{10.3847/1538-4357/aa5cb4}

\bibitem[{{Nguyen} {et~al.}(2018){Nguyen}, {Seth}, {Neumayer}, {Kamann},
  {Voggel}, {Cappellari}, {Picotti}, {Nguyen}, {B{\"o}ker}, \&
  {Debattista}}]{Nguyen:2018}
{Nguyen}, D.~D., {Seth}, A.~C., {Neumayer}, N., {et~al.} 2018, \apj, 858, 118,
  \dodoi{10.3847/1538-4357/aabe28}

\bibitem[{{Nguyen} {et~al.}(2019){Nguyen}, {den Brok}, {Seth}, {Iguchi},
  {Greene}, {Davis}, {Imanishi}, {Takuma}, {Izumi}, \&
  {Cappellari}}]{Nguyen:2019}
{Nguyen}, D.~D., {den Brok}, M., {Seth}, A.~C., {et~al.} 2019, arXiv e-prints,
  arXiv:1902.03813.
\newblock \doarXiv{1902.03813}

\bibitem[{{Novak} {et~al.}(2006){Novak}, {Faber}, \& {Dekel}}]{Novak:2006}
{Novak}, G.~S., {Faber}, S.~M., \& {Dekel}, A. 2006, \apj, 637, 96,
  \dodoi{10.1086/498333}

\bibitem[{{Nowak} {et~al.}(2007){Nowak}, {Saglia}, {Thomas}, {Bender},
  {Pannella}, {Gebhardt}, \& {Davies}}]{Nowak:2007}
{Nowak}, N., {Saglia}, R.~P., {Thomas}, J., {et~al.} 2007, \mnras, 379, 909,
  \dodoi{10.1111/j.1365-2966.2007.11949.x}

\bibitem[{{Onken} {et~al.}(2004){Onken}, {Ferrarese}, {Merritt}, {Peterson},
  {Pogge}, {Vestergaard}, \& {Wandel}}]{Onken:2004}
{Onken}, C.~A., {Ferrarese}, L., {Merritt}, D., {et~al.} 2004, \apj, 615, 645,
  \dodoi{10.1086/424655}

\bibitem[{{Oser} {et~al.}(2012){Oser}, {Naab}, {Ostriker}, \&
  {Johansson}}]{Oser:2012}
{Oser}, L., {Naab}, T., {Ostriker}, J.~P., \& {Johansson}, P.~H. 2012, \apj,
  744, 63, \dodoi{10.1088/0004-637X/744/1/63}

\bibitem[{{Paturel} {et~al.}(2003){Paturel}, {Petit}, {Prugniel}, {Theureau},
  {Rousseau}, {Brouty}, {Dubois}, \& {Cambr{\'e}sy}}]{Paturel:2003}
{Paturel}, G., {Petit}, C., {Prugniel}, P., {et~al.} 2003, \aap, 412, 45,
  \dodoi{10.1051/0004-6361:20031411}

\bibitem[{{Press} {et~al.}(1992){Press}, {Teukolsky}, {Vetterling}, \&
  {Flannery}}]{Press:1992}
{Press}, W.~H., {Teukolsky}, S.~A., {Vetterling}, W.~T., \& {Flannery}, B.~P.
  1992, {Numerical recipes in FORTRAN. The art of scientific computing}

\bibitem[{{Sabra} {et~al.}(2015){Sabra}, {Saliba}, {Abi Akl}, \&
  {Chahine}}]{Sabra:2015}
{Sabra}, B.~M., {Saliba}, C., {Abi Akl}, M., \& {Chahine}, G. 2015, \apj, 803,
  5, \dodoi{10.1088/0004-637X/803/1/5}

\bibitem[{{Saglia} {et~al.}(2016){Saglia}, {Opitsch}, {Erwin}, {Thomas},
  {Beifiori}, {Fabricius}, {Mazzalay}, {Nowak}, {Rusli}, \&
  {Bender}}]{Saglia:2016}
{Saglia}, R.~P., {Opitsch}, M., {Erwin}, P., {et~al.} 2016, \apj, 818, 47,
  \dodoi{10.3847/0004-637X/818/1/47}

\bibitem[{{Sahu} {et~al.}(2019){Sahu}, {Graham}, \& {Davis}}]{Sahu:2019:I}
{Sahu}, N., {Graham}, A.~W., \& {Davis}, B.~L. 2019, \apj, 876, 155,
  \dodoi{10.3847/1538-4357/ab0f32}

\bibitem[{{Salviander} {et~al.}(2007){Salviander}, {Shields}, {Gebhardt}, \&
  {Bonning}}]{Salviander:2007}
{Salviander}, S., {Shields}, G.~A., {Gebhardt}, K., \& {Bonning}, E.~W. 2007,
  \apj, 662, 131, \dodoi{10.1086/513086}

\bibitem[{{Savorgnan} \& {Graham}(2015)}]{Savorgnan:Graham:2015}
{Savorgnan}, G. A.~D., \& {Graham}, A.~W. 2015, \mnras, 446, 2330,
  \dodoi{10.1093/mnras/stu2259}

\bibitem[{{Savorgnan} \& {Graham}(2016)}]{Savorgnan:Graham:2016:I}
{Savorgnan}, G.~A.~D., \& {Graham}, A.~W. 2016, \apjs, 222, 10,
  \dodoi{10.3847/0067-0049/222/1/10}

\bibitem[{{Savorgnan} {et~al.}(2016)}]{Savorgnan:2016:Slopes}
{Savorgnan}, G. A.~D., {et~al.} 2016, \apj, 817, 21,
  \dodoi{10.3847/0004-637X/817/1/21}

\bibitem[{{Schechter}(1980)}]{Schechter:1980}
{Schechter}, P.~L. 1980, \aj, 85, 801, \dodoi{10.1086/112742}

\bibitem[{{S{\'e}rsic}(1963)}]{Sersic:1963}
{S{\'e}rsic}, J.~L. 1963, BAAA, 6, 41

\bibitem[{{Sexton} {et~al.}(2019){Sexton}, {Canalizo}, {Hiner}, {Komossa},
  {Woo}, {Treister}, \& {Hiner Dimassimo}}]{Sexton:2019}
{Sexton}, R.~O., {Canalizo}, G., {Hiner}, K.~D., {et~al.} 2019, \apj, 878, 101,
  \dodoi{10.3847/1538-4357/ab21d5}

\bibitem[{{Shankar} {et~al.}(2013){Shankar}, {Marulli}, {Bernardi}, {Mei},
  {Meert}, \& {Vikram}}]{Shankar:2013}
{Shankar}, F., {Marulli}, F., {Bernardi}, M., {et~al.} 2013, \mnras, 428, 109,
  \dodoi{10.1093/mnras/sts001}

\bibitem[{{Shankar} {et~al.}(2004){Shankar}, {Salucci}, {Granato}, {De Zotti},
  \& {Danese}}]{Shankar:Salucci:2004}
{Shankar}, F., {Salucci}, P., {Granato}, G.~L., {De Zotti}, G., \& {Danese}, L.
  2004, \mnras, 354, 1020, \dodoi{10.1111/j.1365-2966.2004.08261.x}

\bibitem[{{Shankar} {et~al.}(2016){Shankar}, {Bernardi}, {Sheth}, {Ferrarese},
  {Graham}, {Savorgnan}, {Allevato}, {Marconi}, {L{\"a}sker}, \&
  {Lapi}}]{Shankar:2016}
{Shankar}, F., {Bernardi}, M., {Sheth}, R.~K., {et~al.} 2016, \mnras, 460,
  3119, \dodoi{10.1093/mnras/stw678}

\bibitem[{{Shankar} {et~al.}(2019){Shankar}, {Bernardi}, {Richardson},
  {Marsden}, {Sheth}, {Allevato}, {Graziani}, {Mezcua}, {Ricci}, {Penny}, {La
  Franca}, \& {Pacucci}}]{Shankar:2019}
{Shankar}, F., {Bernardi}, M., {Richardson}, K., {et~al.} 2019, \mnras, 485,
  1278, \dodoi{10.1093/mnras/stz376}

\bibitem[{{Shannon} {et~al.}(2015){Shannon}, {Ravi}, {Lentati}, {Lasky},
  {Hobbs}, {Kerr}, {Manchester}, {Coles}, {Levin}, {Bailes}, {Bhat},
  {Burke-Spolaor}, {Dai}, {Keith}, {Os{\l}owski}, {Reardon}, {van Straten},
  {Toomey}, {Wang}, {Wen}, {Wyithe}, \& {Zhu}}]{Shannon:2015}
{Shannon}, R.~M., {Ravi}, V., {Lentati}, L.~T., {et~al.} 2015, Science, 349,
  1522, \dodoi{10.1126/science.aab1910}

\bibitem[{{Silk} \& {Rees}(1998)}]{Silk:Rees:1998}
{Silk}, J., \& {Rees}, M.~J. 1998, \aap, 331, L1.
\newblock \doarXiv{astro-ph/9801013}

\bibitem[{{Thater} {et~al.}(2019){Thater}, {Krajnovic}, {Cappellari}, {Davis},
  {de Zeeuw}, {McDermid}, \& {Sarzi}}]{Thater:Krajnovic:2019}
{Thater}, S., {Krajnovic}, D., {Cappellari}, M., {et~al.} 2019, arXiv e-prints.
\newblock \doarXiv{1902.10175}

\bibitem[{{Thomas} {et~al.}(2016){Thomas}, {Ma}, {McConnell}, {Greene},
  {Blakeslee}, \& {Janish}}]{Thomas:2016}
{Thomas}, J., {Ma}, C.-P., {McConnell}, N.~J., {et~al.} 2016, \nat, 532, 340,
  \dodoi{10.1038/nature17197}

\bibitem[{{Tonry}(1981)}]{Tonry:1981}
{Tonry}, J.~L. 1981, \apjl, 251, L1, \dodoi{10.1086/183681}

\bibitem[{{Tremaine} {et~al.}(2002){Tremaine}, {Gebhardt}, {Bender}, {Bower},
  {Dressler}, {Faber}, {Filippenko}, {Green}, {Grillmair}, {Ho}, {Kormendy},
  {Lauer}, {Magorrian}, {Pinkney}, \& {Richstone}}]{Tremaine:ngc4742:2002}
{Tremaine}, S., {Gebhardt}, K., {Bender}, R., {et~al.} 2002, \apj, 574, 740,
  \dodoi{10.1086/341002}

\bibitem[{{van den Bosch}(2016)}]{van_den_Bosch:2016}
{van den Bosch}, R. C.~E. 2016, \apj, 831, 134,
  \dodoi{10.3847/0004-637X/831/2/134}

\bibitem[{{V{\'e}ron-Cetty} \& {V{\'e}ron}(2010)}]{AGN:Catalogue:2010}
{V{\'e}ron-Cetty}, M.~P., \& {V{\'e}ron}, P. 2010, \aap, 518, A10,
  \dodoi{10.1051/0004-6361/201014188}

\bibitem[{{Volonteri} \& {Ciotti}(2013)}]{Volonteri:2013}
{Volonteri}, M., \& {Ciotti}, L. 2013, \apj, 768, 29,
  \dodoi{10.1088/0004-637X/768/1/29}

\bibitem[{{Walsh} {et~al.}(2015){Walsh}, {van den Bosch}, {Gebhardt},
  {Yildirim}, {G{\"u}ltekin}, {Husemann}, \& {Richstone}}]{Walsh:2015}
{Walsh}, J.~L., {van den Bosch}, R. C.~E., {Gebhardt}, K., {et~al.} 2015, \apj,
  808, 183, \dodoi{10.1088/0004-637X/808/2/183}

\bibitem[{{Williams} {et~al.}(2010){Williams}, {Bureau}, \&
  {Cappellari}}]{Williams:2010}
{Williams}, M.~J., {Bureau}, M., \& {Cappellari}, M. 2010, \mnras, 409, 1330,
  \dodoi{10.1111/j.1365-2966.2010.17406.x}

\bibitem[{{Woo} {et~al.}(2013){Woo}, {Schulze}, {Park}, {Kang}, {Kim}, \&
  {Riechers}}]{Woo:2013}
{Woo}, J.-H., {Schulze}, A., {Park}, D., {et~al.} 2013, \apj, 772, 49,
  \dodoi{10.1088/0004-637X/772/1/49}

\bibitem[{{Woo} {et~al.}(2006){Woo}, {Treu}, {Malkan}, \& {Bland
  ford}}]{Woo:2006}
{Woo}, J.-H., {Treu}, T., {Malkan}, M.~A., \& {Bland ford}, R.~D. 2006, \apj,
  645, 900, \dodoi{10.1086/504586}

\bibitem[{{York} {et~al.}(2000){York}, {Adelman}, {Anderson}, {Anderson},
  {Annis}, {Bahcall}, {Bakken}, {Barkhouser}, {Bastian}, {Berman}, {Boroski},
  {Bracker}, {Briegel}, {Briggs}, {Brinkmann}, {Brunner}, {Burles}, {Carey},
  {Carr}, {Castander}, {Chen}, {Colestock}, {Connolly}, {Crocker}, {Csabai},
  {Czarapata}, {Davis}, {Doi}, {Dombeck}, {Eisenstein}, {Ellman}, {Elms},
  {Evans}, {Fan}, {Federwitz}, {Fiscelli}, {Friedman}, {Frieman}, {Fukugita},
  {Gillespie}, {Gunn}, {Gurbani}, {de Haas}, {Haldeman}, {Harris}, {Hayes},
  {Heckman}, {Hennessy}, {Hindsley}, {Holm}, {Holmgren}, {Huang}, {Hull},
  {Husby}, {Ichikawa}, {Ichikawa}, {Ivezi{\'c}}, {Kent}, {Kim}, {Kinney},
  {Klaene}, {Kleinman}, {Kleinman}, {Knapp}, {Korienek}, {Kron}, {Kunszt},
  {Lamb}, {Lee}, {Leger}, {Limmongkol}, {Lindenmeyer}, {Long}, {Loomis},
  {Loveday}, {Lucinio}, {Lupton}, {MacKinnon}, {Mannery}, {Mantsch}, {Margon},
  {McGehee}, {McKay}, {Meiksin}, {Merelli}, {Monet}, {Munn}, {Narayanan},
  {Nash}, {Neilsen}, {Neswold}, {Newberg}, {Nichol}, {Nicinski}, {Nonino},
  {Okada}, {Okamura}, {Ostriker}, {Owen}, {Pauls}, {Peoples}, {Peterson},
  {Petravick}, {Pier}, {Pope}, {Pordes}, {Prosapio}, {Rechenmacher}, {Quinn},
  {Richards}, {Richmond}, {Rivetta}, {Rockosi}, {Ruthmansdorfer}, {Sand ford},
  {Schlegel}, {Schneider}, {Sekiguchi}, {Sergey}, {Shimasaku}, {Siegmund},
  {Smee}, {Smith}, {Snedden}, {Stone}, {Stoughton}, {Strauss}, {Stubbs},
  {SubbaRao}, {Szalay}, {Szapudi}, {Szokoly}, {Thakar}, {Tremonti}, {Tucker},
  {Uomoto}, {Vanden Berk}, {Vogeley}, {Waddell}, {Wang}, {Watanabe},
  {Weinberg}, {Yanny}, {Yasuda}, \& {SDSS Collaboration}}]{York:2000}
{York}, D.~G., {Adelman}, J., {Anderson}, John~E., J., {et~al.} 2000, \aj, 120,
  1579, \dodoi{10.1086/301513}

\bibitem[{{Yu} {et~al.}(2019){Yu}, {Bian}, {Wang}, {Zhao}, \& {Ge}}]{Yu:2019}
{Yu}, L.-M., {Bian}, W.-H., {Wang}, C., {Zhao}, B.-X., \& {Ge}, X. 2019,
  \mnras, 488, 1519, \dodoi{10.1093/mnras/stz1766}

\end{thebibliography}

\appendix
\vspace*{-1.2cm}
\startlongtable
\begin{deluxetable*}{lcrrcccrr}
\tabletypesize{\footnotesize}
\tablecolumns{9}
\tablecaption{Black Hole Mass versus Velocity Dispersion \label{Extra_fit_parameters} [ $\log(M_{\rm BH}/{\rm M_{\sun}})=\alpha\log(\sigma/200)+\beta$ ] }
\tablehead{
\colhead{ \textbf{Regression} } & \colhead{ \textbf{Minimization} } & \colhead{ \bm{$\alpha$} } & \colhead{ \bm{$\beta$ } } & \colhead{ \bm{$\epsilon$} } & \colhead{ \bm{ $\Delta_{rms | BH}$} } & \colhead{} & \colhead{ \bm{$r$ } } & \colhead{ \bm{$r_s$}}  \\
\colhead{} & \colhead{} & \colhead{} & \colhead{ \textbf{(dex)}} & \colhead{\textbf{(dex)}} & \colhead{\textbf{(dex)}} & \colhead{} & \colhead{} & \colhead{} \\
\colhead{ \textbf{(1)}} & \colhead{\textbf{(2)}} & \colhead{\textbf{(3)}} & \colhead{\textbf{(4)}} & \colhead{\textbf{(5)}} & \colhead{\textbf{(6)}} & \colhead{} & \colhead{\textbf{(7)}} & \colhead{\textbf{(8)}} 
}
\startdata
\multicolumn{9}{c}{ \textbf{Early-Type and Late-Type Galaxies}} \\
\hline
\multicolumn{9}{c}{91 Early-Type Galaxies} \\
\hline
\textsc{bces}$(Bisector)$ & \textit{Symmetric} & $5.71\pm 0.33$ & $8.32\pm 0.05$ & $0.32$ & $0.44$ & \multirow{3}{*}{$\left\}\begin{tabular}{@{}l@{}} \\ \\ \\ \end{tabular}\right.$} & \multirow{3}{*}{$0.86$}  & \multirow{3}{*}{$0.85$}  \\
\textsc{bces}$(M_{\rm BH}| \sigma)$ & $M_{\rm BH}$ & $5.22\pm0.36$ & $8.34\pm0.05$ & $0.32$ & $0.43$ & & \\
\textsc{bces}$(\sigma|M_{\rm BH})$ & $\sigma$ & $6.29\pm0.35$ & $8.29\pm0.06$ & $0.34$ & $0.47$ & & \\
\hline
\multicolumn{9}{c}{46 Late-Type Galaxies} \\
\hline
\textsc{bces}$(Bisector)$ & \textit{Symmetric} & $5.82\pm0.75$ & $8.17\pm0.14$ & $0.57$ & $0.63$ & \multirow{3}{*}{$\left\}\begin{tabular}{@{}l@{}} \\ \\ \\ \end{tabular}\right.$} & \multirow{3}{*}{$0.59$}  & \multirow{3}{*}{$0.49$}  \\
\textsc{bces}$(M_{\rm BH}| \sigma)$ & $M_{\rm BH}$ & $4.07\pm 0.90$ & $7.90\pm 0.17$ & $0.54$ & $0.58$ & & \\
\textsc{bces}$(\sigma | M_{\rm BH})$ & $\sigma$ & $10.06\pm 1.74$ & $8.83\pm 0.30$ & $0.85$ & $0.96$ & & \\
\hline
\multicolumn{9}{c}{Single Regression on (137) Early and Late-Type Galaxies} \\
\hline
\textsc{bces}$(Bisector)$ & \textit{Symmetric} & $6.10\pm0.28$ & $8.27\pm0.04$ & $0.43$ & $0.53$ & \multirow{3}{*}{$\left\}\begin{tabular}{@{}l@{}} \\ \\ \\ \end{tabular}\right.$} & \multirow{3}{*}{$0.86$} & \multirow{3}{*}{$0.87$}  \\
\textsc{bces}$(M_{\rm BH}| \sigma)$ & $M_{\rm BH}$ & $5.50\pm 0.29$ & $8.26\pm0.04$ & $0.42$ & $0.51$ & & \\
\textsc{bces}$(\sigma |M_{\rm BH})$ & $\sigma$ & $6.82\pm 0.32$ & $8.29\pm0.05$ & $0.46$ & $0.58$ & & \\
\hline
\multicolumn{9}{c}{ \textbf{S{\'e}rsic and Core-S{\'e}rsic Galaxies}} \\
\hline
\multicolumn{9}{c}{ 102 S{\'e}rsic Galaxies} \\
\hline
\textsc{bces}$(Bisector)$ & \textit{Symmetric} & $5.75\pm0.34$ & $8.24\pm0.05$ & $0.46$ & $0.55$ & \multirow{3}{*}{$\left\}\begin{tabular}{@{}l@{}} \\ \\ \\ \end{tabular}\right.$} & \multirow{3}{*}{$0.78$} & \multirow{3}{*}{$0.78$} \\
\textsc{bces}$(M_{\rm BH}|\sigma)$ & $M_{\rm BH}$ & $4.86\pm 0.34$ & $8.16\pm0.05$ & $0.45$ & $0.52$ & & \\
\textsc{bces}$(\sigma|M_{\rm BH})$ & $\sigma$ & $7.02\pm0.52$ & $8.34\pm0.07$ & $0.54$ & $0.64$ & & \\
\hline
\multicolumn{9}{c}{35 Core-S{\'e}rsic Galaxies} \\
\hline
\textsc{bces}$(Bisector)$ & \textit{Symmetric} & $8.64\pm 1.10$ & $7.91\pm0.20$ & $0.25$ & $0.46$ & \multirow{3}{*}{$\left\}\begin{tabular}{@{}l@{}} \\ \\ \\ \end{tabular}\right.$} & \multirow{3}{*}{$0.73$} & \multirow{3}{*}{$0.65$} \\
\textsc{bces}$(M_{\rm BH}| \sigma)$ & $M_{\rm BH}$ & $7.74\pm 1.15$ & $8.04\pm0.18$ & $0.25$ & $0.43$ & & \\
\textsc{bces}$(\sigma |M_{\rm BH})$ & $\sigma$ & $9.77\pm1.70$ & $7.74\pm0.31$ & $0.27$ & $0.52$ & & \\
\hline
\multicolumn{9}{c}{\textbf{ Galaxies with and without a disk}} \\
\hline
\multicolumn{9}{c}{93 ES, S0, Sp-Type Galaxies} \\
\hline
\textsc{bces}$(Bisector)$ & \textit{Symmetric} & $5.72\pm0.34$ & $8.22\pm0.06$ & $0.47$ & $0.56$ & \multirow{3}{*}{$\left\}\begin{tabular}{@{}l@{}} \\ \\ \\ \end{tabular}\right.$} & \multirow{3}{*}{$0.79$} & \multirow{3}{*}{$0.78$} \\
\textsc{bces}$(M_{\rm BH}| sigma)$ & $M_{\rm BH}$ & $4.86\pm0.35$ & $8.15\pm0.05$ & $0.45$ & $0.53$ & & \\
\textsc{bces}$(\sigma |M_{\rm BH})$ & $\sigma $ & $6.94\pm0.51$ & $8.32\pm0.07$ & $0.54$ & $0.64$ & & \\
\hline
\multicolumn{9}{c}{ 44 E-Type Galaxies} \\
\hline
\textsc{bces}$(Bisector)$ & \textit{Symmetric} & $6.69\pm0.59$ & $8.25\pm0.10$ & $0.30$ & $0.43$ & \multirow{3}{*}{$\left\}\begin{tabular}{@{}l@{}} \\ \\ \\ \end{tabular}\right.$} & \multirow{3}{*}{$0.82$} & \multirow{3}{*}{$0.80$}  \\
\textsc{bces}$(M_{\rm BH}|\sigma)$ & $M_{\rm BH}$ & $6.05\pm0.67$ & $8.32\pm0.10$ & $0.29$ & $0.41$ & & \\
\textsc{bces}$(\sigma|M_{\rm BH})$ & $\sigma$ & $7.47\pm0.69$ & $8.16\pm0.12$ & $0.32$ & $0.47$ & & \\
\hline
\multicolumn{9}{c}{\textbf{ Galaxies with and without a bar}} \\
\hline
\multicolumn{9}{c}{ 50 Barred Galaxies} \\
\hline
\textsc{bces}$(Bisector)$ & \textit{Symmetric} & $5.30\pm0.54$ & $8.14\pm0.10$ & $0.45$ & $0.53$ & \multirow{3}{*}{$\left\}\begin{tabular}{@{}l@{}} \\ \\ \\ \end{tabular}\right.$} & \multirow{3}{*}{$0.65$}  & \multirow{3}{*}{$0.61$}  \\
\textsc{bces}$(M_{\rm BH}| \sigma)$ & $M_{\rm BH}$ & $3.97\pm0.59$ & $7.97\pm0.10$ & $0.43$ & $0.49$ & & \\
\textsc{bces}$(\sigma |M_{\rm BH})$ & $\sigma$ & $7.86\pm 1.30$ & $8.48\pm 0.19$ & $0.61$ & $0.71$ & &  \\
\hline
\multicolumn{9}{c}{87 Non-Barred Galaxies} \\
\hline
\textsc{bces}$(Bisector)$ & \textit{Symmetric} & $6.16\pm0.42$ & $8.28\pm 0.06$ & $0.40$ & $0.51$ & \multirow{3}{*}{$\left\}\begin{tabular}{@{}l@{}} \\ \\ \\ \end{tabular}\right.$} & \multirow{3}{*}{$0.86$} & \multirow{3}{*}{$0.86$}  \\
\textsc{bces}$(M_{\rm BH}| \sigma)$ & $M_{\rm BH}$ & $5.57\pm0.43$ & $8.30\pm0.06$ & $0.40$ & $0.49$ & & \\
\textsc{bces}$(\sigma |M_{\rm BH})$ & $\sigma$ & $6.88\pm0.45$ & $8.25\pm0.07$ & $0.44$ & $0.55$ & &  \\
\hline
\multicolumn{9}{c}{\textbf{ Galaxies with and without an AGN}} \\
\hline
\multicolumn{9}{c}{ 41 AGN host Galaxies} \\
\hline
\textsc{bces}$(Bisector)$ & \textit{Symmetric} & $6.26\pm0.49$ & $8.21\pm0.09$ & $0.55$ & $0.63$ & \multirow{3}{*}{$\left\}\begin{tabular}{@{}l@{}} \\ \\ \\ \end{tabular}\right.$} & \multirow{3}{*}{$0.83$}  & \multirow{3}{*}{$0.79$}  \\
\textsc{bces}$(M_{\rm BH}| \sigma)$ & $M_{\rm BH}$ & $5.37\pm0.51$ & $8.16\pm0.09$ & $0.53$ & $0.60$ & & \\
\textsc{bces}$(\sigma |M_{\rm BH})$ & $\sigma$ & $7.48\pm 0.66$ & $8.28\pm 0.10$ & $0.63$ & $0.72$ & & \\
\hline
\multicolumn{9}{c}{96 Galaxies without AGN} \\
\hline
\textsc{bces}$(Bisector)$ & \textit{Symmetric} & $5.92\pm0.31$ & $8.30\pm 0.05$ & $0.37$ & $0.48$ & \multirow{3}{*}{$\left\}\begin{tabular}{@{}l@{}} \\ \\ \\ \end{tabular}\right.$} & \multirow{3}{*}{$0.87$} & \multirow{3}{*}{$0.88$}  \\
\textsc{bces}$(M_{\rm BH}| \sigma)$ & $M_{\rm BH}$ & $5.43\pm0.33$ & $8.29\pm0.05$ & $0.37$ & $0.46$ & & \\
\textsc{bces}$(\sigma |M_{\rm BH})$ & $\sigma$ & $6.51\pm0.33$ & $8.30\pm0.05$ & $0.39$ & $0.51$ & & \\
\enddata
\tablecomments{
Columns:
(1) Type of regression performed.
(2) The coordinate direction in which the offsets from the regression line is minimized.
(3) Slope of the regression line.
(4) Intercept of the regression line.
(5) Intrinsic scatter in the $\log(M_{\rm BH})$-direction \citep[using Equation 1 from][]{Graham:Driver:2007}.
(6) Total root mean square (rms) scatter in the $\log(M_{\rm BH})$-direction.
(7) Pearson correlation coefficient.
(8) Spearman rank-order correlation coefficient.
}
\end{deluxetable*}
\(\)

\startlongtable
\begin{deluxetable*}{lcrrccrr}
\tabletypesize{\footnotesize}
\tablecolumns{8}
\tablecaption{Regression Lines Including All 143 Galaxies With Velocity Dispersions\label{Total_sample_para}
 [ $\log(M_{\rm BH}/{\rm M_{\sun}})=\alpha\log(\sigma/200)+\beta$ ] }
\tablehead{
\colhead{ \textbf{Category} } & \colhead{ \textbf{Number} } & \colhead{ \bm{$\alpha$} } & \colhead{ \bm{$\beta$ } } & \colhead{ \bm{$\epsilon$} } & \colhead{ \bm{ $\Delta_{rms | BH}$} }  & \colhead{ \bm{$r$ } } & \colhead{ \bm{$r_s$}}  \\
\colhead{} & \colhead{} & \colhead{} & \colhead{ \textbf{(dex)}} & \colhead{\textbf{(dex)}} & \colhead{\textbf{(dex)}} & \colhead{} & \colhead{} \\
\colhead{ \textbf{(1)}} & \colhead{\textbf{(2)}} & \colhead{\textbf{(3)}} & \colhead{\textbf{(4)}} & \colhead{\textbf{(5)}} & \colhead{\textbf{(6)}} & \colhead{\textbf{(7)}} & \colhead{\textbf{(8)}} 
}
\startdata
Early-Type Galaxies & 95 & $5.05\pm 0.26$ & $8.37\pm 0.04$ & $0.33$ & $0.44$ & $0.90$  & $0.87$ \\
Late-Type Galaxies & 48 & $4.47\pm 0.80$ & $8.04\pm 0.15$ & $0.67$ & $0.70$ & $0.56$  & $0.46$  \\
\hline
All Galaxies & 143 & $5.29\pm 0.32$ & $8.30\pm0.04$ & $0.50$ & $0.58$ & $0.86$ & $0.86$  \\
\hline
S{\'e}rsic Galaxies & 108 & $4.83 \pm 0.35$ & $8.22\pm0.06$ & $0.52$ & $0.59$ & $0.80$ & $0.77$ \\
Core-S{\'e}rsic Galaxies & 35 & $8.50\pm 1.10$ & $7.91\pm0.20$ & $0.25$ & $0.46$ & $0.73$ & $0.65$ \\
\hline
Galaxies with a disk (ES, S0, Sp-types) & 98 & $4.90\pm 0.38$ & $8.21\pm 0.06$ & $0.54$ & $0.60$ & $0.79$ & $0.76$ \\
Galaxies without a disk (E-type) & 45 & $5.41\pm 0.66$ & $8.40\pm 0.10$ & $0.31$ & $0.42$ & $0.88$ & $0.82$  \\
\hline
Barred Galaxies & 52 & $4.05 \pm 0.54$ & $8.01\pm0.10$ & $0.45$ & $0.51$ & $0.74$  & $0.66$  \\
Non-Barred Galaxies & 91 & $5.46\pm 0.34$ & $8.36\pm 0.06$ & $0.48$ & $0.55$ & $0.86$ & $0.86$  \\
\hline
AGN host Galaxies & 42 & $5.23\pm 0.75$ & $8.20\pm0.08$ & $0.62$ & $0.67$ & $0.82$  & $0.81$  \\
Galaxies without AGN & 101 & $5.26\pm 0.28$ & $8.34\pm 0.05$ & $0.44$ & $0.52$ & $0.87$ & $0.87$  \\
\enddata
\tablecomments{
Columns:
(1) Subclass of galaxies.
(2) Number of galaxies in a subclass.
(3) Slope of the line obtained from the \textsc{BCES(Bisector)} regression.
(4) Intercept of the line obtained from the \textsc{BCES(Bisector)} regression.
(5) Intrinsic scatter in the $\log(M_{\rm BH})$-direction \citep[using Equation 1 from][]{Graham:Driver:2007}.
(6) Total root mean square (rms) scatter in the $\log(M_{\rm BH})$ direction.
(7) Pearson correlation coefficient.
(8) Spearman rank-order correlation coefficient.
}
\end{deluxetable*}
\(\)

\startlongtable
\begin{deluxetable*}{lcrrcccrr}
\tabletypesize{\footnotesize}
\tablecolumns{9}
\tablecaption{Galaxy Luminosity versus Velocity Dispersion\label{L_sigma_parameters} [ $\log(L)=\alpha\log(\sigma/200)+\beta$ ] }
\tablehead{
\colhead{ \textbf{Regression} } & \colhead{ \textbf{Minimization} } & \colhead{ \bm{$\alpha$} } & \colhead{ \bm{$\beta$ } } & \colhead{ \bm{$\epsilon$} } & \colhead{ \bm{ $\Delta_{rms | L}$} } & \colhead{} & \colhead{ \bm{$r$ } } & \colhead{ \bm{$r_s$}}  \\
\colhead{} & \colhead{} & \colhead{} & \colhead{ \textbf{(dex)}} & \colhead{\textbf{(dex)}} & \colhead{\textbf{(dex)}} & \colhead{} & \colhead{} & \colhead{} \\
\colhead{ \textbf{(1)}} & \colhead{\textbf{(2)}} & \colhead{\textbf{(3)}} & \colhead{\textbf{(4)}} & \colhead{\textbf{(5)}} & \colhead{\textbf{(6)}} & \colhead{} & \colhead{\textbf{(7)}} & \colhead{\textbf{(8)}} 
}
\startdata
\multicolumn{9}{c}{ \textbf{V-band}} \\
\hline
\multicolumn{9}{c}{97 Core-S\'ersic ETGs} \\
\hline
\textsc{bces}$(Bisector)$ & \textit{Symmetric} & $4.86\pm 0.54$ & $8.52\pm 0.07$ & $0.30$ & $0.37$ & \multirow{3}{*}{$\left\}\begin{tabular}{@{}l@{}} \\ \\ \\ \end{tabular}\right.$} & \multirow{3}{*}{$0.52$}  & \multirow{3}{*}{$0.53$}  \\
\textsc{bces}$(L| \sigma)$ & $L$ & $3.38\pm0.48$ & $8.70\pm0.06$ & $0.28$ & $0.32$ & & \\
\textsc{bces}$(\sigma|L)$ & $\sigma$ & $8.55\pm1.53$ & $8.08\pm0.19$ & $0.44$ & $0.58$ & & \\
\hline
\multicolumn{9}{c}{80 S\'ersic ETGs} \\
\hline
\textsc{bces}$(Bisector)$ & \textit{Symmetric} & $2.44\pm0.18$ & $8.41\pm0.04$ & $0.28$ & $0.31$ & \multirow{3}{*}{$\left\}\begin{tabular}{@{}l@{}} \\ \\ \\ \end{tabular}\right.$} & \multirow{3}{*}{$0.73$}  & \multirow{3}{*}{$0.69$}  \\
\textsc{bces}$(L| \sigma)$ & $L$ & $1.93\pm 0.18$ & $8.35\pm 0.04$ & $0.27$ & $0.29$ & & \\
\textsc{bces}$(\sigma | L)$ & $\sigma$ & $3.30\pm 0.36$ & $8.51\pm 0.05$ & $0.35$ & $0.38$ & & \\
\hline
\multicolumn{9}{c}{\textbf{ 3.6 $\, \mu \rm m$}} \\
\hline
\multicolumn{9}{c}{ 24 Core-S\'ersic ETGs} \\
\hline
\textsc{bces}$(Bisector)$ & \textit{Symmetric} & $5.16\pm0.53$ & $8.56\pm0.08$ & $0.00$ & $0.19$ & \multirow{3}{*}{$\left\}\begin{tabular}{@{}l@{}} \\ \\ \\ \end{tabular}\right.$} & \multirow{3}{*}{$0.86$}  & \multirow{3}{*}{$0.76$}  \\
\textsc{bces}$(L| \sigma)$ & $L$ & $5.48\pm0.70$ & $8.51\pm0.11$ & $0.00$ & $0.20$ & & \\
\textsc{bces}$(\sigma |L)$ & $\sigma$ & $4.86\pm 0.47$ & $8.60\pm 0.07$ & $0.00$ & $0.18$ & & \\
\hline
\multicolumn{9}{c}{42 S\'ersic ETGs} \\
\hline
\textsc{bces}$(Bisector)$ & \textit{Symmetric} & $2.97\pm0.43$ & $8.72\pm 0.07$ & $0.33$ & $0.36$ & \multirow{3}{*}{$\left\}\begin{tabular}{@{}l@{}} \\ \\ \\ \end{tabular}\right.$} & \multirow{3}{*}{$0.61$} & \multirow{3}{*}{$0.61$}  \\
\textsc{bces}$(L| \sigma)$ & $L$ & $2.10\pm0.40$ & $8.68\pm0.06$ & $0.32$ & $0.33$ & & \\
\textsc{bces}$(\sigma |L)$ & $\sigma$ & $5.04\pm0.92$ & $8.81\pm0.09$ & $0.49$ & $0.53$ & & \\
\hline
\multicolumn{9}{c}{24 LTGs (All S\'ersic)} \\
\hline
\textsc{bces}$(Bisector)$ & \textit{Symmetric} & $2.10\pm0.41$ & $8.90\pm 0.09$ & $0.17$ & $0.20$ & \multirow{3}{*}{$\left\}\begin{tabular}{@{}l@{}} \\ \\ \\ \end{tabular}\right.$} & \multirow{3}{*}{$0.70$} & \multirow{3}{*}{$0.68$}  \\
\textsc{bces}$(L| \sigma)$ & $L$ & $1.64\pm0.44$ & $8.83\pm0.10$ & $0.16$ & $0.18$ & & \\
\textsc{bces}$(\sigma |L)$ & $\sigma$ & $2.89\pm0.42$ & $9.03\pm0.08$ & $0.21$ & $0.25$ & & \\
\enddata
\tablecomments{
Columns:
(1) Type of regression performed.
(2) The coordinate direction in which the offsets from the regression line is minimized.
(3) Slope of the regression line.
(4) Intercept of the regression line.
(5) Intrinsic scatter in the $\log(L)$-direction \citep[using Equation 1 from][]{Graham:Driver:2007}.
(6) Total root mean square (rms) scatter in the $\log(L)$-direction.
(7) Pearson correlation coefficient.
(8) Spearman rank-order correlation coefficient.
}
\end{deluxetable*}
\(\)

\end{document}